\RequirePackage{rotating}
\documentclass[fleqn,usenatbib]{mnras}

\usepackage[T1]{fontenc}
\usepackage{units}

\DeclareRobustCommand{\VAN}[3]{#2}
\let\VANthebibliography\thebibliography
\def\thebibliography{\DeclareRobustCommand{\VAN}[3]{##3}\VANthebibliography}

\usepackage{graphicx}	
\usepackage{amsmath}	
\usepackage{amssymb}	
\usepackage{newtxtext,newtxmath}
\usepackage{multirow}
\usepackage{lscape}
\usepackage{makecell}
\usepackage{subfigure}
\usepackage{float}
\usepackage{newtxtext,newtxmath}
\usepackage{lineno}
\setcitestyle{notesep={ }}
\makeatletter
\newcommand{\labtext}[2]{%
  \@bsphack
  \csname phantomsection\endcsname 
  \def\@currentlabel{#1}{\label{#2}}%
  \@esphack
}

\newcommand{\Msun}{{{\rm M}_\odot}}
\newcommand{\redcheck}{{\color{black}\checkmark}}
\newcommand{\myhash}{\#}

\setcitestyle{notesep={}}

\makeatother

\title[Search for Periodic Variability in Blazars]{Search for Periodic Variability in $\gamma$-ray Blazars Using \textit{Fermi}-LAT}
\author[P. Pe\~nil et al.]{
P. Pe\~nil,$^{1}$\thanks{E-mail: ppenil@clemson.edu}
M. Ajello,$^{1}$\thanks{E-mail: majello@clemson.edu}
S. Buson,$^{2,3}$\thanks{E-mail: sara.buson@uni-wuerzburg.de}
A. Dom\'inguez,$^{4}$\thanks{E-mail: alberto.d@ucm.es}
J.R. Westernacher-Schneider,$^{5}$
A. Rico,$^{1}$
\newauthor
S. Adhikari,$^{1}$
J. Zrake,$^{1}$
\\
$^{1}$Department of Physics and Astronomy, Clemson University, Kinard Lab of Physics, Clemson, SC 29634-0978, USA\\
$^{2}$Julius-Maximilians-Universität Würzburg, Fakultät für Physik und Astronomie, Emil-Fischer-Str. 31, D-97074 Würzburg, Germany\\
$^{3}$Deutsches Elektronen-Synchrotron DESY, Platanenallee 6, 15738 Zeuthen, Germany\\
$^{4}$IPARCOS and Department of EMFTEL, Universidad Complutense de Madrid, E-28040 Madrid, Spain\\
$^{5}$Westgate Research, PO Box 181, Cardiff, ON K0L 1M0, Canada
}

\date{Accepted 2025 July 4. Received 2025 July 1; in original form 2025 April 4}

\pubyear{2024}

\begin{document}
\label{firstpage}
\pagerange{\pageref{firstpage}--\pageref{lastpage}}
\maketitle
\begin{abstract}
Blazars are known to exhibit variability across a broad range of timescales. This behavior can include periodicity in their $\gamma$-ray emission, whose clear detection remains an ongoing challenge, partly due to the inherent stochasticity of the processes involved and also the lack of adequately well-sampled light curves. In this study, we perform a systematic search for periodicity in a selected sample of 24 $\gamma$-ray blazars using twelve years of Fermi-LAT data. The sample comprises the most promising candidates selected from a previous study, extending the light curves by three additional years, expanding the analyzed energy range from $>$1~GeV to $>$0.1~GeV to improve photon statistics, and enhancing the time-series analysis methodology. We incorporate upper-limit flux points in the analysis rather than discarding them, thereby preserving the temporal structure in the light curves. A suite of seven complementary time-series analysis methods is employed to ensure statistical robustness, including autoregressive models, representing a methodological advancement over the prior work. A further improvement is the explicit estimation of the look-elsewhere effect, which allows us to assess the global significance of any detected signals. The study is also supported by additional statistical treatments employed to minimize false detections and strengthen the reliability of the results. Our analysis reveals a hint of periodicity in PG 1553+113 with a global significance of $\approx$1.8$\sigma$. For the remaining sources in the sample, the re-evaluation of previously reported periodicities indicates that they are statistically consistent with arising from stochastic variability. 
\end{abstract}

\section{Introduction} \label{sec:intro}
Active Galactic Nuclei (AGN), i.e., galaxies with an accreting supermassive black hole (SMBH) in the center, are among the most luminous persistent objects in the Universe \citep[e.g.,~][]{soltan_1982, cavaliere_1989, wiita_lecture}. The complex interplay between the black hole hosted at the AGN core and the surrounding circumnuclear gas leads to the formation of an accretion disk and may influence the properties of the host galaxy \citep[e.g., the stellar mass of the inner bulge, ][]{haring_galaxy}. A small fraction of AGN is distinguished by the presence of a pair of highly collimated, relativistic jets. These jets typically form in opposite directions, perpendicular to the rotational plane of the SMBH-accretion disk system. When one of the jets is closely aligned with the line of sight, the object is referred to as a blazar. The observed emission from blazars is strongly dominated by the jet, spans energies ranging from radio to $\gamma$-rays, and can be highly variable on time scales ranging from seconds to years \citep[e.g.,][]{urry_variability,urry_multiwavelengh}.

During their cosmological evolution galaxies can merge \citep[e.g.,][]{jiang_merging}, frequently at moderate/high redshifts \citep[e.g.,][]{rieger_2007, merger_galaxies} when they are also more gas-rich \citep[e.g.,][]{tacconi_gas}. In the process of an AGN merger, one expects that the two SMBHs hosted at the respective centers eventually meet and form a binary system with a typical separation of about kiloparsec \citep{begelman_1980}. Such a binary system is referred to as a supermassive black hole binary (SMBHB). As a consequence of dynamical friction, the SMBHs can converge to the center of mass of the galaxy and shrink their separation down to sub-parsec distances \citep[e.g.,][]{dosopoulou_friction}. Finally, the last stage of merging is the coalescence of the SMBHBs into a single SMBH \citep[e.g.,][]{colpi_merge}, resulting in one of the loudest sources of gravitational waves in the Universe \citep[e.g.,][]{enoki_gravitational}.

The identification of SMBHB systems with sub-parsec separation is an important topic of astrophysics, but remains challenging since they cannot generally be spatially resolved and are rare. Limits on the nanohertz stochastic gravitational wave background from the Pulsar Timing Array \citep[PTA,][]{holgado_pta, rieger_2019} suggest that only a minor fraction ($\lesssim\,$0.01–0.1\%) of blazars could be harboring SMBHBs with sub-parsec separations \citep[e.g.,][]{liu_binary}. Such SMBHBs could imprint peculiar time-variability patterns in their emitted radiation with periods of one year.

One of the strategies employed to pinpoint candidate SMBHBs has been the search for periodic modulation, also known as quasi-periodic oscillations, in AGN light curves (LCs). The detection of periodicity in blazars has been limited by the lack of continuous and sensitive long-term sampling. Recently, the Large Area Telescope (LAT) on board the \textit{Fermi Gamma-ray Space Telescope} has alleviated this issue by scanning the entire sky regularly with high sensitivity \citep[][]{fermi_lat}. During more than a decade of continuous monitoring, since {\it Fermi} was launched in 2008, the LAT has collected a wealth of valuable data, offering the unprecedented opportunity to monitor flux variations from thousands of blazars in the GeV band. Densely sampled, unbiased, high-quality LCs are now accessible for all sources in the LAT catalogs. The fourth $Fermi$-LAT source catalog \citep[4FGL,][]{abdollahi_4fgl, ballet_4fgl} is based on the first 8 years of LAT observations and includes more than 5000 sources, among which about 3000 are confidently associated with blazar counterparts. Recently, the 4FGL-DR3 catalog (Data Release 3) has been published, which is based on the first 12 years of LAT observations \citep[][]{abdollahi_4fgl_dr3}.

Searching for periodicity in a time series (an ordered sequence of fluxes as a function of time) requires measuring the power of the time series at each frequency and identifying the dominant frequency, if any. This information is captured in periodograms, which are obtained through a variety of techniques such as Lomb-Scargle \citep[][]{lomb_1976, scargle_1982} and wavelets \citep[][]{foster_wwz, wavelet_torrence}. For the most part, studies in the literature that search for periodicity in blazar $\gamma$-ray LCs have mainly focused on a few selected objects and typically have employed at most two or three time-series algorithms \citep[e.g.,][]{zhang_pks2155, tavani_pdm_pg_1553, bhatta_s5_0716}. These previous studies found a few candidates with periodicity \citep[e.g., PG~1553+113 or OJ 287, ][, respectively]{ackermann_pg1553, valtonen_oj287} but due to the stochastic nature of emission in blazars, these findings remain under debate \citep[e.g.,][]{covino_negation}. As a pioneering investigation in \citet[][, P20, hereafter]{penil_2020}, we performed a comprehensive search for periodicities in blazar $\gamma$-ray LCs, targeting $\sim$2,000 blazars and adopting ten of the most used methods for time-series analysis. Our effort discovered 24 periodic blazar candidates with local (pre-trial) significance above $2.5\sigma$. 

In the work presented here, we reanalyze the 24 candidates from P20, employing new data that cover three additional years of \textit{Fermi}-LAT observations. We apply a similar periodicity-detection pipeline as the one described by P20 and include new statistical methods. Specifically, we also use the autoregressive models Autoregressive Integrated Moving Average (ARIMA) and Autoregressive Fractionally Integrated Moving Average (ARFIMA) to deal with the stochastic uncertainty introduced by noise in the periodicity search. Finally, we also take into account the look-elsewhere effect to provide robust significance for the detected periods.

The paper is organized as follows. In $\S$\ref{sec:fermidata}, the blazar sample and data reduction methodology are presented. Then, $\S$\ref{sec:methodology} details the periodicity analysis methodology, and $\S$\ref{sec:results} shows and discusses our results. In $\S$\ref{sec:discussion}, there is a description of the potential interpretation of the results focusing on the two sources with the most significant periodicity. We summarize the findings in $\S$\ref{sec:summary}.

\section{gamma-ray sample} \label{sec:fermidata}
\subsection{{\it Fermi}-LAT data reduction}
The $Fermi$-LAT data are reduced with the Python package \texttt{fermipy} \citep{Wood:2017yyb}. The following procedure is adopted for each source in the sample. We select photons of the \texttt{Pass 8 SOURCE} class \citep[][]{atwood_source_class, bruel_pass8}, in a region of interest (ROI) of 15$^\circ$ $\times$ 15$^\circ$ square, centered at the target. To minimize the contamination from $\gamma$ rays produced in the Earth’s upper atmosphere, a zenith angle cut of $\theta < 90^{\circ}$ is applied. We also applied the standard data quality cuts ($ \rm DATA\_QUAL > 0) \&\& (LAT\_CONFIG == 1$) and removed time periods coinciding with solar flares and $\gamma$-ray bursts detected by the LAT. The ROI model includes all 4FGL-DR2 catalog sources \citep[][]{4FGL_dr2} located within 20$^{\circ}$ from the ROI center, as well as the Galactic and isotropic diffuse emission \footnote{\url{https://fermi.gsfc.nasa.gov/ssc/data/access/lat/BackgroundModels.html}} (\texttt{gll\_iem\_v07.fits} and \texttt{iso\_P8R3\_SOURCE\_V2.txt}). 

We perform a binned analysis in the 0.1-800 GeV energy range, using 10 bins per decade in energy and 0.1$^{\circ}$ spatial bins, and adopting the \texttt{P8R3\_SOURCE\_V2} instrument response functions. First, a maximum likelihood analysis is performed over the full-time range considered here, i.e., 2008 Aug 04 15:43:36 UTC to 2020 Dec 10 00:01:26 UTC. In the fit, we model the sources in the ROI, adopting the spectral shapes and parameters reported in 4FGL. We allow the normalization and spectral index of the target source to vary, as well as the normalizations of all sources within 3$^{\circ}$ of the ROI center and the isotropic and Galactic diffuse components. Since our data span a different integration time with respect to 4FGL, our first results are checked for potential newly detected sources with an iterative procedure. To this aim, a test statistic (TS) map is produced. The TS is defined as $2\log(L/L_0)$, where
\textit{$L$} is the likelihood of the model with a point source at a given position and \textit{$L_0$} is the likelihood without the source. A TS value of 25 corresponds to a statistical significance of $\gtrsim4.0\sigma$~\citep[according to the prescription adopted in][]{mattox1996, abdollahi_4fgl}. A TS map is produced by including a putative point source at each pixel of the map and evaluating its significance over the current best-fit model. The test source is modeled with a power-law spectrum where only the normalization is allowed to vary in the fit, whereas the photon index is fixed at 2. We look for significant peaks (TS$>$25) in the TS map, with a minimum separation of 0.5$^{\circ}$  from existing sources in the model, and add a new point source to the model at the position of the most significant peak found. Then, the ROI is fitted again, and a new TS map is produced. This process is iterated until no more significant excesses are found, generally leading to the addition of two point sources. 

To produce the LCs, we split the data for each source into 28-day bins and perform a full likelihood fit in each time bin. For the likelihood fits of the time bins, the best-fit ROI model obtained from the full-time interval analysis is adopted. We first attempt a fit allowing variation in the normalizations of the target and of all sources in the inner $3^\circ$ of the ROI, along with the diffuse components. If the fit does not converge, the number of free parameters is progressively restricted in the fit in an iterative way until the fit successfully converges. We begin this iterative process by fixing sources in the ROI that are weakly detected (i.e., with TS$<$4). Next, we fix sources with TS$<$9. Then, we fix sources up to $1^\circ$ from the ROI center and those with TS$<$25. Finally, we fix all parameters except the target source’s normalization. We consider the target source to be detected when TS$>$1 in the corresponding time bin. Whenever this condition is not fulfilled, a 95\% confidence upper limit is reported from the likelihood distribution \citep[][]{penil_2025_trends}. These points are denoted by down arrows in the plots of the LCs (Figure \ref{fig:lc_candidates_low}).

\subsection{Source selection}
The blazar sample in P20 was analyzed using 9 years of \textit{Fermi}-LAT observations (see Table \ref{tab:candidates_list}). In this work, we reanalyze the 24 periodicity candidates similarly pinpointed in P20 but with several improvements:

\begin{enumerate}
\item Extend the observing time to 12 years, from August 2008 until December 2020. 
\item Expand the energy range from $\rm{>1}$ GeV to $\rm{>0.1}$ GeV, thus increasing the photon statistics, improving the signal-to-noise ratio, and dramatically reducing the number of upper limits in the LCs. For example, S4~1144+40 had a fraction of 40\% upper limits in P20, which is reduced to $<$1\% here. 
\item Retain the information for the LC points with low statistics (i.e., non-detection) by substituting the upper-limit data with the flux value that maximizes the likelihood function for that time bin \citep{penil_2025_trends}. The likelihood function is scanned in the same way for bins that result in a positive detection and for bins where typically an upper limit is reported (instead of the best fit flux with uncertainties). Here we use the best-fit flux (and its uncertainty) regardless of the statistical significance (TS) in that bin.   
\end{enumerate}

To evaluate whether the substitution of upper-limit data points introduces any bias into the periodicity analysis, we conduct a test using the two sources in our sample with the highest percentage of gaps in their LCs (see Table~\ref{tab:candidates_list}): 87GB164812.2+524023 (12\%) and MG2J130304+2434 (9.9\%). In this test, we remove all upper-limit points from the LCs. We then apply two of the methods from our analysis framework, the Lomb-Scargle Periodogram and Phase Dispersion Minimization, both of which are suitable for analyzing irregularly sampled time series.

The results obtained from the gappy LCs are consistent with those derived from the original versions, where upper limits are retained through substitution. Specifically, we obtain the same periodicities and similar values for the associated test statistics in both cases. These findings align with those reported in P20, where it was shown that the influence of observational gaps becomes statistically significant when the gap fraction reaches or exceeds $\sim$50\%. Therefore, for moderate levels of missing data, our substitution strategy appears to be a robust approach that preserves the integrity of the periodicity search.

Finally, we use the same 28-day binning for the LCs. This time binning provides an adequate compromise between a computationally manageable analysis and sensitivity to long-term variations (of the order of a year).

\section{Methodology} \label{sec:methodology}
\subsection{Overview of Methodology}\label{sec:overview}
Periodicity searches are limited in part by noise. Many Galactic and extragalactic astrophysical sources show erratic brightness fluctuations with steep power spectra \citep[e.g.,~][]{gao_power_law}. In this context, the noise is defined as random variations in the source emission. Noise is classified according to the power-law index $\beta$ of the power spectral density \citep[PSD, $\propto f^{-\beta}$ where $f$ is frequency,][]{rieger_2019}. The PSD measures the power in a signal as a function of the frequency. White, pink, and red noises are characterized by indices of $\beta\!=\!0$, $\beta\!=\!1$, and $\beta\!=\!2$, respectively \citep[e.g.,~][]{tarnopolski20}.

The statistical significance of any putative periodicities must be evaluated in the context of this stochastic noise. We, therefore, identify two challenges : (1) to search for and identify periodicities in the large dataset of LCs and (2) to determine the statistical significance of the periodicity given the characteristics of the noise appropriate for each source.

To search for the periodic signals, we have developed a pipeline that applies a number of standard periodicity-search algorithms to the dataset. The details of the algorithms are given in $\S$\ref{sec:methods} and $\S$\ref{sec:arima}. Each algorithm searches for periods across a given frequency range and produces a test statistic from which potentially interesting periodic signals can be identified. The test statistic for each algorithm is computed by evaluating the amplitude of the periodicity against an alternative (null) hypothesis that can be rapidly evaluated. Tables \ref{tab:methods_results} and \ref{tab:mcmc_bayesian_arfima_results} list the results of this pipeline with the period and test-statistic value for the most promising periodicity candidates. Details are given in the following sections.

We evaluate the statistical significance of any potentially interesting periodicities identified by the pipeline using a Monte Carlo procedure in which a large number of artificial LCs are generated using the method of \citet{emma_lc} using a stochastic model whose parameters have been fitted to the observed LC, as described in $\S$\ref{sec:sig_correction}. The statistical significance found from this procedure is presented in Table \ref{tab:methods_results}. In $\S$\ref{sec:global_sig}, we also evaluate the impact of the ``look-elsewhere'' effect on our results.

Finally, we search for long-term ($\sim$years) periodicity since we are looking for blazars to be SMBHB candidates in the gas-driven regime (see $\S$\ref{sec:discussion}). Consequently, we search for periods in the [1-6] year range. While a study of shorter periods (e.g.,~$\sim$months) would suffer less contamination due to red noise, other challenges appear. First, the smaller bins contain fewer $\gamma$-ray photons, so the statistical uncertainty per bin is larger. Second, in the context of searching for binaries, the residence time at a given orbital period $t_{\rm orb}$ scales as $t_{\rm orb}^{8/3}$ in the gravitational wave-driven regime \citep[][]{haiman_2009}. Thus, considering month-long periods rather than year-long periods would reduce the expected number of binaries in the sample by more than 99\%. On the other hand, this also implies that any periodicity found shorter than $\sim$year-long is more likely to be due to other processes, e.g., disk and/or jet procession from a single SMBH. These alternative processes are interesting in their own right, and we intend to explore shorter periods in future works.

\subsection{Periodicity search methods}\label{sec:methods}

Careful considerations must be taken in a periodicity-search analysis. In general, each method has specific properties, e.g., accuracy, computational time, and sensitivity to irregular time series. Therefore, selecting just one time-series method is arguably arbitrary since all have limitations and advantages \citep{methods_critica_PKS_0735,lomb_vdp}. Consequently, we want to profit from the strengths of the many different algorithms in the literature in detecting periodicities, compare their results systematically, and provide the community with a comprehensive search for periodicity in which reasonable algorithms have been used, helping to other researchers in their analysis.
In light of this, we employed in P20 ten of the most widely used methods for periodicity identification to reduce the impact of their individual limitations and take advantage of their individual strengths. These ten methods were organized in a pipeline described in P20, to which we refer the reader for further details.

However, this pipeline has some caveats. In P20, we used the bootstrap and Fisher's method of randomization \citep[][]{linnel_pdm} to infer the test statistics, which assume white noise as a null hypothesis. This assumption is not correct since the intrinsic variability in blazars is red-noise-like \citep[][]{vaughan_criticism}. To solve this, we substitute these techniques by generating artificial LCs based on a power-law approach using the technique presented by \citet{timmer_koenig_1995}. The generated artificial LCs have the same standard deviation, median flux, and sampling as the original LCs. An LC produced with this technique is based on a Gaussian-like LC generation, but the $\gamma$-ray flux is log-normal distributed \citep[e.g.,][]{shah_lognormality}. \citet{shah_expontiate} obtains the log-normal distribution by exponentiating the LC. Furthermore, this technique is fast, which is a requirement for the first analysis stage of the pipeline. For the remaining methods, we use the strategy described by \citet{emma_lc}, which allows the generation of LCs with the same PSD and probability density function as real blazar LCs. Following \citet{emma_lc}, we generate LCs by applying the bending-power law approach \citep{chakraborty_bending_power_law}, defined by the expression:

\begin{equation} \label{eqn:bending} 
  P(\nu) = A \left( 1 + \left\{ \frac{\nu}{\nu_{b}} \right\}^{\alpha} \right)^{-1} + C, 
\end{equation} 

where $A$ is the normalization ($rms^{2}/year^{-1}$), the spectral index is $\alpha$, and the bending frequency is $\nu_{b}$. This approach provides realistic models of blazars' variability on timescales from weeks to years \citep{chakraborty_bending_power_law}. 

Poisson noise is a form of statistical noise that arises due to the inherent randomness in the counting of photons, particularly prevalent in the high-energy range, such as gamma rays. This type of noise leads to fluctuations in the count rate of detected photons, resulting in variability in the observed LC of a blazar. Given that each photon arrival is a random event, the LC will naturally exhibit statistical variations, even if the source's actual flux remains constant. Poisson noise contributes a constant (white noise) to the power spectrum. The value of this constant is estimated using the following expression:

\begin{equation} 
 C=\frac{2D}{N\bar{x}^2}\bar{\Delta{x^2}},
\end{equation} 

Where $D$ is the total time duration of the LC, $\bar{x}$ is the mean flux, $\bar{\Delta{x^2}}$ represents the average squared error of the flux uncertainties, and $N$ the number of data in the LC. We incorporate Gaussian-distributed random errors, matching the characteristics observed in the actual data to the artificial LCs. This inclusion serves to represent the presence of Poisson noise.

In addition to that, we removed the REDFIT from the pipeline \citep[][]{redfit}. REDFIT assumes an AR(1) (first-order autoregressive process) to model the red noise process, which is inappropriate for blazars \citep[][]{vaughan_bayesian}. Additionally, the test statistics estimation assumes a single, best fit, which overestimates the test statistics \citep[e.g.,][]{vaughan_bayesian, vaughan_criticism}.

Therefore, in this work, we use the methods listed below\footnote{The computational times required to perform atomic operations for each method are as follows: LSP (1.1$\mathrm{x}10^{-3}$ $s$), GLSP (6.5$\mathrm{x}10^{-4}$ $s$), PDM (2.9$\mathrm{x}10^{-3}$ $s$), CWT (2.2$\mathrm{x}10^{-3}$ $s$), and MCMC Sine (7.1$\mathrm{x}10^{-1}$ $s$). These results are obtained using the LC of PG 1553+113. All computations were carried out on a system equipped with an Intel(R) Core(TM) i7-5500U CPU running at 2.40 GHz (dual-core, 4 logical processors) and 16 GB of RAM. 
}:

\begin{enumerate}
    \item Lomb-Scargle periodogram \citep[LSP,][]{lomb_1976, scargle_1982} which is superposed on a red noise spectrum to obtain the test statistics \citep{power_law}.
    \item Generalized Lomb-Scargle periodogram \citep[GLSP,][]{lomb_gen}. 
    \item Phase Dispersion Minimization \citep[PDM,][]{pdm}. 
    \item Continuous Wavelet Transform \citep[CWT,][]{wavelet_torrence}. 
    \item Markov Chain Monte Carlo Sinusoidal Fitting \citep[MCMC Sine,][]{emcee}. This method allows us to obtain a model of the oscillating signal, characterizing the offset, amplitude, period, and phase. This information is used to evaluate the performance of the other methods against the noise (see $\S$\ref{sec:eval}).  
\end{enumerate}

The test statistics for the peaks in the periodicity analysis are determined by generating 1,000,000 artificial LCs using the method described in \citet{emma_lc} for both the GLSP and CWT methods. The test statistic is defined as the fraction of simulated LCs in which the power at any frequency within the studied range exceeds the highest peak observed in the original LC. This process limits the statistical significance to a maximum of $\approx$4.9$\sigma$. For the PDM method, due to computational constraints, 150,000 artificial LCs are generated, limiting the statistical significance at $\approx$4.5$\sigma$.

Additionally, the choice of the number of artificial LCs directly affects the precision of the test statistics estimation. While the large sample size for GLSP and CWT ensures more accurate significance levels, the computational demands of PDM limit its precision, though it still provides reliable estimates for most practical purposes. 

Finally, we use new methods described in the following sections in this current work.

\subsection{Autoregressive models}\label{sec:arima}
In this work, we include autoregressive models to improve our analysis pipeline. These models are proposed as efficient methods for astronomical data analysis, particularly to infer any periodic behavior \citep{scargle_1981, caceres_arima}. To analyze the {\it Fermi}-LAT LCs, the ARIMA \citep[][]{chatfield_arima} and ARFIMA \citep[][]{feigelson_arima} are applied. These autoregressive models are suitable for analyzing astronomical time series since they allow modeling a wide variety of LC properties (e.g., irregular or quasi-periodic, constant mean or variable mean) or combinations of stochastic and deterministic behaviors \citep[][]{feigelson_arima}. These characteristics make these models more robust against stochastic noise, helping to infer periodic behaviors more accurately \citep[][]{caceres_arima}. 

Both models are based on an auto-correlation (correlation of the LC with itself) where one looks for dependencies of the current values (of the LC) on past values. In this approach, the current values are modeled as the sum of two linear combinations of past values (with $p$ and $q$ terms, respectively) that model an autoregressive and stochastic process, respectively 
\citep[see][for details]{feigelson_arima}. The parameter $d$ in ARIMA represents how stationary the LC is, i.e., whether a time series has a constant mean and variance (in which case, $d$=0). In ARFIMA, the parameter $d$ represents the type of process. For instance,  0$<d<$1 corresponds to a long memory process, while -0.5$<d<$0 is a short memory process. This parameter is in the 0-0.5 range for a stationary time series. A stochastic process is modeled as ARFIMA(1,d,0) \citep[][]{xu_arfima_lc}, and ARIMA(0,0,1) \citep[][]{zhang_periodicity_arima}.

In our methodology, the ARFIMA model can be applied to stationary time series. ARFIMA can model non-stationary series, but it is not recommended due to potential inconsistencies in its applications \citep[][]{arfima_baillie_1996}. ARIMA models are preferred for non-stationary time series because they are specifically designed to address this characteristic through integer differencing \citep[see,~][]{caceres_arima}. Stationarity is measured using the augmented Dickey-Fuller unit root test \citep[][]{dickey_fuller}, which tests the null hypothesis that a time series is non-stationary. For the purposes of this study, we reject the null hypothesis if the p-value is $\rm{\leq0.05}$. 

The periodicity search with these methods begins with selecting the best-fit ARFIMA/ARIMA model using Akaike's Information Criterion \citep[AIC,~][]{akaike_criterio}. The AIC is a statistical estimator that evaluates a model that fits the data. AIC is used to compare models to determine which one performs the best fit for the data. After that, the residuals are obtained from the original LC and the selected ARFIMA/ARIMA model. Then, we compute the Autocorrelation Function (ACF) on these residuals (see Figure \ref{fig:arfima}). We use the same criterion presented in the literature by using a $\rm{\geqslant2\sigma}$ threshold for test statistics \citep[e.g.,~][]{zhang_pks0301, zhang_periodicity_arima, yang_carma}. Finally, we use the Ljung-Box test to evaluate the quality of fit between the LC and its ARFIMA/ARIMA model \citep{ljung_test}. The Ljung-Box test checks whether or not the autocorrelations for the residuals are non-zero, denoting the lack (or not) of model fit. We consider the fit adequate when the p-value is at least 0.05. Higher p-values mean that the ARFIMA/ARIMA model fits the LC better.

To implement the ARFIMA/ARIMA analysis, we use the R packages \texttt{stats}\footnote{\url{https://www.rdocumentation.org/packages/stats/versions/3.6.2/topics/arima}} and \texttt{arfima}\footnote{\url{https://www.rdocumentation.org/packages/forecast/versions/8.13/topics/arfima}}. This R functionality is accessible from a Python environment via the package \texttt{rpy2}\footnote{\url{https://rpy2.github.io/doc/latest/html/index.html}}.

With ARIMA and ARFIMA\footnote{The computational times required to perform atomic operations for each method are as follows: ARIMA (2.2 $s$), and ARFIMA (2.1 $s$). These results are obtained using the LC of PG 1553+113. All computations were carried out on a system equipped with an Intel(R) Core(TM) i7-5500U CPU running at 2.40 GHz (dual-core, 4 logical processors) and 16 GB of RAM.}, the total number of methods used in the analysis is seven.

\section{Correction and the Calibration of the Test Statistics}

\subsection{Global significance}\label{sec:global_sig}
In our periodicity analysis, there was no prior knowledge of the frequencies of the potential signal. In these conditions, it is statistically more rigorous to employ a ``global significance'' \citep[e.g.,][]{benkhali_power_spectrum}. This significance accounts for the look-elsewhere effect, which is the ratio between the probability of observing the excess at some fixed value and the probability of observing it anywhere in the value range considered in the analysis \citep{gross_vitells_trial}. The ``global significance'' is obtained by applying a correction to the ``local significance'', which is the test statistics of the periodicity in an LC at a specific value obtained for each method. This correction is approximated by
\begin{equation}\label{eq:trial}
p_{\mathrm{global}}=1-(1-p_{\mathrm{local}})^{N},
\end{equation}
where $N$ is the trial factor. In our study, we have to consider two potential issues. One is that we do not know the frequency for each source a priori, so we must search for the highest peak in each periodogram. The second issue is that we do not know a priori which sources exhibit periodic behavior, so we must also select them from the periodograms. Consequently, searching $P$ independent periods (frequencies) in each of the periodograms of $B$ blazars, the number of trials is 
\begin{equation}
\label{eq:p}
 N=P \times B.    
\end{equation}
 We do not consider the number of methods in the trials since we present all the results for all the methods equally in our tables for each blazar, avoiding picking the highest significant result according to a single method. 
  
In P20, we analyzed $\approx$2000 blazars; however, after eliminating blazars for which $\geqslant$50$\%$ of the LCs consist of upper limits, the number of blazars included in our sample was 351. There is no unique way to determine $P$ in a given periodogram. For instance, \citet{lomb_gen} set $P$ based on the frequency range and resolution considered in the periodicity analysis. In our periodograms, we have followed this approach, considering 100 periods (to have a good balance between computational time and resolution in the periodograms). However, for a 12-year LC with $\approx$12 samples/year (for approximately $M=144$ points in the LC), our period range ([1-6] years) corresponds to 11 independent frequencies sampled. These frequencies are obtained from the expression,

\begin{equation}\label{eq:nyquist}
f_{j}= \frac{1}{\delta T}*\frac{j}{M},
\end{equation}
where $\delta T = 1/12$ and $j = 1$,…,$12$, in our case.

A complementary approach to determine $P$ is to perform Monte Carlo simulations \citep[see, e.g.][]{cumming_indep_freq}. Following this method, we estimate $P$ by computing the discrete Fourier transforms (DFTs) of $10^{8}$ simulated LCs using the technique of \citet[][]{timmer_koenig_1995}.

The distribution of PSD powers is that of a $\chi^{2}$ with two degrees of freedom, which can be used to compute the ``local significance'' for each PSD. The ``global significance'' is computed from the distribution of the highest peak PSD powers for the $10^{8}$ simulated LCs. 

\begin{figure}
	\includegraphics[width=\columnwidth]{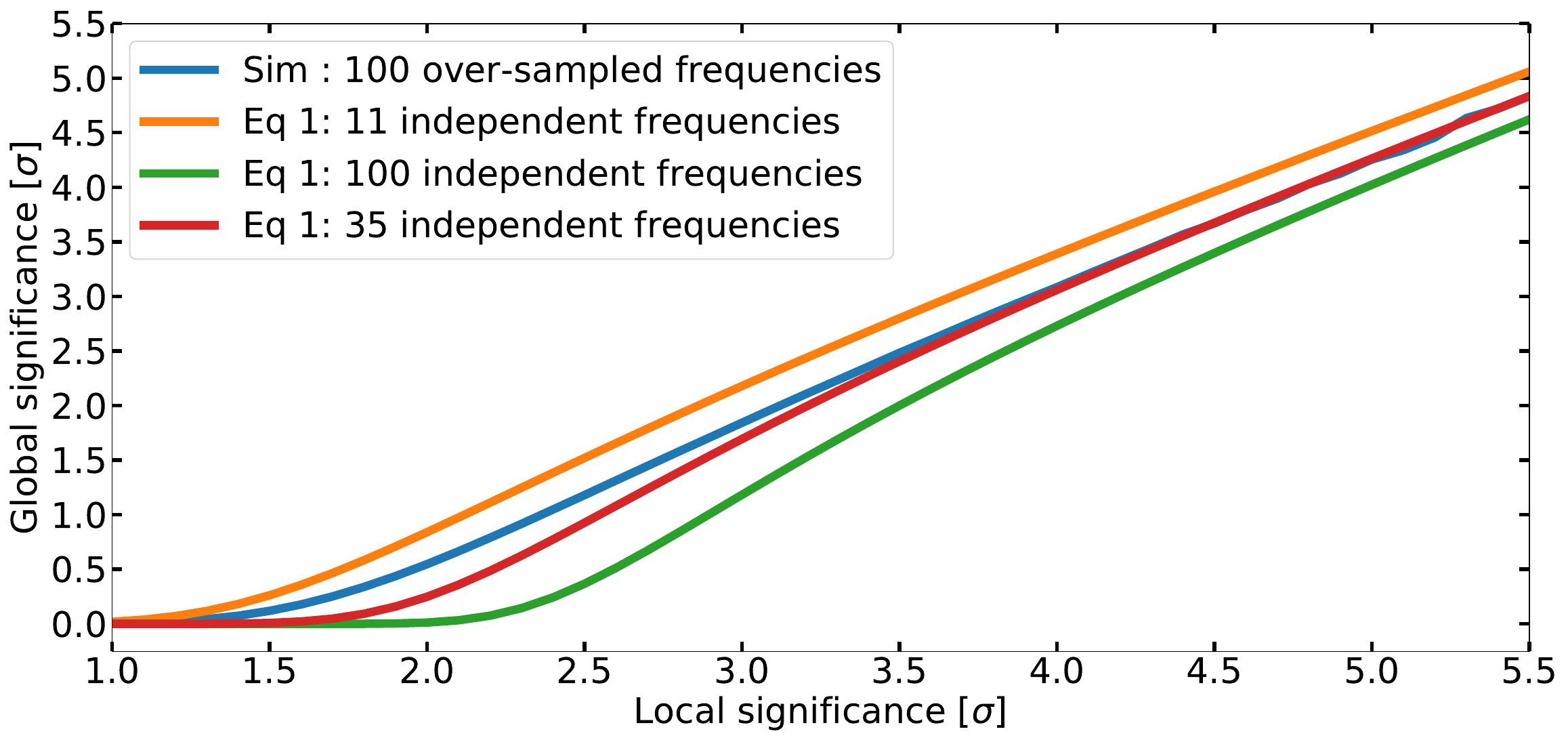}
	\caption{Relation to estimate the number of oversampled frequencies $P$ needed in equation~\ref{eq:p} to compute the trial factor.
 ``Eq 1'' denotes the results of applying the equation \ref{eq:trial} for a specific number of independent frequencies.} \label{fig:trials}
\end{figure}

To reduce the effects of both red noise leak (transfer of power from low to high frequencies that can conceal a PSD) and aliasing (that can flatten the low-frequency PSD), the LCs are oversampled \citep{psresp_uttley} by a factor of 10 (for a total of 1440 time bins). Then, the steps above are repeated for each over-sampled PSD, searching it and extracting the power of the highest peak within the 100 frequencies of interest. Finally, we calculate the experimental global and local significance for the over-sampled case (see blue curve in Figure \ref{fig:trials}).

We select the $P$ to correct the ``local significance'' $\geq4\sigma$ to correct precisely the results of the potential high-significance candidates. Figure \ref{fig:trials} shows that adopting $P=35$ gives the best agreement with the over-sampled analysis for the relationship between the ``global significance'' and the ``local significance''. Consequently, we choose $P=35$ for this work, resulting in a trial factor of 12,285 ($351$$\times$$35$). The resulting ``global significances'' for the high test statistics are:
\begin{enumerate}
\item $\approx$2.8$\sigma$ for test statistics of $\approx$5$\sigma$
\item $\approx$1.8$\sigma$ for a test statistics of $\approx$4.5$\sigma$
\item $<$1$\sigma$ for test statistics $<$4.5$\sigma$.
\end{enumerate}

To evaluate the influence of $P$ (number of independent frequencies) on the ``global significance'', we perform a test. In this test, we fix the $B$ and vary $P$ to observe how it affects the ``global significance''. We use PG 1553+113 as a benchmark, which has a ``local significance'' of 4.5$\sigma$, corresponding to a ``global significance'' of $\approx$1.8$\sigma$. The resulting ``global significances'' for this``local significance'' are:

\begin{enumerate}
\item For $P=100$, it is $\approx$1.3$\sigma$
\item For $P=75$, it is $\approx$1.4$\sigma$
\item For $P=50$, it is $\approx$1.6$\sigma$
\item For $P=25$, it is $\approx$1.9$\sigma$
\item For $P=11$, it is $\approx$2.3$\sigma$
\end{enumerate}

These results illustrate the role of $P$ in determining the ``global significance''. As $P$ increases, the ``global significance'' systematically decreases.

\subsection{Power-spectral index estimation}\label{sec:psi}
The PSD can be fitted by a power law according to the expression $\propto\rm f^{-\beta}$, where $\rm f$ is the frequency and $\beta$ is the power-law index \citep[e.g.,~][]{gao_power_law,tarnopolski20}. 
This form for the PSD denotes that $\gamma$-ray variability in blazars is stochastic \citep[e.g.,~][]{sobolewska_slopes}. Additionally, estimates of the power-law index provide information about the nature of the variability, denoting the role of the accretion disk in the emission of the jets \citep[e.g.,~][]{bhatta_s5_0716}.

The power-spectral indices are estimated using the maximum likelihood and MCMC (ML-MCMC) \footnote{We use the Python package emcee}. 

The results obtained for the candidates are shown in Table \ref{tab:slopes_poisson}. The estimated indices for the power spectrum in most of our candidates are in the range of [0.9$-$1.5]. This range is compatible with the values proposed in the literature  \citep[e.g.,~][]{nakagawa_slopes, sobolewska_slopes, kushwaha_slopes, bhatta_s5_0716}. However, the obtained indices are $\pm$30\% different than those reported by \citet{yang_carma}. \citet{benkhali_power_spectrum} reported different power spectral indices related to several of our blazars, except for PKS~2155$-$304 with a compatible $\rm\beta\approx1$. In Table \ref{tab:slopes_poisson}, the mean of the indices for all candidates is $\sim$1.2 with a standard deviation of 0.4. These indices are compatible with those from \citet{bhatta_s5_0716}. 

\begin{figure}
	\includegraphics[width=\columnwidth]{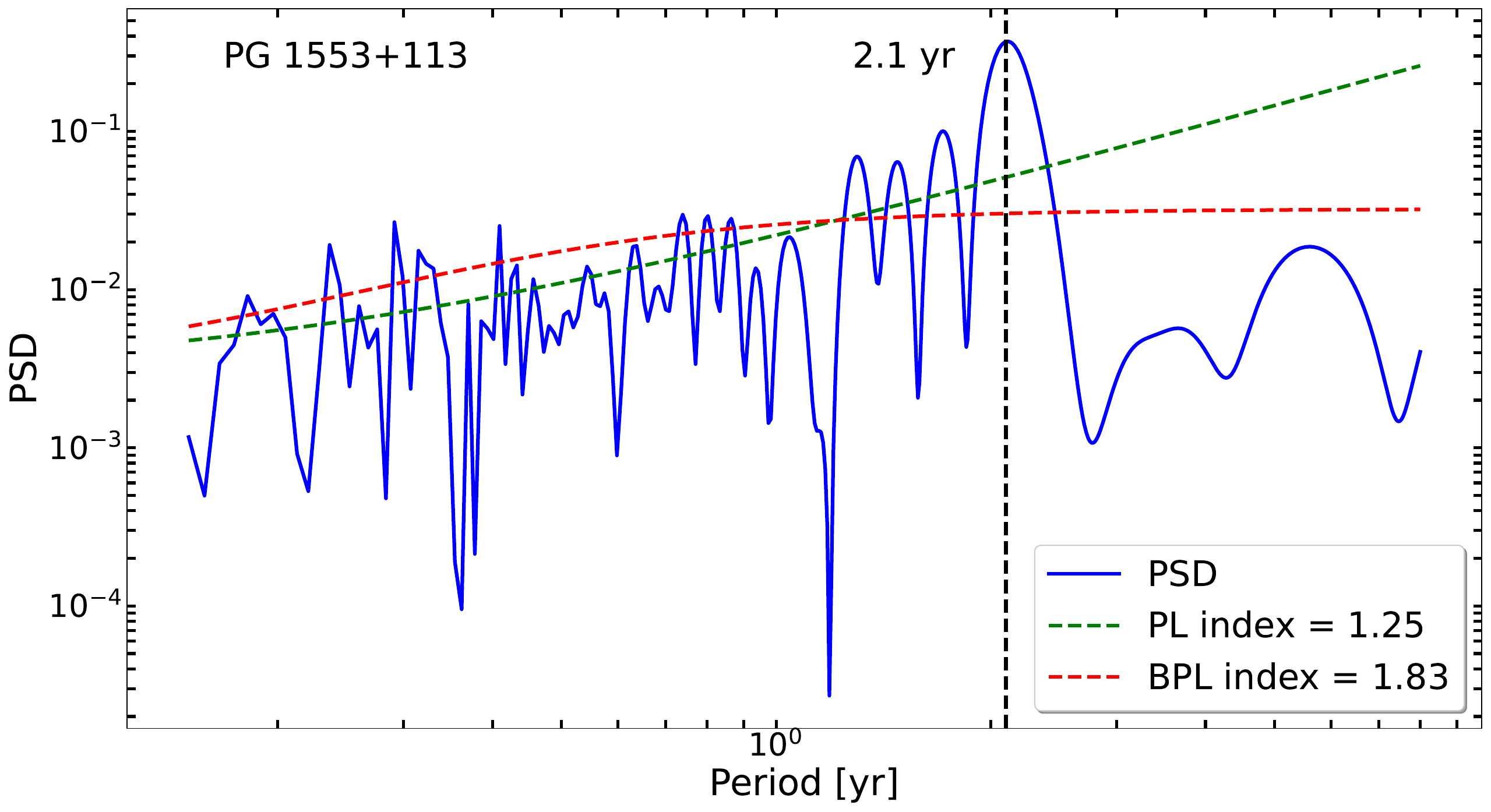}
	\caption{PSD fits for PG 1553+113 according to the PL and BPL approaches. }
	\label{fig:slope_mcmc}
\end{figure}

We utilize the AIC for each PSD model to evaluate which one provides a more accurate fit. We employ the concept of relative likelihood of models (RLM) to compare the different PSD models. This comparison is conducted by considering a p-value$\rm{\leq0.05}$ and applying the following expression:
\begin{equation} \label{eqn:rml} 
\exp\left(\frac{\text{AIC}_{\text{min}} - \text{AIC}_{i}}{2}\right)
\end{equation}

Here, $\text{AIC}_{\text{min}}$ represents the minimum AIC value among the models, and $\text{AIC}_{i}$ represents the AIC value of the other model in consideration. As a result, the bending power-law (BPL) model consistently provides a better fit to the PSD in all cases, as shown in Table \ref{tab:slopes_poisson}. 

Most of the power-law indices shown in Table \ref{tab:slopes_poisson} are consistent with a pink-noise process (spectral index $\approx$1), which may be caused by disk modulations that materialize as jet modulations. \citep[][]{bhatta_variablity}. This process can be associated with short time-scale variability in the periodic baseline \citep[associated with sporadic flaring activity, which might indicate instabilities and turbulence in the accretion flow through the disc or in the jet, ][]{abdo_variability}. Additionally, a pink-noise process can be associated with long-term variability coherent on time scales on the order of decades \citep{bhatta_s5_0716}. For example, a process that can produce stochastic emission at both short and long timescales is magnetic reconnection \citep{bhatta_s5_0716}. In addition to that, variable accretion with large accretion episodes ($\approx$50$–$200\,$\Msun$  yr$^{-1}$) followed by smaller ones can also explain PSDs with indices of $\approx$1 \citep{jolley_mass_accretion}.

\citet{czerny_accretion_disk} present a modeling of the relationship between instabilities in the accretion disk and variability, in which the time scales of these instabilities range from hours to a few years for blazars with BH masses of $10^8$$-$$10^9\Msun$. For instance, these accretion episodes can be associated with a steep power-law index (red noise, with a spectral index $\approx$2), which produces a dominance of the long-term variability in the LC \citep[e.g.,][]{kunjaya_avalanches}. This scenario may be applicable to PKS 0208$-$512 (with a power-law index of 1.6$\pm$0.4). 

Alternatively, \citet{models_variability} presents a model to interpret the different timescales observed in the multiwavelength variability of blazars. The variability of the synchrotron emission is powered by a stochastic process with a duration from minutes to years. Regarding the variability of the inverse Compton emission, it results from two stochastic processes, which are linearly overlapping with a duration from one to thousands of days. These two processes are the dissipation in the magnetic field of the jet and inhomogeneities in the number of photons for the inverse-Compton process.  

Finally, \citet[][ and references therein]{rieger_2019} propose a model of the dominant radiation processes. Specifically, the power-law index of the PSD depends on the component that dominates the high-energy emission—external Compton or synchrotron self-Compton (with an index of $\gtrsim 2.0$ signifying the former process). 

Furthermore, we employ the ML-MCMC to estimate the index according to the bending power law described in Eq. \ref{eqn:bending} (see Figure \ref{fig:slope_mcmc}). The resulting indices are reported in Table \ref{tab:slopes_poisson}. The indices are steeper than those reported in the power law case, compatible with the values presented in \citet{gaur_bending_power_law}.

\subsection{Test Statistics Correction}\label{sec:sig_correction}

We also perform a correction to the test statistics of the results of the high-significance candidates denoted by a test statistics $\geq$3$\sigma$ (see $\S$\ref{sec:periodic} and Table \ref{tab:methods_results}). This correction is implemented to assess the implications of modifying the assumed PSD models and analyzing the influence on the test statistics. We perform this correction to evaluate the effects of changing the assumed PSD (power law and bending power law). We follow the methodology outlined in $\S$\ref{sec:methods} to compute the updated test statistics.

In this correction, we use the PSD parameters and their uncertainties of Table \ref{tab:slopes_poisson} obtained with the ML-MCMC method. Specifically for the power law, we consider three different models, combining the PSD indices and normalization values: the index and the normalization minus the corresponding uncertainties, the index and the normalization, and the index and the normalization plus the corresponding uncertainties. Regarding the bending power law, we perform the correction using the PSD indices, the bending frequencies, and normalization values shown in Table \ref{tab:slopes_poisson}, with their uncertainties. We also considered three models: index, bending frequency, and normalization minus their corresponding uncertainties; index, bending frequency, and normalization; and index, bending frequency, and normalization plus their corresponding uncertainties. 

The results for this correction are shown in Table \ref{tab:correction_power_law_new} and Table \ref{tab:correction_bending_power_law_new}. As they show, a ``redder'' PSD, characterized by a higher power at lower frequencies, leads to reduced significance levels in the detection of long-frequency periods \citep[][]{benkhali_power_spectrum}. The methods with the largest corrections are LSP and CWT, while the blazar with the largest corrections is PG 1553+113.

Furthermore, a key distinction between the LSP and the CWT lies in their treatment of time: LSP evaluates the periodic content over the entire time series globally, while CWT provides a time-localized view, capturing how the dominant frequencies vary throughout the observation period. This time-frequency resolution, while beneficial for identifying transient or evolving signals, can also make CWT more sensitive to the variability in the data, such as noise fluctuations or transient events. As a result, the peak frequency identified by CWT may deviate slightly from the true underlying periodicity, producing a broader frequency response and, thus, a lower associated test statistic.  These combined factors help explain why CWT tends to yield lower significance values compared to LSP in our analysis. This can be observed in Table \ref{tab:methods_results}, where in 83\% of the cases, CWT results in bigger uncertainty in the period, with 100\% of the uncertainty reported by the LSP in 44\% of such cases.

\subsection{Evaluation of methods against noise} \label{sec:eval}
We now evaluate the periodicity detection methods when periodicity is present, but the oscillating fluxes are contaminated by noise \citep{lomb_vdp}. These methods are LSP, GLSP, PDM, and CWT. We consider four different noises: ``white noise'', ``pink noise'' (generating random power-law indices in the range [$0.8 - 1.2$]), ``red noise'' (with power-law indices in the range [$1.8 - 2.2$]), and ``broken power-law'' models. 

To implement this scenario, we define a sinusoidal function
\begin{equation} 
\phi(t) = O + A\sin \bigg(\frac{2 \pi t}{T} + \theta \bigg),
\end{equation}
where the parameters are the offset ($O$), the amplitude ($A$), the period ($T$), and the phase ($\theta$). The offset, amplitude, and phase are randomly selected from the values obtained for each of the 24 candidates from the Markov Chain Monte Carlo sinusoidal fitting method from P20. Specifically, the ranges are $[2-15]\times 10^{-8}$ ph cm$^{-2}$ s$^{-1}$ for the offset, $[4-40]\times 10^{-8}$ ph cm$^{-2}$ s$^{-1}$ for the amplitude, and $[0-2\Pi]$ for the phase. We also evaluate any bias in the detection by defining a periodicity range based on the ones from our sample (that is, $[1.5 - 4.5]$ years). From here, we check if a period is more likely to be detected than others.

We simulate 100,000 sinusoidal LCs for each period-noise combination. Both pink and red noise LCs are produced by applying the Python package \texttt{colorednoise}\footnote{\url{https://github.com/felixpatzelt/colorednoise}}, which is based on \citet{timmer_koenig_1995}. The noisy LCs from broken power-law models are generated by the Python package of \citet{connolly_code}, which is derived from \citet{emma_lc}.

\begin{figure*}
	\centering
	\includegraphics[scale=0.22]{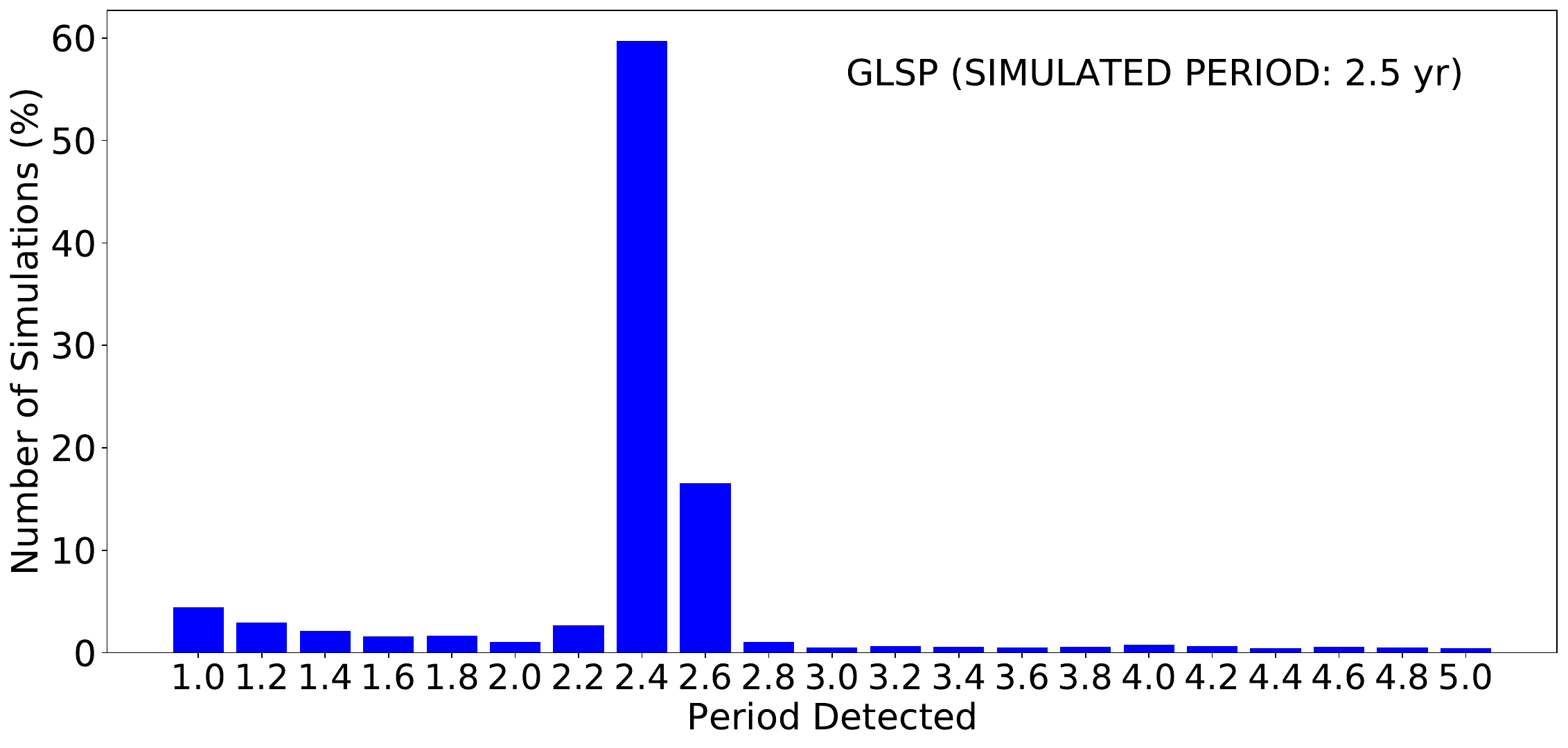}
	\includegraphics[scale=0.22]{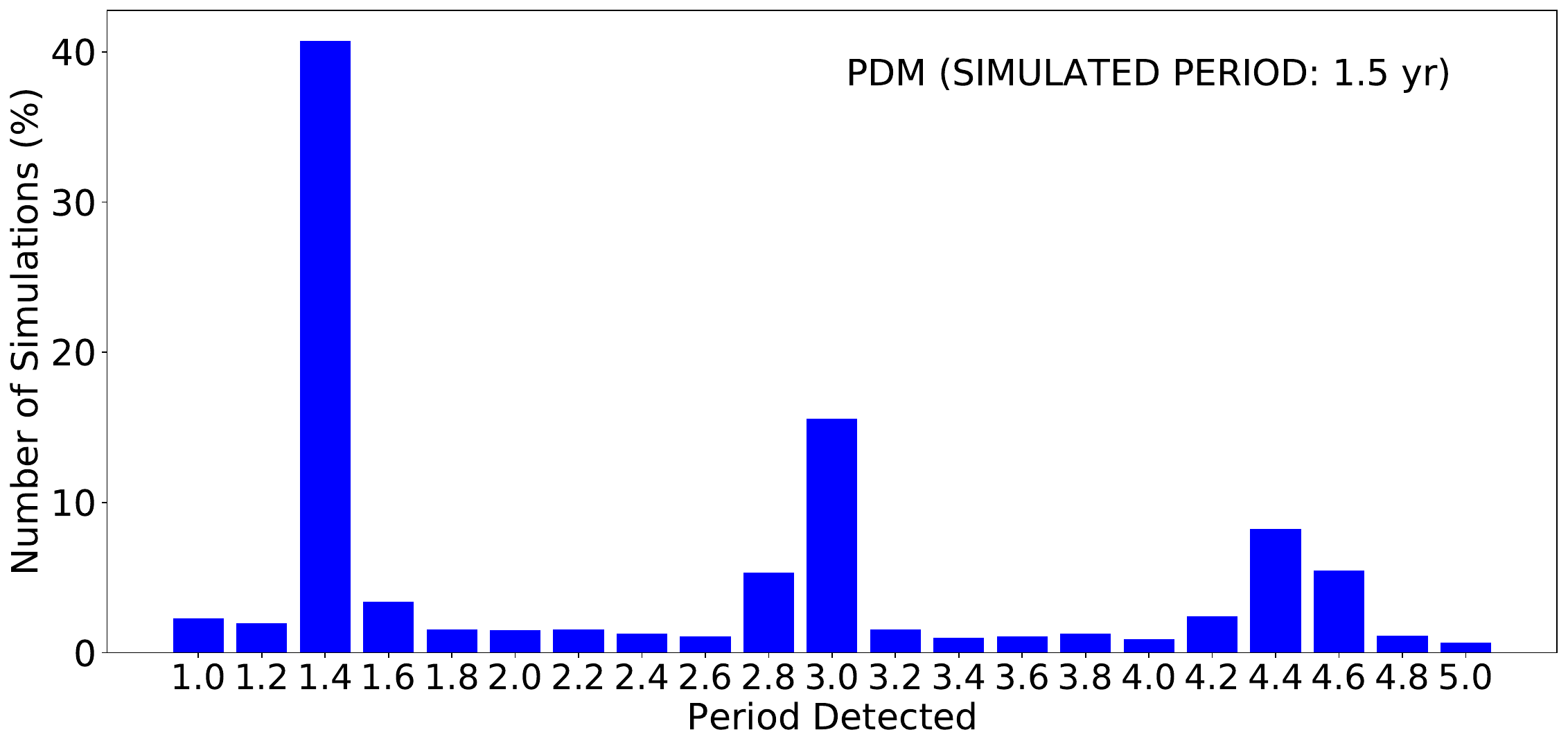}
	\includegraphics[scale=0.22]{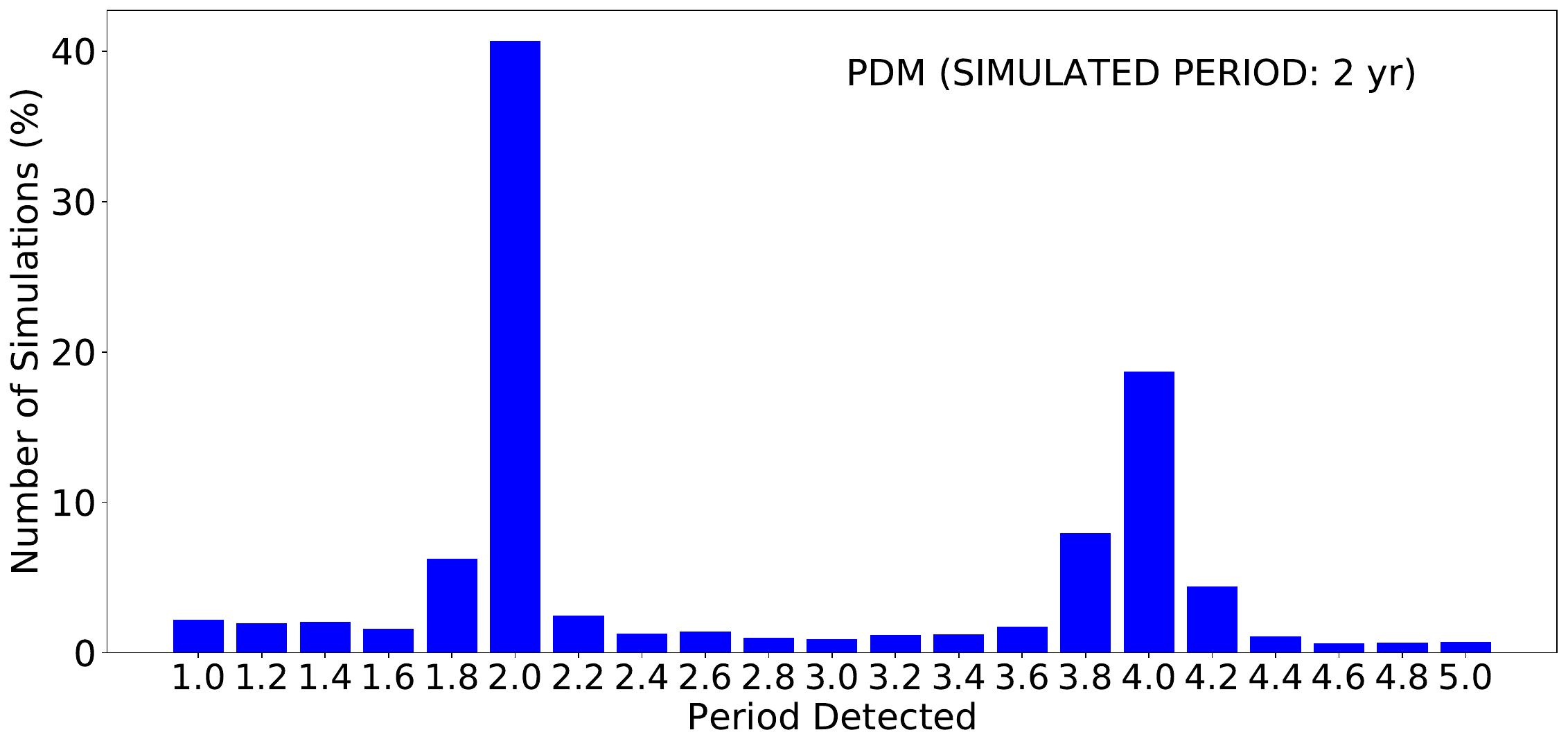}
	\caption{Results of evaluating the false-positive detection rates for the various periodicity-detection methods. {\it Top}: The results for the GLSP and PDM  methods in the case of a sinusoidal LC with periods of 2.5 yr (left) and 1.5 yr (right) that are contaminated with red noise. These results indicate that these methods are robust for periodicity detection against red noise. {\it Bottom}: The results of the PDM simulations for periods of 2 yr contaminated with red noise. The harmonic effects observed by P20 are also shown.}
	\label{fig:methods_exp2}
\end{figure*}

The worst results are obtained for white noise and the broken power-law model, with average detection rates of 5\%-20\%. The method that is the most sensitive to noise is LSP, having the lowest detection rate for all noise patterns. The methods are most robust for fluxes that are contaminated with red noise. In particular, the methods that perform the best are CWT and GLSP, with detection rates of $\approx$25\%-65\% (see Figure \ref{fig:methods_exp2}). Any bias in detecting a specific period is not observed. However, there is a tendency for detection to be more sensitive to noise when the periods are on the higher end of our range, i.e.,~[3.5-4.5] yr. This is expected since fewer cycles would be included in the data. The exception is pink noise, which results in roughly constant detection rates for all the periods for each method (detection rates $\approx$12\%-17\%). Finally, the PDM simulations show the harmonic effects observed in P20. Harmonic effects are only observed when the sinusoidal LCs are contaminated by red noise (see Figure~\ref{fig:methods_exp2}).

To further assess the likelihood of false detections, we conduct an additional test aimed at evaluating the spurious detection rate associated with each method. This involves generating 100,000 synthetic LCs following the procedure outlined in \citet{emma_lc}. The resulting period–significance distributions allow us to identify any potential biases toward particular period ranges, which could artificially align with the detected period. This analysis helps determine whether the period–significance pair derived from the original data may arise purely from stochastic variability, leading the method to register a false detection. Additionally, this test also provides a way to estimate the probability that a given periodicity analysis method produces a spurious detection. By quantifying how often similar period-significance combinations appear in purely stochastic data, we can obtain the method’s susceptibility to misidentifying noise-induced features as significant periodic signals. We also compute the percentage of cases in which synthetic LCs produce the same period-significance combination as observed in the real data. An occurrence is defined as a case where the recovered period falls within the uncertainty interval of the period found in the original LC, and the associated significance is equal to or higher than that of the original detection. As a reference case, we apply this test to PG 1553+113. The procedure is repeated using both noise models discussed in $\S$\ref{sec:sig_correction}, the power-law and the bending power-law PSDs.

For the power-law PSD, the detection rates (defined as $\geq$3$\sigma$) are as follows: 0.02\% for the LSP, 0.31\% for the PDM, 0.03\% for the GLSP, and 0.01\% for the CWT. These values indicate that, under a PL noise assumption, false-positive detections of $\geq 3\sigma$ can still occur with low probability. None of the methods showed any coincidence between the detected period-significance pairs and those obtained from the original data.

When assuming a bending power-law power PSD, the rates of spurious $\geq3\sigma$ detections are: 0.0\% for LSP, 0.2\% for PDM, 0.1\% for GLSP, and 0.0\% for CWT. None of the methods showed any coincidence between the detected period-significance pairs and those obtained from the original data.

Finally, we repeat the test using pure red-noise LCs with a power spectral index of 2. The resulting detection rates are 0.01\% for LSP, 0.24\% for PDM, 0.02\% for GLSP, and 0.0\% for CWT, again reinforcing the robustness of the original detections against red-noise contamination.

These results show that the tested methods, LSP, PDM, GLSP, and CWT, are unlikely to yield significant detections from purely stochastic signals. The very low occurrence rates of spurious $\geq$3$\sigma$ detections, combined with the zero coincidence of both period and significance with those obtained from real data, suggest that such detections are not easily produced by random fluctuations or artifacts resulting from some bias of the used methods.

\begin{table*}
\centering
\caption{Comparison of periods and local test statistics for the blazars analyzed in P20. The test statistics and periods reported here are the averages of those obtained in the different methods employed in the periodicity-search analysis (to allow a comparison with P20). From {\it top} to the {\it bottom}: $>$3$\sigma$ candidates and low-significance candidates.  The candidates are sorted according to the median of their test statistics. Note that this median significance does not have an actual statistical meaning; it is used as an arbitrary way to combine all test statistics to sort the candidates. The blazars are characterized by their \textit{Fermi}-LAT 4FGL source name, equatorial coordinates (deg), AGN type, redshift, association name, period (in years), and test statistics obtained by \citet{penil_2020}, and the average period (in years) and test statistics obtained in this work. We include the percentage of upper limits (UL) in the LC. We also included the global significance in the last column resulting from the method described in $\S$\ref{sec:global_sig}. Note that some sources have two significant periods (organized by the peak amplitude), denoted by $\star$.\label{tab:candidates_list}}
{%
\begin{tabular}{ccccccccccc}
\hline
\hline
        4FGL Source Name & RA(J2000) & Dec(J2000) & Type & Redshift & Association Name & UL (\%) & P20 Period & Period & Global \\
        &  &  &  &  & & & [yr] (S/N)  & [yr] (S/N) & (S/N) \\
        \hline
        J1555.7+1111 & 238.93169 & 11.18768 & bll & 0.433 & PG 1553+113 & 0\% & 2.2 ($>$4.0$\sigma$) &  2.2 (4.5$\sigma$) & 1.8 $\sigma$\\
        J2158.8$-$3013 & 329.71409 & -30.22556 & bll & 0.116 & PKS 2155$-$304 & 0\% & 1.7 ($>$3.0$\sigma$) & 1.7 (3.3$\sigma$) & $\approx$0$\sigma$ \\
        \hline
	J0811.3+0146 & 122.86418 & 1.77344 & bll & 1.148 & OJ 014 & 0.6\% & 4.3 ($>$3.5$\sigma$) & 4.1 (2.9$\sigma$) & $\approx$0$\sigma$ \\
        J0457.0$-$2324 & 74.26096 & -23.41384 & fsrq & 1.003 & PKS 0454$-$234 & 0\% & 2.6 ($>$2.5$\sigma$) &  3.6 (2.8$\sigma$) & $\approx$0$\sigma$ \\      
        J0721.9+7120$\star$ & 110.48882 & 71.34127 & bll & 0.127 & S5 0716+714 & 0\% & \makecell{2.8 ($>$2.5$\sigma$) \\ 0.9 ($>$2$\sigma$)} &  \makecell{2.7 (2.8$\sigma$) \\ 0.9 (2.0$\sigma$)} & $\approx$0$\sigma$ \\
        J0043.8+3425 & 10.96782 & 34.42687 & fsrq & 0.966 & GB6 J0043+3426 & 4.9\% & 1.8 (4.0$\sigma$) & 1.9 (2.7$\sigma$) & $\approx$0$\sigma$ \\
        J0521.7+2113 & 80.44379 & 21.21369 & bll & 0.108 & TXS 0518+211 & 0\% & 2.8 ($>$3.0$\sigma$) & 3.1 (2.6$\sigma$) & $\approx$0$\sigma$ \\
        J1649.4+5238 & 252.35208 & 52.58336 & bll & -- & 87GB 164812.2+524023 & 12\% & 2.7 ($>$2.5$\sigma$) & 2.8 (2.2$\sigma$) & $\approx$0$\sigma$ \\
        J0449.4$-$4350 & 72.36042 & -43.83719 & bll & 0.205 & PKS 0447$-$439 & 0\% & 2.5 (3.0$\sigma$) & 1.9 (2.1$\sigma$) & $\approx$0$\sigma$ \\
        J0428.6$-$3756 & 67.17261 & -37.94081 & bll & 1.11 & PKS 0426$-$380 & 0\% & 3.4 (3.0$\sigma$) & 3.6 (2.1$\sigma$) & $\approx$0$\sigma$ \\ 
        J0303.4$-$2407 & 45.86259 & -24.12074 & bll & 0.266 & PKS 0301$-$243 & 0\% & 2.0 (3.0$\sigma$) & 2.1 (2.0$\sigma$) & $\approx$0$\sigma$ \\ 
        J1146.8+3958 & 176.73987 & 39.96861 & fsrq & 1.089 & S4 1144+40 & 0.6\% & 3.3 ($>3.0$$\sigma$) & 3.3 (1.9$\sigma$) & $\approx$0$\sigma$ \\
        J1248.2+5820 & 192.07728 & 58.34622 & bll & -- & PG 1246+586 & 0\% & 2.0 (3.0$\sigma$) & 2.1 (1.9$\sigma$) & $\approx$0$\sigma$ \\  
        J0252.8$-$2218 & 43.20377 & -22.32386 & fsrq & 1.419 & PKS 0250$-$225 & 1.2\% & 1.2 ($>$2.5$\sigma$) & 1.2 (1.7$\sigma$) & $\approx$0$\sigma$ \\     
	J2258.0$-$2759$\star$ & 344.50485 & $-$27.97588 & fsrq & 0.926 & PKS 2255$-$282 & 8.1\% & 1.3 ($>$3.5$\sigma$) & \makecell{2.8 (1.7$\sigma$) \\ 1.4 (1.1$\sigma$)} & $\approx$0$\sigma$ \\	 
        J1903.2+5541 & 285.80851 & 55.67557 & bll & -- & TXS 1902+556 & 0\% & 3.8 ($>$2.5$\sigma$) & 3.3 (1.5$\sigma$) & $\approx$0$\sigma$ \\
        J0501.2$-$0157 & 75.30886 & -1.98359 & fsrq & 2.291 & S3 0458$-$02 & 1.8\% & 1.7 ($>$2.5$\sigma$) & 3.8 (1.4$\sigma$) & $\approx$0$\sigma$ \\	
  	J1303.0+2435 & 195.75454 & 24.56873 & bll & 0.993 & MG2 J130304+2434 & 9.9\% & 2.0 ($>$2.5$\sigma$) & 2.1 (1.2$\sigma$) & $\approx$0$\sigma$ \\
        J2056.2$-$4714$\star$ & 314.06768 & -47.23386 & fsrq & 1.489 & PKS 2052$-$47 & 0\% & 1.7 ($>$2.5$\sigma$) & \makecell{3.1 (1.1$\sigma$) \\ 1.7 (2.2$\sigma$)} & $\approx$0$\sigma$ \\
        J0818.2+4223 & 124.56174 & 42.38367 & bll &  0.530	& S4 0814+42 & 0\% & 2.2 (3.5$\sigma$) & 2.2 (0.8$\sigma$) & $\approx$0$\sigma$ \\ 
        J0211.2+1051 & 32.81532 & 10.85811 & bll & 0.2 & MG1 J021114+1051 & 0.6\% & 1.7 ($>$3.5$\sigma$) & 2.9 (0.8$\sigma$) & $\approx$0$\sigma$ \\ 
	J0102.8+5825 & 15.71134 & 58.41576 & fsrq & 0.644 & TXS 0059+581 & 4.3\% & 2.1 (3.0$\sigma$) & 4.0 (0.8$\sigma$) & $\approx$0$\sigma$ \\		
        J1454.5+5124 & 223.63225 & 51.413868 & bll & -- & TXS 1452+516 & 0.6\% & 2.1 ($>$3.5$\sigma$) & 2.1 (0.7$\sigma$) & $\approx$0$\sigma$ \\ 
	J0210.7$-$5101 & 32.68952 & -51.01695 & fsrq & 1.003 & PKS 0208$-$512 & 0\% & 2.6 ($>$3.0$\sigma$) & 3.8 (0.1$\sigma$) & $\approx$0$\sigma$ \\
\hline
\hline
\end{tabular}%
}
\end{table*}

\section{Results and Discussion} \label{sec:results}
We run the LCs of our 24 blazars through our improved pipeline. The results of the analysis are listed in Table \ref{tab:methods_results} and Table \ref{tab:mcmc_bayesian_arfima_results}.   

\begin{figure}
	\includegraphics[width=\columnwidth]{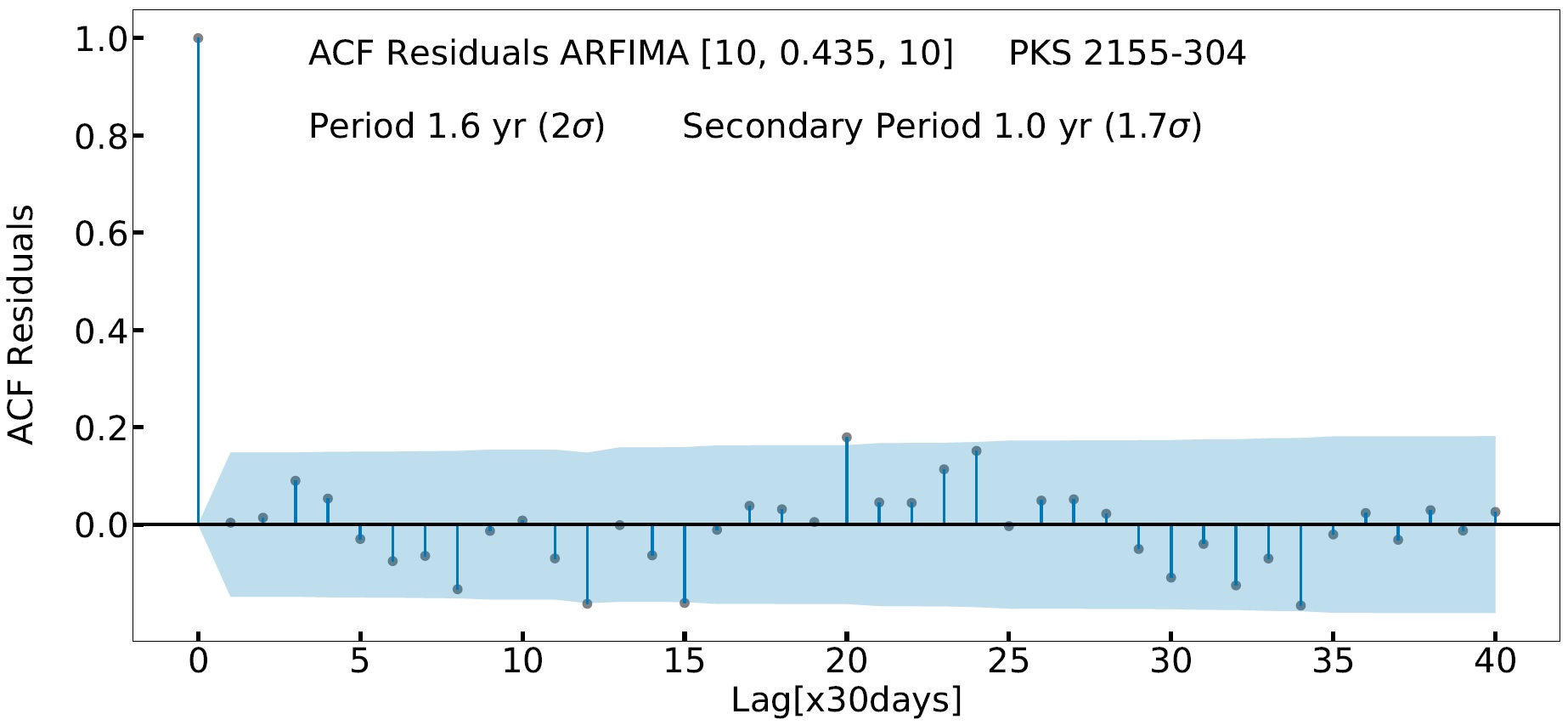}
	\caption{Correlation of the residuals (between the LC and the ARFIMA [p=10, d=0.453, q=10] model) for PKS 2155$-$204. The colored area represents the 2$\sigma$ confidence level. The peak at zero lag represents the autocorrelation of the residual with itself.}
	\label{fig:arfima}
\end{figure}

To sort the blazars, we quantify the periodicity test statistics by computing the median test statistics across all methods. 

\subsection{High-significance periodicity candidates} \label{sec:periodic}

\begin{figure*}
	\centering
	\includegraphics[scale=0.2295]{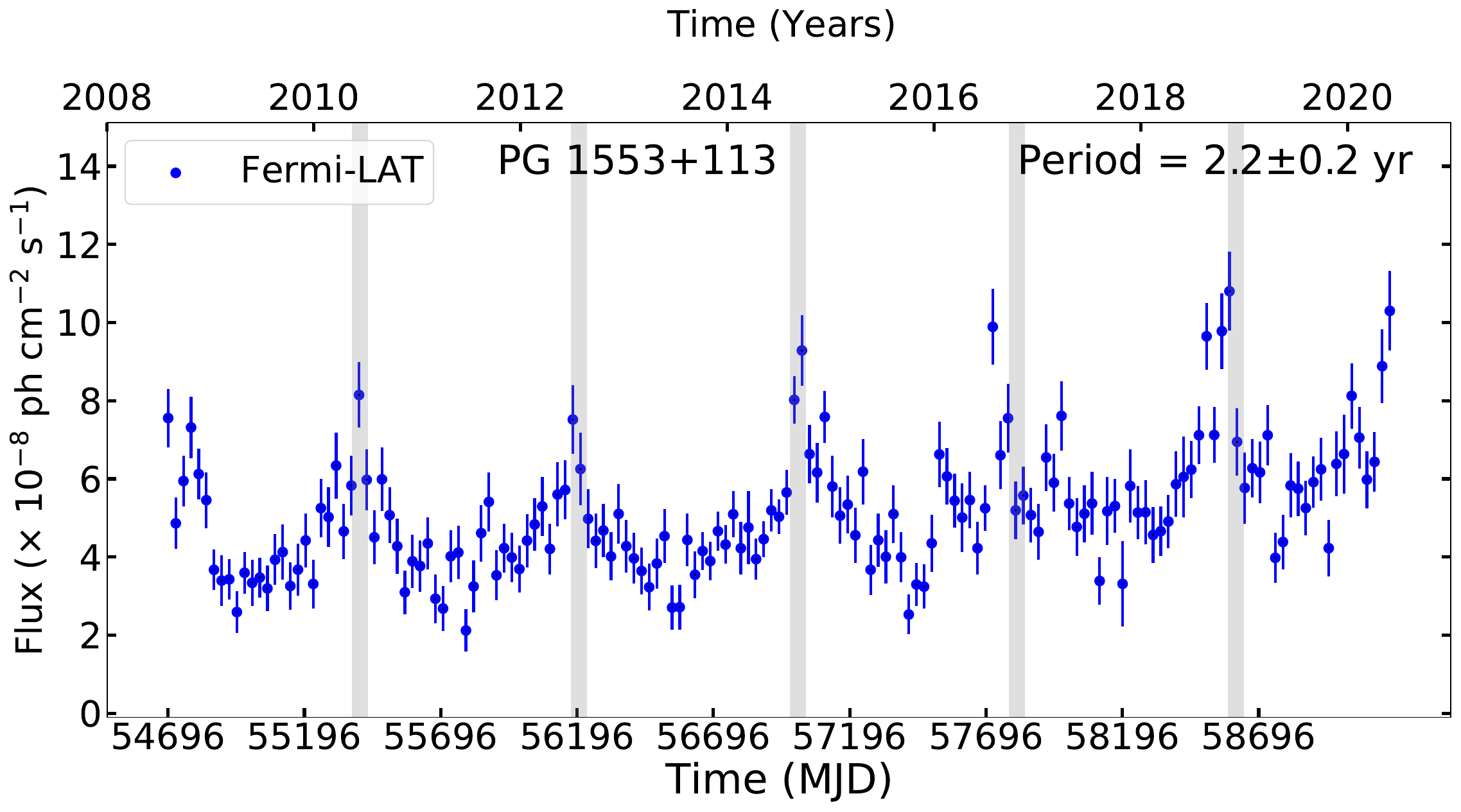}
  	\includegraphics[scale=0.2295]{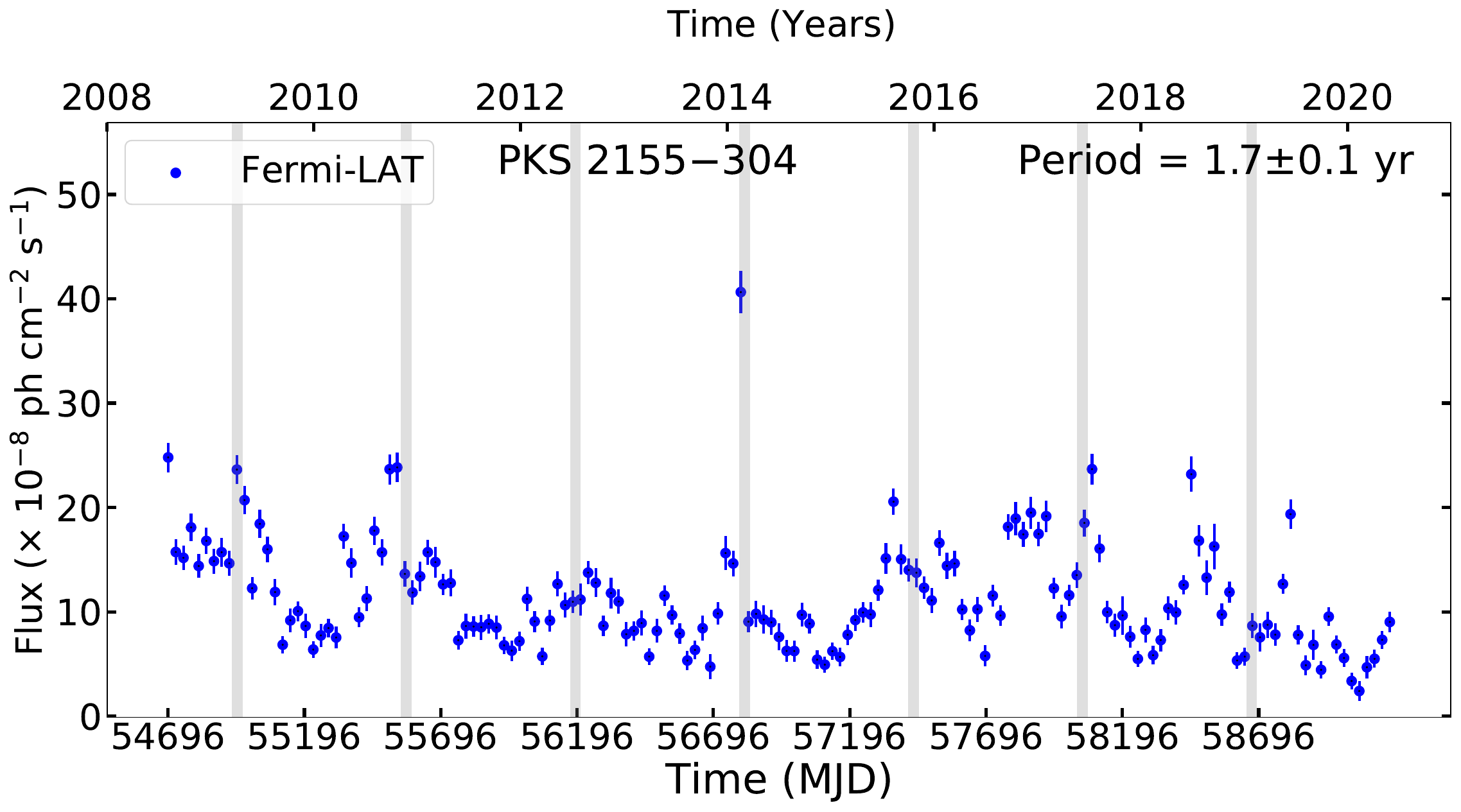}
	\caption{Light curves of the $>$4$\sigma$ periodicity sample. The gray vertical bars approximate high-flux periods suggested by the period inferred by the methodology for the given blazar. The width of the gray bars indicates the uncertainty in the periodic signal. Note that it is difficult to identify the periodic emission of S5~0716+714 by eye because this source may have two detected periods (see later).}
	\label{fig:lc_candidates}
\end{figure*}

The first group consists of the five blazars with periodicity detections at a median test statistics of $>$3\,$\sigma$, before the correction described in $\S$\ref{sec:global_sig} is applied. This sample includes 2 sources PG~1553+113 and PKS~2155$-$304 (Figure~\ref{fig:lc_candidates}). Two different scenarios arise when applying the ARFIMA/ARIMA methods:

\begin{enumerate}
	\item The methods find a significant period that is similar (compatible period) to those found via the methods listed in $\S$\ref{sec:methodology} and Table \ref{tab:methods_results} (e.g., PKS~2155$-$304).	
	\item The modeling provides a compatible period at $\rm{\geq1.5\sigma}$ (PG~1553+113, the period of $\approx$2 yr for the latter is observed as a secondary peak).
\end{enumerate}

Regarding PG~1553+113 and PKS~2155$-$304, we confirm our previous period detection \citep[e.g., compatible with][respectively]{tavani_pdm_pg_1553, zhang_pks2155}. 

\citet{covino_negation} searched for blazar periodicity in a sample of 10 blazars, including the 2 from our high-significance sample, and they did not find evidence for periodicity. In this case, the different results could be associated with the analysis methodology employed in \citet{covino_negation}. \citet{benkhali_power_spectrum} also reported mild support for periodicity in some of our high-significance candidates. 

Our results are, in general, not consistent with those by \citet[][ except for PG 1553+113]{yang_carma}. The reason can be that these authors analyze most of the candidates in P20 using the Continuous-time Autoregressive Moving Average (CARMA). However, according to \citet{feigelson_arima}, the CARMA method assumes weak stationary behavior in LCs without considering any trends in the flux. However, ARFIMA treats such behaviors in the LCs \citep[][]{feigelson_arima}. \citet{yang_carma} justify the use of CARMA on the basis that the LCs they analyzed are irregularly sampled (due to the presence of upper limits). However, moderately irregular measurements can be treated as evenly-spaced time series with missing data; hence, the ARIMA/ARFIMA models can also be applied and are better than the CARMA models for analyzing blazar LCs.

In \citet{alba_ssa}, the same period is obtained for both objects, with local significances of 4.8$\sigma$\footnote{The maximum significance to be obtained according to the number of artificial LCs used to estimate it.} and 4.5$\sigma$, for PG 1553+113 and PKS 2155$-$304, respectively. In this study, a novel method for periodicity analysis is employed: Singular Spectrum Analysis \citep[][]{ssa_greco, SSA_algorithm}. This method decomposes the time series into components (oscillatory, trend, and noise), allowing periodicity analysis while mitigating contaminating factors that can distort period search. This approach may explain the difference in local significance inference, as it effectively reduces the impact of distortion factors.

Similarly, in \citet{ren_s5_1044+71}, a comparable period is identified for both blazars, with local significances of 3.0$\sigma$ and 2.2$\sigma$, respectively. These discrepancies in local significance may arise from differences in the methodologies used to estimate them. Specifically, in \citet{ren_s5_1044+71}, the PSD is modeled with a smooth bending power-law plus white noise, as described in \citet{vaughan_bayesian}. However, the paper does not present the parameters and plots for each PSD fit, preventing a direct comparison with our estimations.

\subsection{Low-significance periodicity candidates} \label{sec:low}

The rest of the sample includes 9 blazars with periods detected at $\rm{\geqslant2\sigma}$ (OJ 014, PKS 0454$-$234, S5~0716+714, GB6 J0043+3426, TXS 0518+211, 87GB 164812.2+524023, PKS 0447-439, PKS 0426$-$380, PKS 0301$-$243; see Table \ref{tab:candidates_list} and Figure \ref{fig:lc_candidates_low}). We discard 13 candidates with a median test statistics of $\rm{<2.0\sigma}$ 
(S4 1144+40, PG 1246+586, PKS 0250$-$225, PKS 2255$-$282, TXS 1902+556, S3 0458$-$02, MG2 J130304+2434, PKS 2052$-$47, S4 0814+42, MG1 J021114+1051, TXS 0059+581, TXS 1452+516, PKS 0208$-$512). 

The periods of the blazars presented previously in the literature are mostly compatible with our corresponding results \citep[e.g., PKS 0426$-$380 and  MG2~J130304+2434,][, respectively]{zhang_pks0426, zhang_pks0301}. In the case of S5~0716+714, excluding the ARFIMA/ARIMA analysis, we confirm the period of $\approx$2.7 yr found in P20 and in the recent analysis by \citet{bhatta_s50714_optical}. This period differs from the $\approx$ 0.9 yr period reported by \citet{prokhorov_set}, \citet{li_S5_0716_714}, \citet{sandrinelli_S5_0716_71}, and \citet{bhatta_s5_0716}. However, we report the $\approx$0.9 yr period as a secondary peak (see Table \ref{tab:methods_results}). These disagreements may be due to the fact that we use more telescope time \citep[e.g., $\sim$8 yr in ][]{li_S5_0716_714, bhatta_s5_0716} or by the technique employed to infer test statistics \citep[e.g.,][]{prokhorov_set, bhatta_s5_0716}. For PKS~0454$-$234, \citet{bhatta_s5_0716} do not find any significant detection (note that we use more telescope time, and the test statistics is obtained differently).

In general, \citet{alba_ssa} reported a higher local significance for the previously studied objects, namely 4.8$\sigma$ for OJ 014 and TXS 0518+211. This increased significance is attributed to the analysis of a ``clean'' LC free of potential distorting factors.

We note that not all of the candidates presented by P20 increased the test statistics with the additional 3 years of data. For example, large drops in test statistics were found for TXS 0059+581, TXS 1452+516 or PKS 0208$-$512 (see Table~\ref{tab:candidates_list}). This fact might indicate that the inferred period could not be genuine \citep[e.g.,][]{liu_pg1302}.

This reduction of significance with more years could be due to different factors. For instance, the flare at the end of the LC of PKS 0208$-$512 affects the periodicity analysis. A flare adds a strong non-periodic feature to the LC, increasing the overall variability and introducing additional power across a broad range of frequencies. This distorts the periodogram and can obscure the previously detected periodic signal. As a result, the coherence of the signal weakens, and its statistical significance decreases. This scenario could also apply to TXS 0059+581. Although the flare was present in the 9-year LC, with the extra years, it could be more evident that the modulation is not sustained, and the signal is likely dominated by transient variability rather than a persistent periodic process. Consequently, the extension of the time baseline reveals that the previously obtained periodicity may have been biased by the episodic presence of the flare, rather than reflecting a stable periodic pattern.

In other scenarios (i.e., PG 1246+586, TXS 1452+516), the addition of three years of data can contribute to diluting the apparent coherence of the signal. This results in a higher dominance of red noise. As the baseline increases, low-frequency fluctuations inherent to red noise gain more statistical weight, often reducing the contrast between a potential periodic signal and stochastic variability. Consequently, the test statistics of the obtained period decrease as the signal becomes less distinguishable from the background noise structure.

\subsection{Lower energy counterparts} 
For some of the blazars in our sample, $\sim$year-long periods have previously been reported at lower energies. In this section, we review an incomplete segment of the relevant literature, primarily focusing on PG~1553+113 and PKS~2155$-$304, which arguably are the most popular and promising candidates. \cite{covino_gaussian} also summarize studies that have reported periodic LCs for PG~1553+113 and PKS~2155$-$304 in $\gamma$-ray and optical bands. Another recent discussion on the optical emission of PKS~2155$-$304 and S5 0716+714 can be found in section 3.3.1 of \citet{bhatta_s5_0716}.

\citet{ackermann_pg1553} first identified a $\approx \unit[2.1]{yr}$ periodicity in the optical, radio, and $\gamma$-ray bands for PG~1553+113. Their $\unit[2.1]{yr}$ period agrees with our results. A later study by \citet{covino_gaussian} used Gaussian process modeling and a longer temporal baseline to search for periodicity in PG 1553+113. This study found that the quality of the fit in both the optical and $\gamma$-ray bands can be improved by including a component with a similar period to the one found in our results ($\approx \unit[2.2]{yr}$).

To our knowledge, the first positive identification of $\sim$year-long periods for PKS~2155$-$304 was found in the optical band by \citet{fan_pks2155}. The periods they reported were approximately $\unit[4.2]{yr}$ and $\unit[7]{yr}$. A later study by \citet{zhang_pks2155_2014} used a much longer temporal baseline for analyzing optical data but did not recover these multi-year periods as primary periodicities. Instead, \citet{zhang_pks2155_2014} reported a period of $\approx \unit[0.87]{yr}$, and suggested that the $\unit[4.2]{yr}$ and $\unit[7]{yr}$ periods reported by~\citet{fan_pks2155} may be respectively 4th and 7th harmonics of the primary $\unit[0.87]{yr}$ period. The optical periodicity of $\approx \unit[0.87]{yr}$ for PKS~2155$-$304 is approximately half of the $\gamma$-ray periodicity of $\approx \unit[1.7]{yr}$ that we report in this work. \citet{sandrinelli_pks2155} also reported a detection of the $\approx \unit[0.87]{yr}$ periodicity in the optical band, as well as a tentative detection of a $\sim$year-long periodicity (similar to the $\approx \unit[1.7]{yr}$ periodicity in $\gamma$-rays that we report here). Their findings were reiterated in \citet{sandrinelli_redfit}. There is some indication of the $\unit[0.87]{yr}$ period in Figure 2 of the publication by \citet{covino_gaussian}, but that period was not the focus of their work. \citet{covino_gaussian} explain the seemingly discrepant periodicity measurements for PKS~2155$-$304 as possibly arising from differences in the noise models considered in the data analysis.

There are indications from some studies of consistent periodicities in the optical and $\gamma$-ray bands. For example, a recent analysis by \citet{bhatta_s50714_optical} found similar periods in the optical and $\gamma$-ray bands for two of our high-significance sources: $\approx \unit[1.7]{yr}$ for PKS~2155$-$304 and $\approx \unit[2.7]{yr}$ for S5 0716+714. In particular, they did not find evidence for the $\unit[0.87]{yr}$ optical period for PKS~2155$-$304 reported in \cite{zhang_pks2155_2014}. They argued that consistent periods in different bands are evidence of the periodicity being genuine.

\section{Periodicity Interpretation} \label{sec:discussion}
A variety of scenarios have been proposed in the literature to explain periodic emissions in blazars. Possible interpretations of long-term periodicity can be grouped into two categories: (1) periodicity in single black hole systems and (2) periodicity in binary black hole systems.

\subsection{Periodicity in single black hole systems}
In the first category, the emission mechanism is interpreted in the context of a single-SMBH system, where the periodic emission is produced by modulations in the accretion flow or by jet precession. For example, \citet{mohan_blobs} propose a scenario in which plasma inhomogeneities from the accretion disk propagate into the jet and generate periodic emission due to helical motions along magnetic surfaces within the jet. In models proposed in \citet{villata_jet_1999} and \citet{ostorero_jet_2004}, the precession of the jet axis induces periodic variation in the line-of-sight emission.

Alternatively, periodicity could emerge from the overall modulation of the accretion power, which is (sensibly) assumed to be reflected in the jet power. \cite{gracias_modulation_disk} invoked such scenario for the periodicity of PKS~2155$-$304 \citep[see][]{hess_pks2155}. The lognormal shape of the flux PDF suggests an underlying multiplicative process \citep[e.g.,][]{shah_lognormality}; as such, it favors an accretion-disk origin for the periodicity over an additive process such as jets-in-jet \citep[e.g.,][]{distribution_uttley, rieger_2019}. Since accretion disks in the vicinity of SMBHs are expected to be dominated by radiation pressure, they are subject to a variety of instabilities \citep{lightman_eardley1974}, which can lead to limit-cycle behavior that conceivably appears as quasi-periodic emission over a few cycles \citep{FKR}. Recent 3-dimensional simulations of a single SMBH give a concrete example of opacity-driven light curve variability lasting for $2-6$ cycles, encompassing a span of years to decades \citep{jiang_opacity}. Over a few cycles, such variability may appear to be quasi-periodic.

\subsection{Periodicity in binary black hole systems}
The second category of proposed mechanisms invokes an SMBHB. There are two variations of this scenario: (i) perturbations caused by the secondary SMBH destabilize the accretion flow of the primary SMBH, modulating the accretion rate and, as a consequence, the luminosity of the blazar \citep{sandrinelli_redfit}; (ii) the periodic emission may be due to jet precession, ultimately caused by the orbiting SMBHs \citep{sobacchi_binary_2017}. The SMBHB hypothesis has been applied to explain the LC of PG 1553+113 \citep[e.g.,][]{cavaliere_binary_2017, sobacchi_binary_2017, tavani_pdm_pg_1553}.

Major galaxy mergers are found more frequently at a moderate redshift of about $z\sim1$ when galaxies are also more gas-rich \citep[e.g.,][]{tacconi_gas, koss_gass_rich}. In turn, the number of SMBHBs should increase with increasing redshift and increasing mass \citep{volontery_binary}. Most of our candidates reside at redshifts  $z\gtrsim1$ (see Table~\ref{tab:candidates_list}).

In order to discriminate between different hypotheses, the periodicity search may be complemented by spectroscopy. Two different approaches can be useful to infer the presence of an SMBHB by using optical emission lines: radial velocity shifts and line shapes. Regarding radial velocity shifts, the lines are expected to oscillate about their rest-frame wavelength on the time scale corresponding to the orbital period \citep[e.g.,][, typically one targets the $\rm{H\beta}$ and $\rm{Mg{II}}$ lines]{liu_binary}. The other technique to infer close binary systems is to observe double-peaked profiles in the emission lines \citep[e.g. H$\beta$,][]{kovacevic_shape_lines}.

\subsection{Application of the binary hypothesis to PG~1553+113 and PKS~2155$-$304} \label{sec:bin}

Here we discuss the application of the binary hypothesis to two blazars in our sample with the highest-significance periodicity: PG 1553+113 and  PKS 2155$-$304. These candidates are arguably the most promising because they are bright and have the greatest number of cycles ($\approx 6$ for PG 1553+113, $\approx 7$ for PKS 2155$-$304), and our autoregression analysis (ARFIMA/ARIMA) reports compatible periods with the methods of Table~\ref{tab:mcmc_bayesian_arfima_results}.

\citet{covino_gaussian} used Gaussian processes to analyze data from optical and previous {\it Fermi}-LAT observations. They found that adding a periodic component increased the statistical quality of fit for both candidates PG 1553+113 and PKS 2155$-$304 (although it is less convincing for PG 1553+113). A large number of cycles also helps in addressing the concerns raised by \citet{vaughan_criticism} regarding spurious periodicity in LCs with only a few putative periods present. Also, note that the test statistics we report has increased with the addition of 3 years of data; in contrast, adding new data for quasar PG 1302$-$102 resulted in a decrease in test statistics for the periodicity, which was seen as evidence against the authenticity of the periodicity \citep{liu_pg1302}. The idea was that if the periodicity were genuine, one would generally expect the test statistics of the detected period to become larger with new data. We note that the latest upper bound on the binary fraction of blazars from the Pulsar Timing Array \citep[at most $\approx 1/1000$ blazars host binaries with orbital periods less than 5 years,][]{holgado_pta} is compatible with the presence of few true SMBHBs in the \emph{Fermi}-LAT catalog \citep[][]{holgado_pta}.

For PKS 2155$-$304, there is a disparity in the mass estimates for the SMBH. The $M_{\rm BH} - L_{\rm bulge}$ relation suggests a central black hole mass of $M_{\rm BH} \gtrsim 2\times10^8 M_\odot$. On the other hand, VHE variability on time scales of $\lesssim$ 200 seconds suggest a black hole mass of $M_{\rm BH} \lesssim 4\times10^7 M_\odot$ \citep{rieger_2010}. Variability in the X-ray band supports a lower mass estimate \citep{hayashida_xray, czerny_xray}. \cite{rieger_2010} argued that the discrepancy in mass estimates could be explained if PKS 2155$-$304 hosts an SMBHB: the $M_{\rm BH} - L_{\rm bulge}$ relation measures the binary mass while the fast variability reflects the mass of the secondary SMBH. This scenario requires that the jet emission comes mostly from the secondary SMBH, which is \emph{prima facie} consistent with recent simulations of accreting unequal-mass binaries \citep{duffell_massratio}. These simulations show that the secondary can accrete as much as $10\times$ faster than the primary. 
The different mass estimates constitute a bound on the SMBHB mass ratio$-q$ $\lesssim 0.2$, which implies that the secondary accretes at $\approx 3-10\times$ the rate of the primary \citep[as seen in the simulations of ][]{duffell_massratio}. However, the physics of jet launching is rather complicated and not fully understood, so caution is warranted with circumstantial evidence such as this. Caution is also warranted when applying 2-dimensional thin-disk simulation results such as from \citet{duffell_massratio}, since that setup uses a variety of approximations. We refer to \citet{rieger_2019} and references therein for a recent overview of the binary hypothesis for PKS 2155$-$304. Attempts to explain the short time-scale variability without a binary are given in \citet{begelman_variability} and \citet{levinson_variability}.

For PG 1553+113, variability on a 2400-second time scale leads to an SMBH mass estimate of $M_{\rm BH} \approx 4 \times 10^{7}$ - $8 \times 10^{8} M_{\odot}$, depending on the black hole’s spin \citep[with higher spins implying larger masses, ][]{dhiman_variability}. However, an estimate based on the $M_{\rm BH}$$-$$L_{\rm bulge}$ relation is not possible because the host galaxy has not yet been identified. As a basis for discussion, let us consider the orbital parameters from the binary model proposed by \citet{cavaliere_binary_2017}: primary SMBH mass $5\times10^8 M_\odot$, mass ratio $q=0.1$, eccentricity $e\approx 0.2$. In the model, mild non-zero eccentricity is required for the secondary SMBH to perturb the presumed jet from the primary SMBH on the orbital time scale. Thin-disk simulations find two values of the eccentricity that are stable under gas-driven binary evolution, $e = 0$ (circular) and $e \approx 0.45$ \citep{zrake_eccentricity, dorazio_eccentricity}. Thus, one may infer that the $e=0.2$ binary has been circularizing for some time due to gravitational wave emission. Consistency then requires that the evolution of the binary is already dominated by the emission of gravitational waves (i.e., the orbital radius is less than the ``gas decoupling'' orbital radius). Let the binary semi-major axis be $a$ and its total mass be $M$. We can estimate the orbital period when gas-driven evolution is on par with gravitational wave-driven evolution by assuming post-Newtonian orbital evolution \citep{peters_1964} and using $d \log a/d \log M = -l$. $l$ is the ``accretion eigenvalue'', which determines the rate at which the binary semi-major axis evolves (and whether the binary is expanding or shrinking). Here, $l\sim \mathcal{O}(1)$ is determined by gas-accretion physics from thin-disk simulations. In following this procedure, we can set $a$, the rate of gas-driven evolution, to that of the gravitational-wave—driven evolution and solve for the orbital period. A positive value of $l$ corresponds to a gas-driven inspiral, which is reasonable for large disk Mach numbers \citep[$\gtrsim 25$;][]{tiede_mach} typically expected in AGN disks. Let us consider an accretion rate equal to the Eddington limit (assuming a typical efficiency of $\epsilon=0.1$). Even if we use a very large value $l=10$ (which results in a shorter orbital period at decoupling), we find that the orbital period at decoupling is nearly 6 years. Thus, under these standard assumptions, the proposed binary seems to have transitioned into the gravitational wave-driven regime with an orbital period of $\approx 1.5$ years (in the source frame). It would be interesting to carefully compute the orbital evolution history to see whether the proposed eccentricity of $e=0.2$ at the current orbital separation is consistent with the equilibrium value prior to the time of decoupling. 

Lastly, simulations of accreting binaries have generally shown two possible main periodicities in accretion rates$-$that of the binary's orbital period, $T_{\rm binary}$, and that of an overdensity (called a ``lump'') orbiting at larger radii with a period $\approx (5-10)\times T_{\rm binary}$. The lump period is expected to dominate for mass ratios $q\gtrsim 0.2$ \citep{duffell_massratio, farris_2014, dorazio_massratio} and eccentricities $e\lesssim 0.1$ \citep{miranda_eccentricity, zrake_eccentricity}. If the periodic LCs revealed in this study correspond to the lump period in a circumbinary disk, then the SMBHBs are orbiting faster and are thus closer to merging. Future measurements of gravitational waves from SMBHBs could reveal whether the electromagnetic emission is being modulated by the lump or the binary orbital period \citep{dorazio_pg1302}.

\section{Summary} \label{sec:summary}
In this work, we analyze the $\gamma$-ray LCs of the most promising 24 periodicity candidates presented by P20. Relative to P20, these LCs are extended from 9 to 12 years of {\it Fermi}-LAT data and the energies are expanded from $>$1~GeV to $>$0.1~GeV, increasing the signal-to-noise ratio. In addition, the upper limits are treated in a way that allows us to retain information about the flux variations. We also employ an improved version of our periodicity-search pipeline, composed of seven different methods to infer periods and the associated test statistics. This pipeline is improved by including the ARIMA/ARFIMA autoregressive algorithm. As a result, we have obtained that 1553+113 with periodic $\gamma$-ray emission detected at a ``global significance'' level of  $\approx1.8\sigma$.


\section{Acknowledgements}
The \textit{Fermi}-LAT Collaboration acknowledges generous ongoing support
from a number of agencies and institutes that have supported both the
development and the operation of the LAT as well as scientific data analysis.
These include the National Aeronautics and Space Administration and the
Department of Energy in the United States, the Commissariat \`a l'Energie Atomique and the Centre National de la Recherche Scientifique / Institut National de Physique Nucl\'eaire et de Physique des Particules in France, the Agenzia Spaziale Italiana and the Istituto Nazionale di Fisica Nucleare in Italy, the Ministry of Education,Culture, Sports, Science and Technology (MEXT), High Energy Accelerator Research Organization (KEK) and Japan Aerospace Exploration Agency (JAXA) in Japan, and the K.~A.~Wallenberg Foundation, the Swedish Research Council and the Swedish National Space Board in Sweden.

Additional support for science analysis during the operations phase is gratefully acknowledged from the Istituto Nazionale di Astrofisica in Italy and the Centre National d'\'Etudes Spatiales in France. This work was performed in part under DOE Contract DE-AC02-76SF00515.

P.P and M.A acknowledge funding under NASA contract 80NSSC20K1562. S.B. acknowledges financial support by the European Research Council for the ERC Starting grant MessMapp, under contract no. 949555, and by the German Science Foundation DFG, research grant “Relativistic Jets in Active Galaxies” (FOR 5195, grant No. 443220636). A.D. is thankful for the support of the Ram{\'o}n y Cajal program from the Spanish MINECO, Proyecto PID2021-126536OA-I00 funded by MCIN / AEI / 10.13039/501100011033, and Proyecto PR44/21‐29915 funded by the Santander Bank and Universidad Complutense de Madrid.

This work was supported by the European Research Council, ERC Starting grant \emph{MessMapp}, S.B. Principal Investigator, under contract no. 949555, and by the German Science Foundation DFG, research grant “Relativistic Jets in Active Galaxies” (FOR 5195, grant No. 443220636).

Finally, the authors acknowledge S. Fegan for useful discussions, which helped to improve the paper.

\section*{Data Availability}

All the data used in this work are publicly available or available on request to the responsible for the corresponding observatory/facility.

\section{Software}
\begin{enumerate}
        \item arfima R-software \citep{arima_hyndman_2008, arima_hyndman_2024},
	\item astroML \citep{astroml},
	\item astropy \citep{astropy_2013, astropy_2018, astropy_2022}, 
	\item colorednoise \url{https://github.com/felixpatzelt/colorednoise},
	\item emcee \citep {emcee}, 
	\item fermipy software package \citep{Wood:2017yyb},
	\item PyAstronomy \citep{PyAstronomy},	  
        \item PyCWT\footnote{We have modified the sources of the code to estimate the test statistics with  \citet{connolly_code}}  \url{https://pypi.org/project/pycwt/},
		\item rpy2 \url{https://rpy2.github.io/doc/latest/html/index.html},
	\item stats R-software \citep{r_manual},  
	\item statsmodels \citep[][]{stats_scipy_2010},, 
	\item SciPy \citep {SciPy},
	\item Simulating light curves \citep{connolly_code},
\end{enumerate}

\bibliographystyle{mnras}
\bibliography{literature} 
\appendix
\setcounter{table}{0}
\setcounter{figure}{0}
\renewcommand{\thetable}{A\arabic{table}}
\renewcommand{\thefigure}{A\arabic{figure}}
\section*{Appendix}\label{sec:appendix}

\subsection{Light Curves}
Figure \ref{fig:lc_candidates_low} reports the LCs of the low-significance blazars. 
\begin{figure*}
	\centering
         \includegraphics[scale=0.2195]{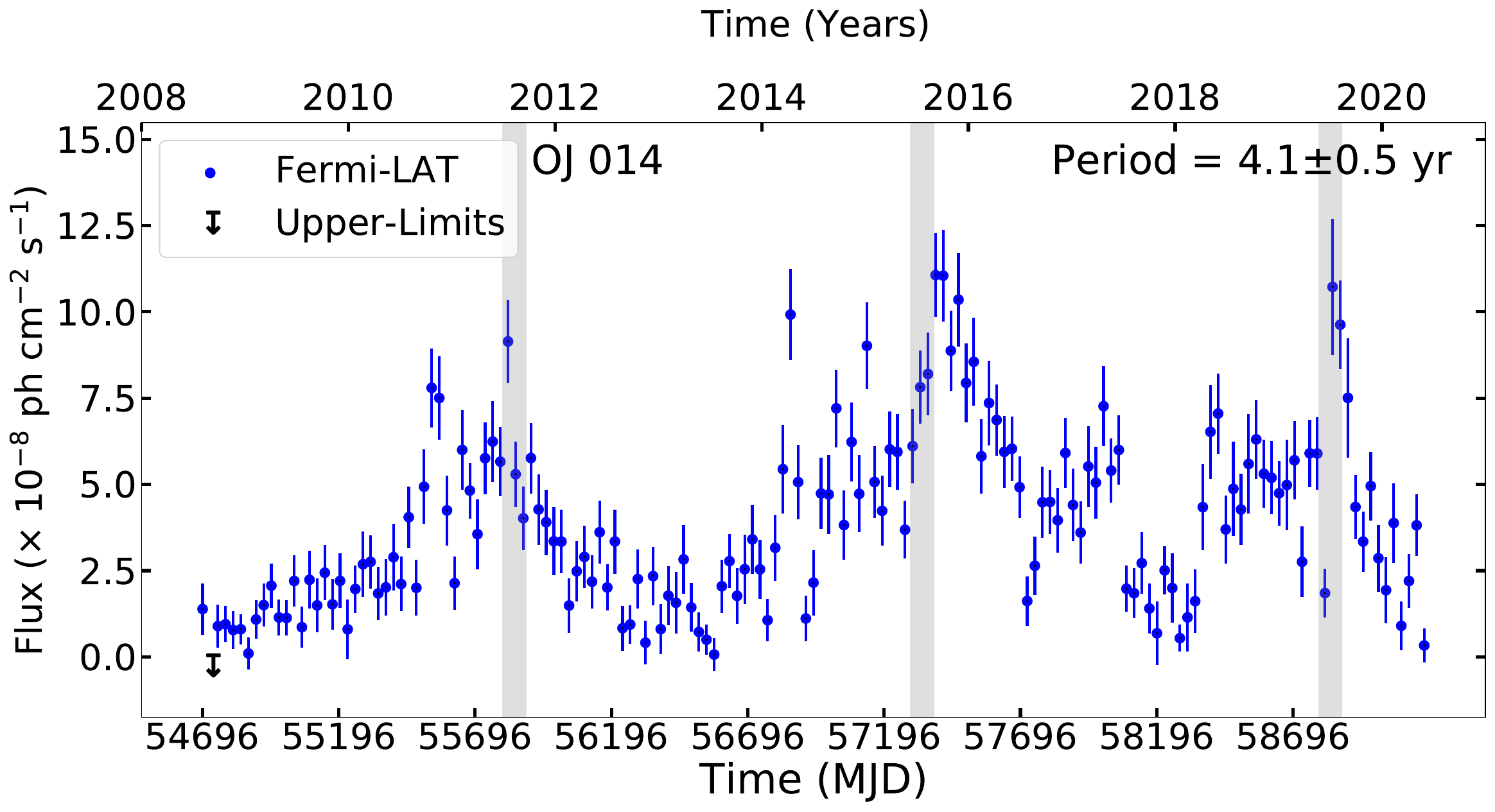}
	\includegraphics[scale=0.2115]{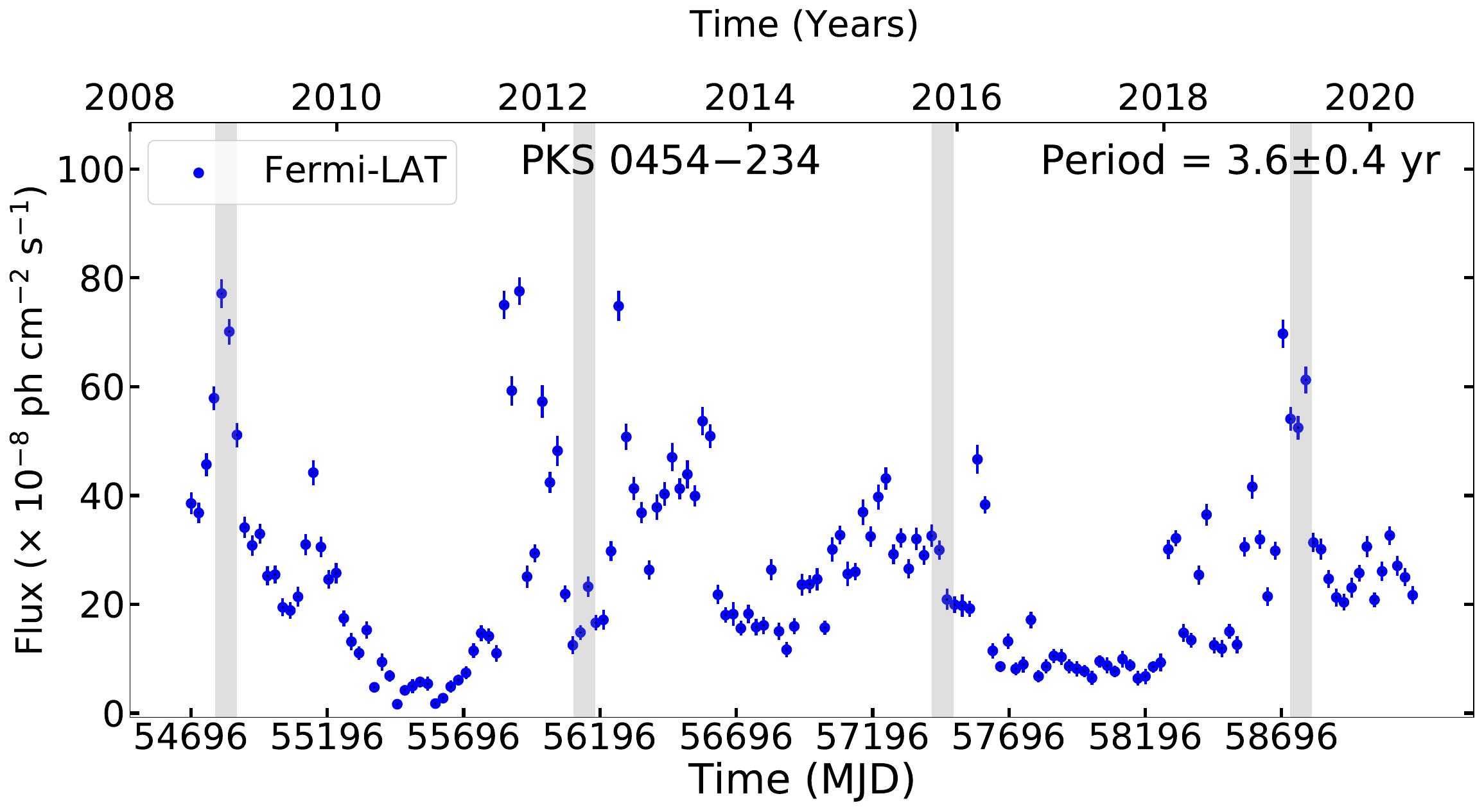}
	\includegraphics[scale=0.2195]{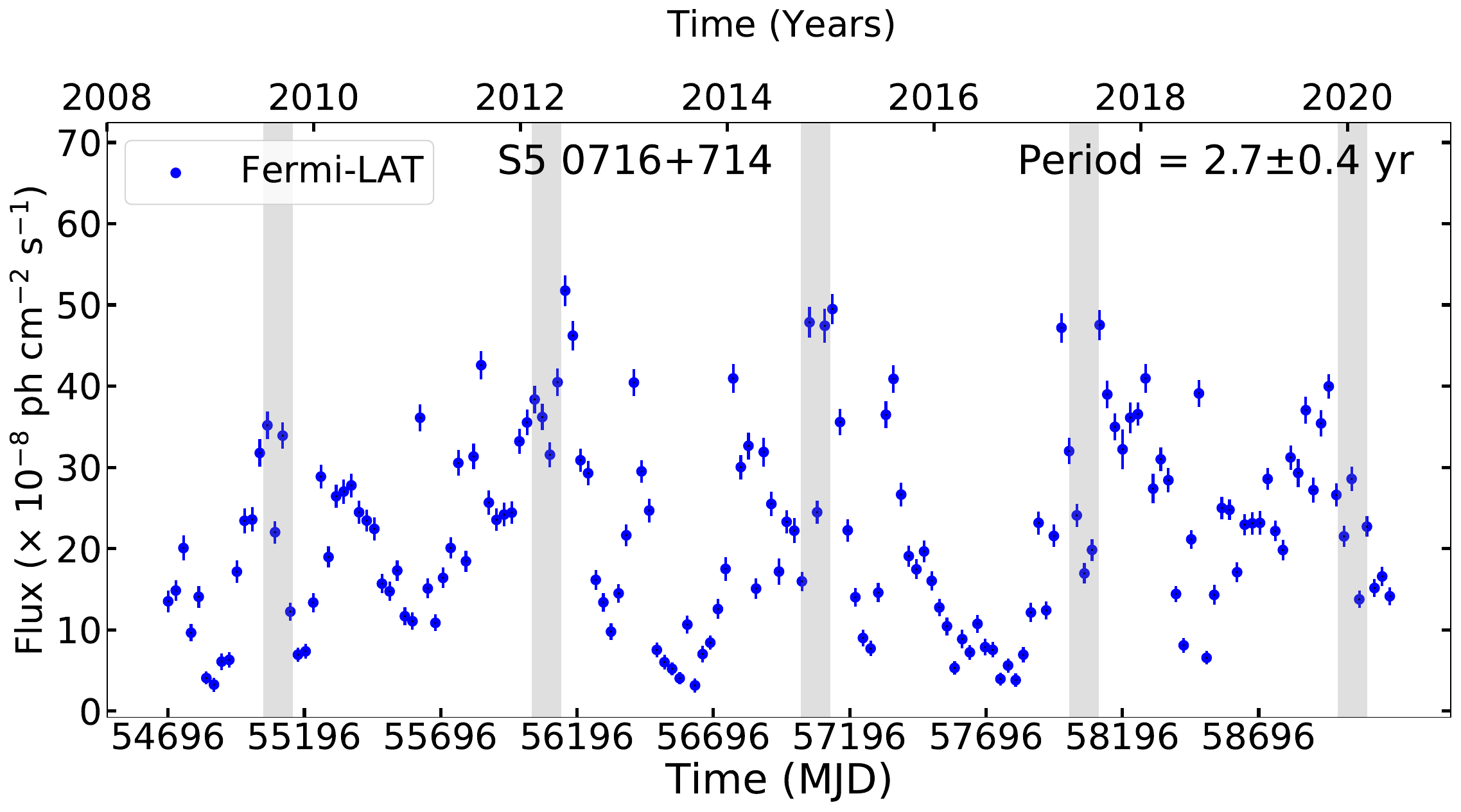}
         \includegraphics[scale=0.2195]{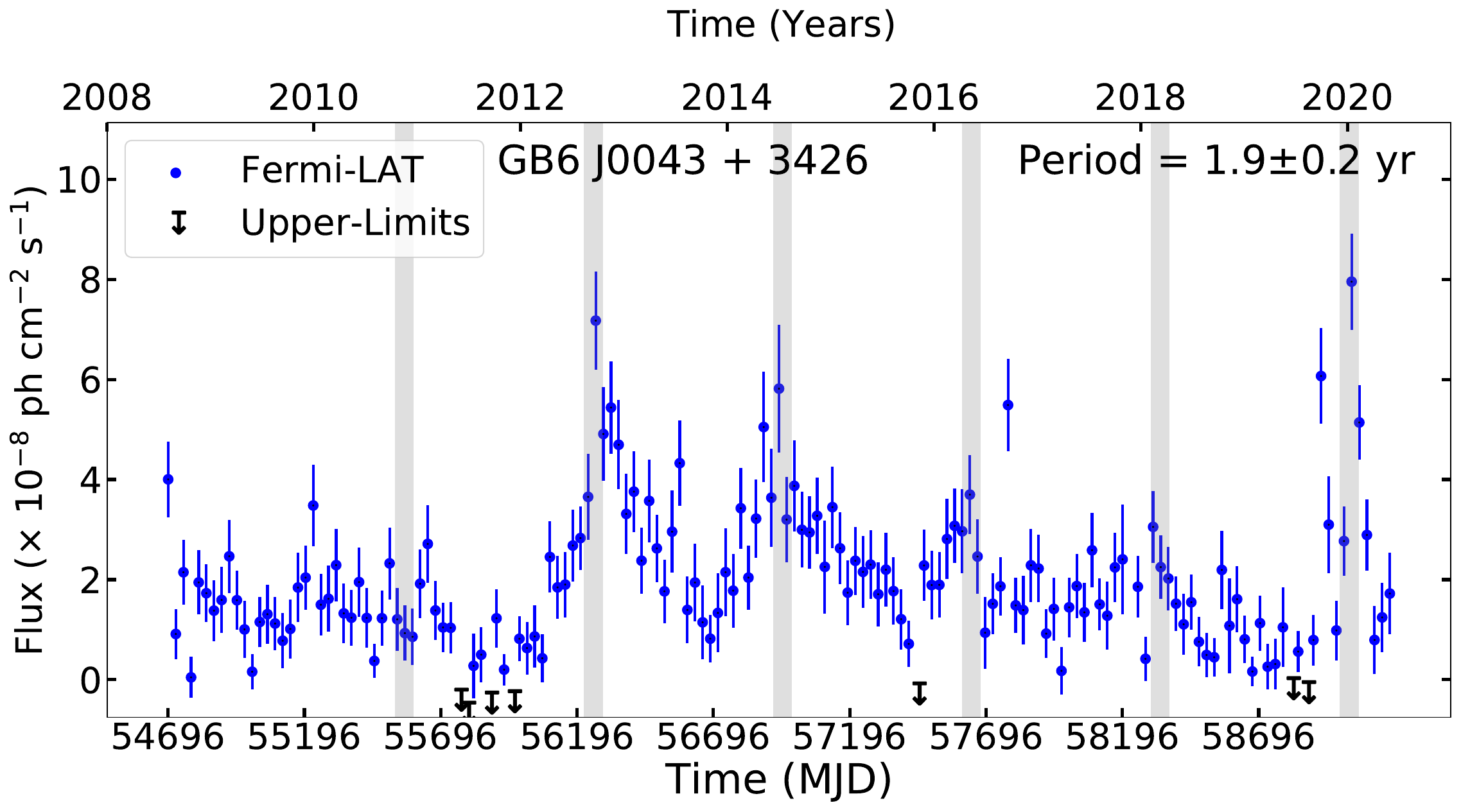}
         \includegraphics[scale=0.2195]{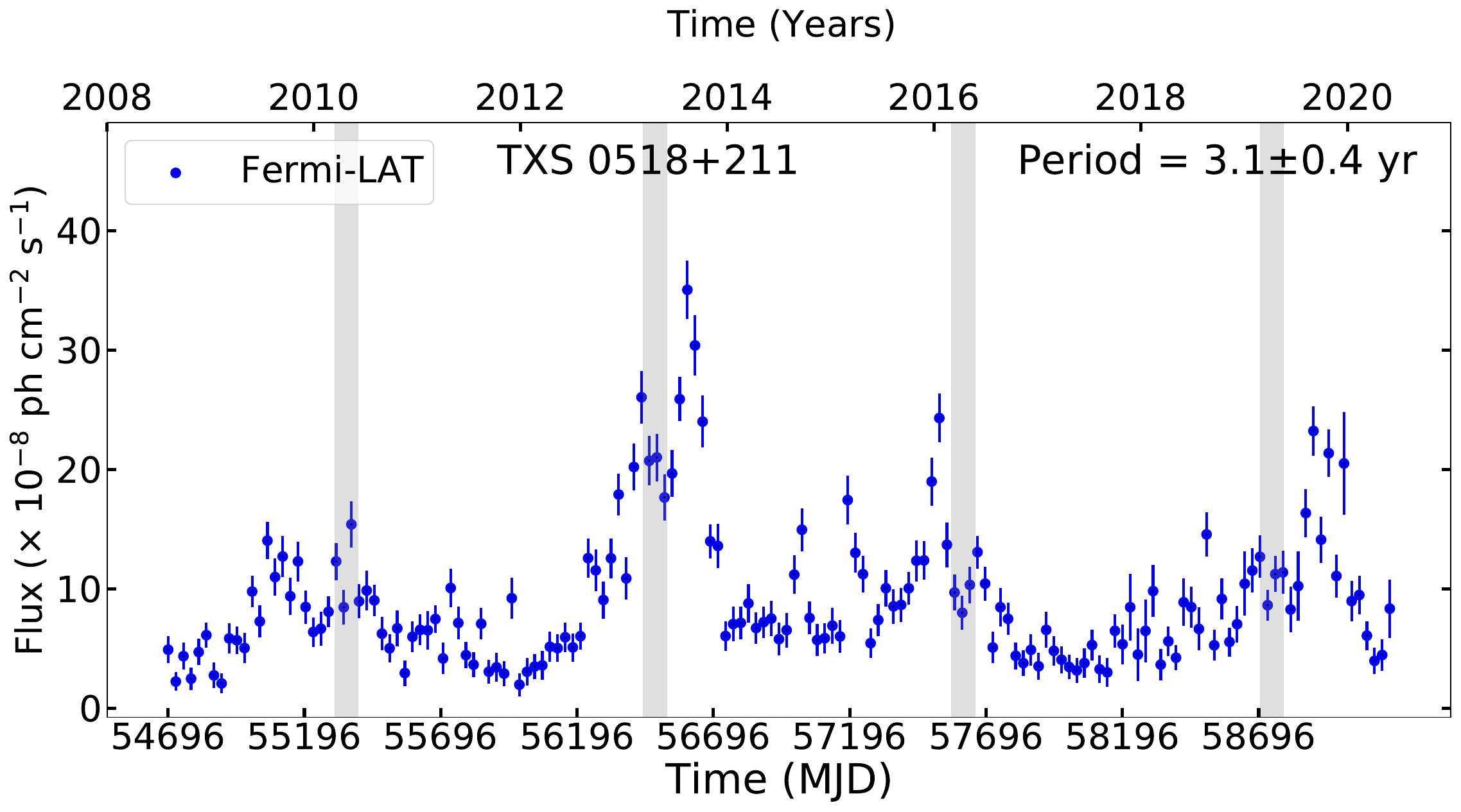}
         \includegraphics[scale=0.2195]{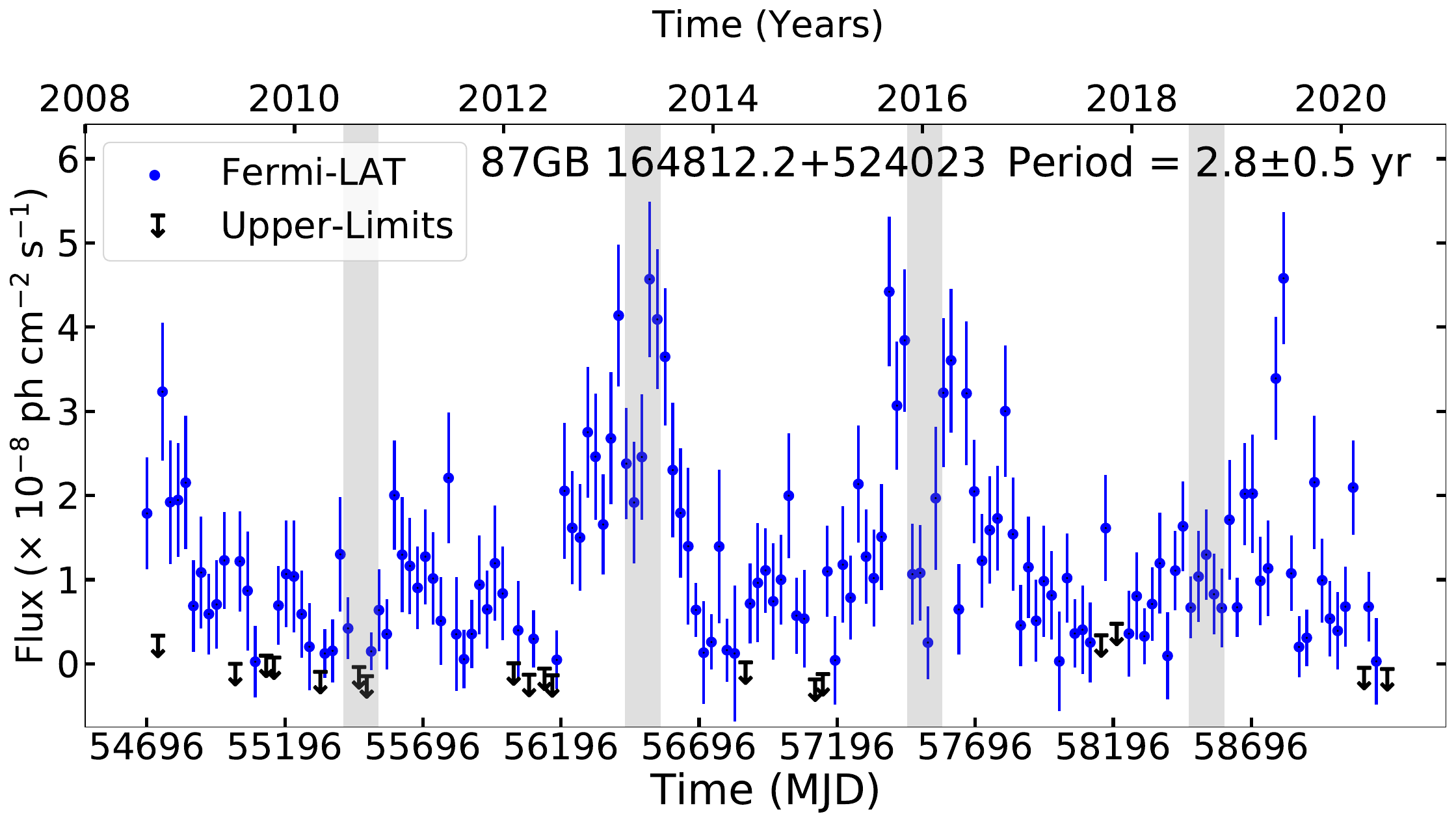}         
         \includegraphics[scale=0.2195]{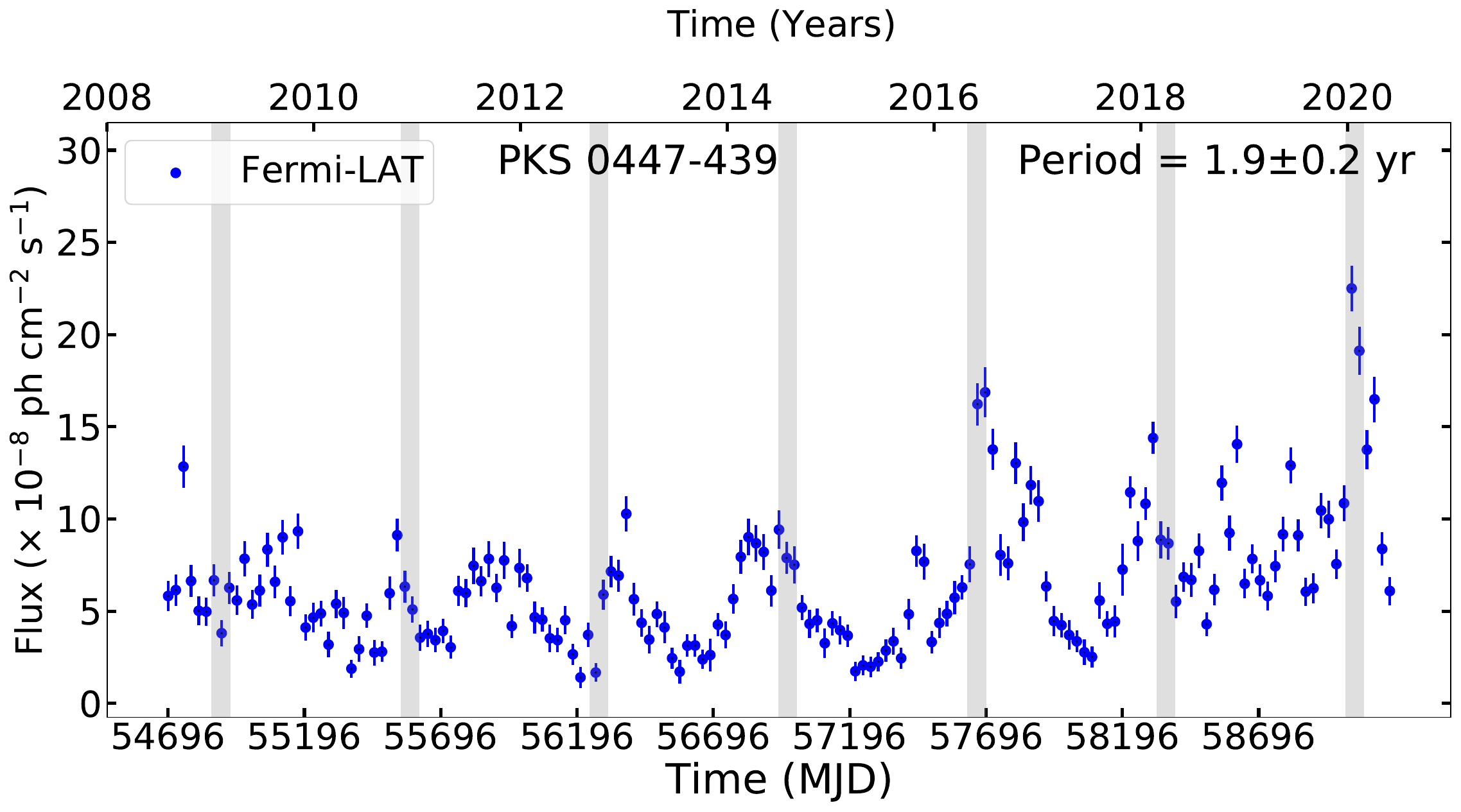}
	\includegraphics[scale=0.2165]{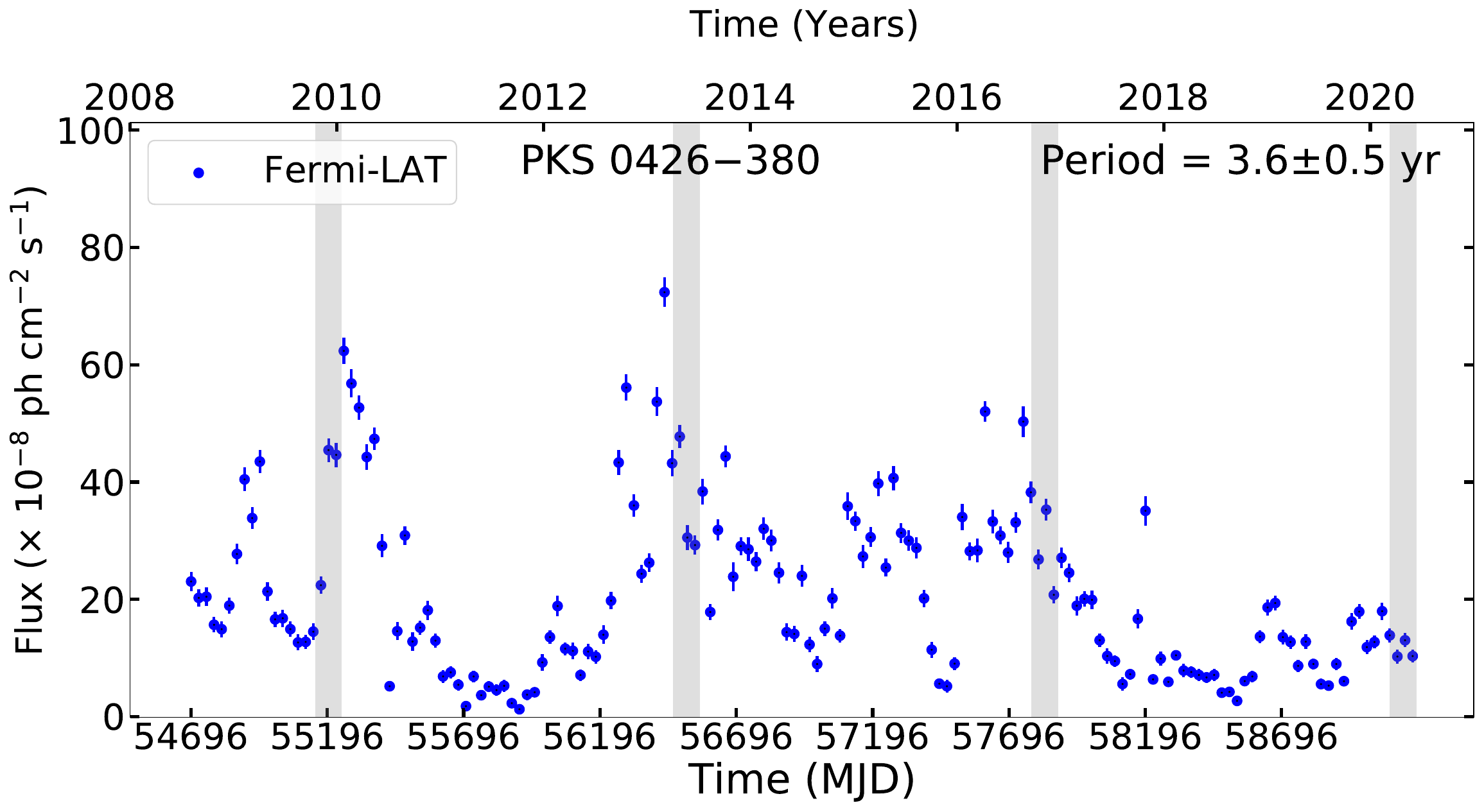}
	\caption{Light curves of the low-significance blazars presented in Table \ref{tab:candidates_list}. The gray vertical bars approximate high-flux periods suggested by the period inferred by the methodology for the given blazar. The width of the gray bars indicates the uncertainty in the periodic signal.}
	\label{fig:lc_candidates_low}
\end{figure*}
\setcounter{figure}{1}
\begin{figure*}
	\centering	
         \includegraphics[scale=0.2195]{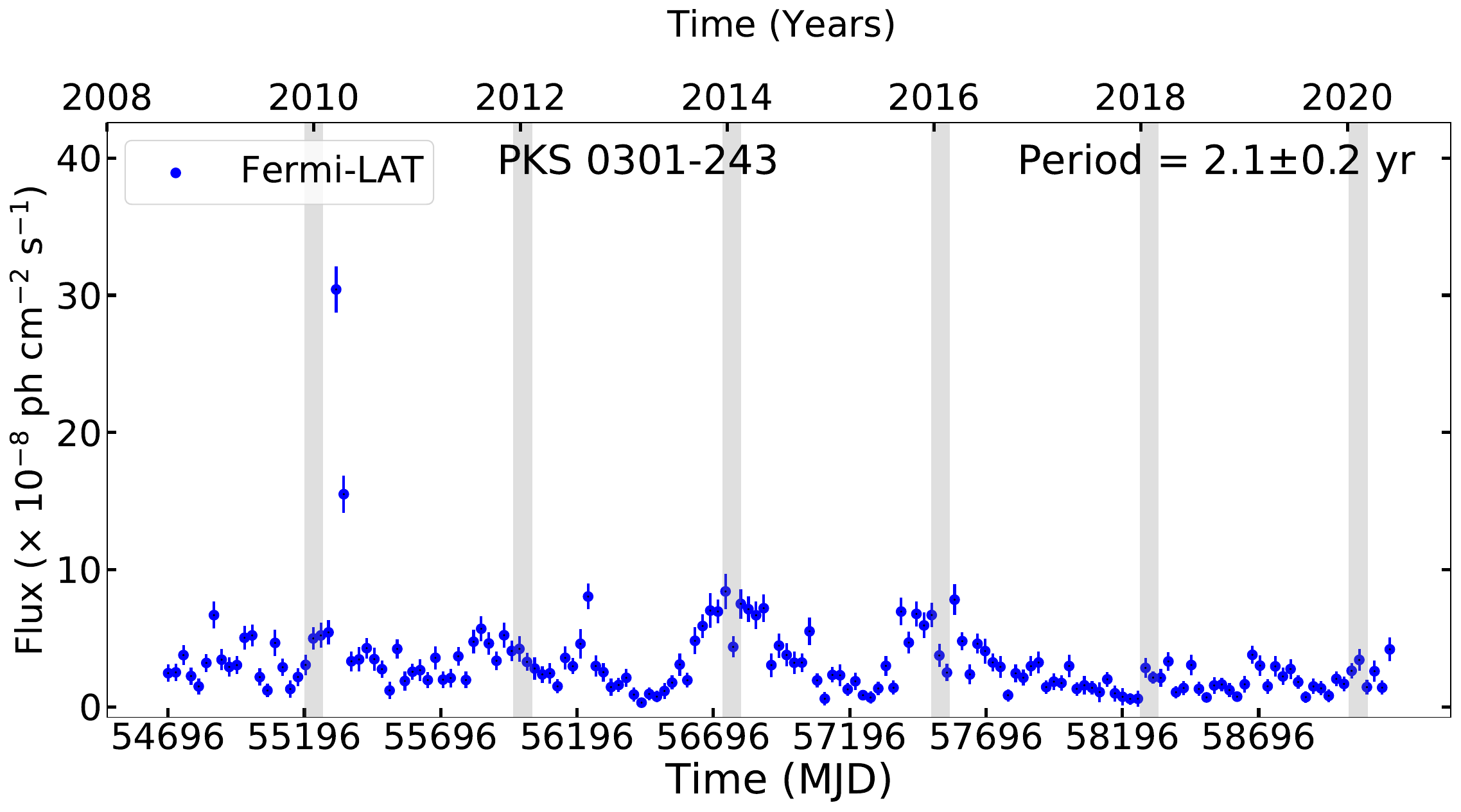}
        \includegraphics[scale=0.2195]{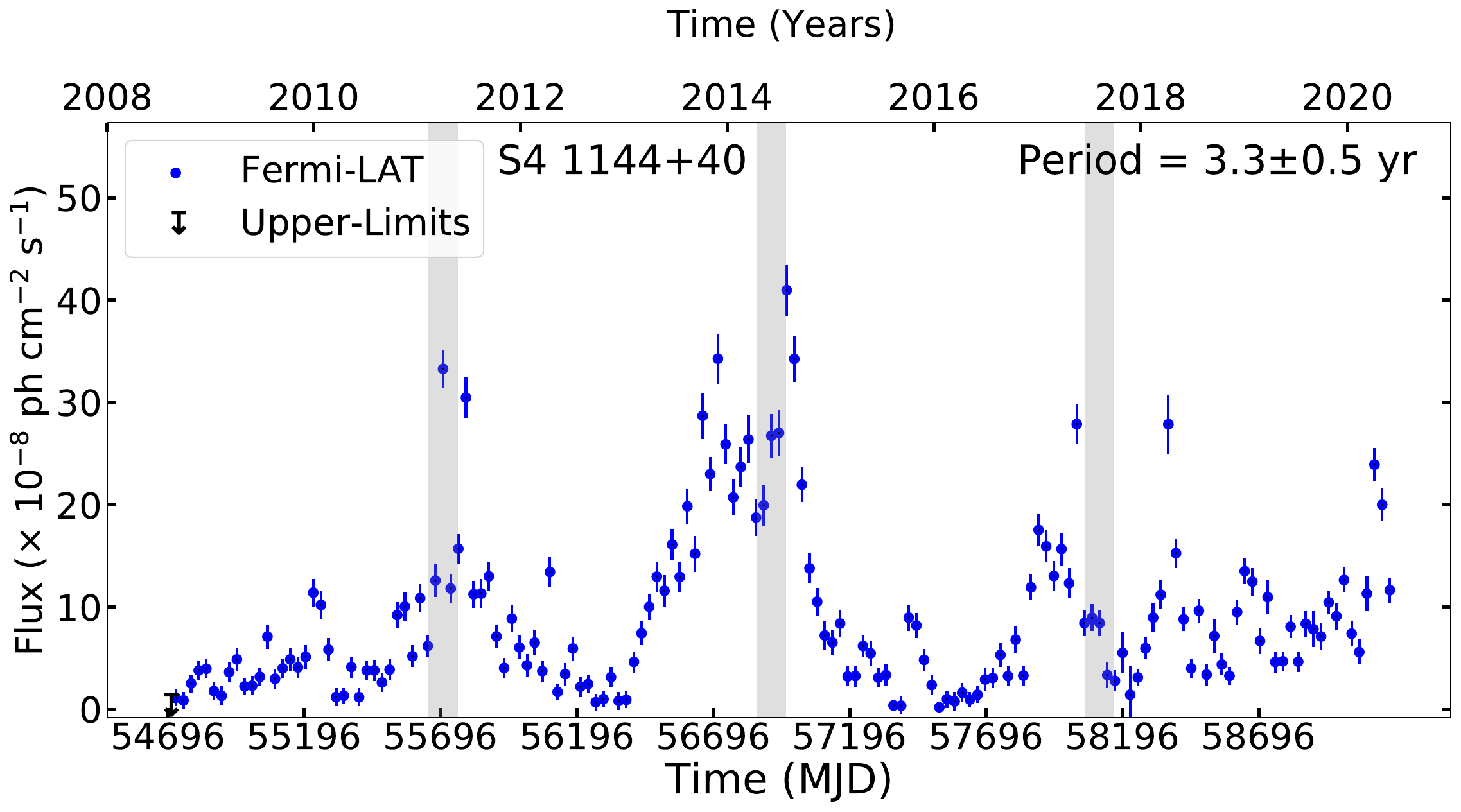}
        \includegraphics[scale=0.2195]{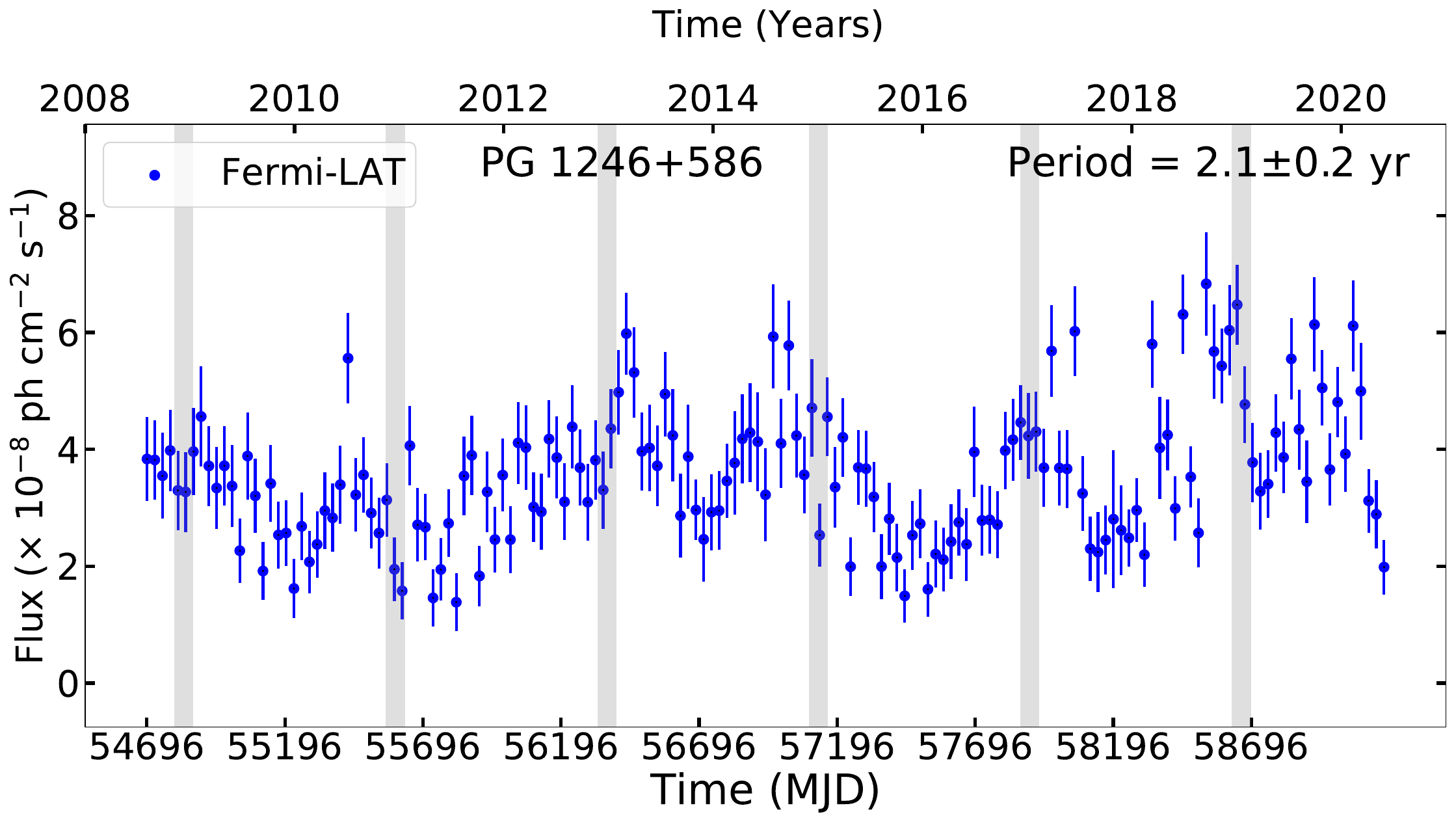}        
        \includegraphics[scale=0.2195]{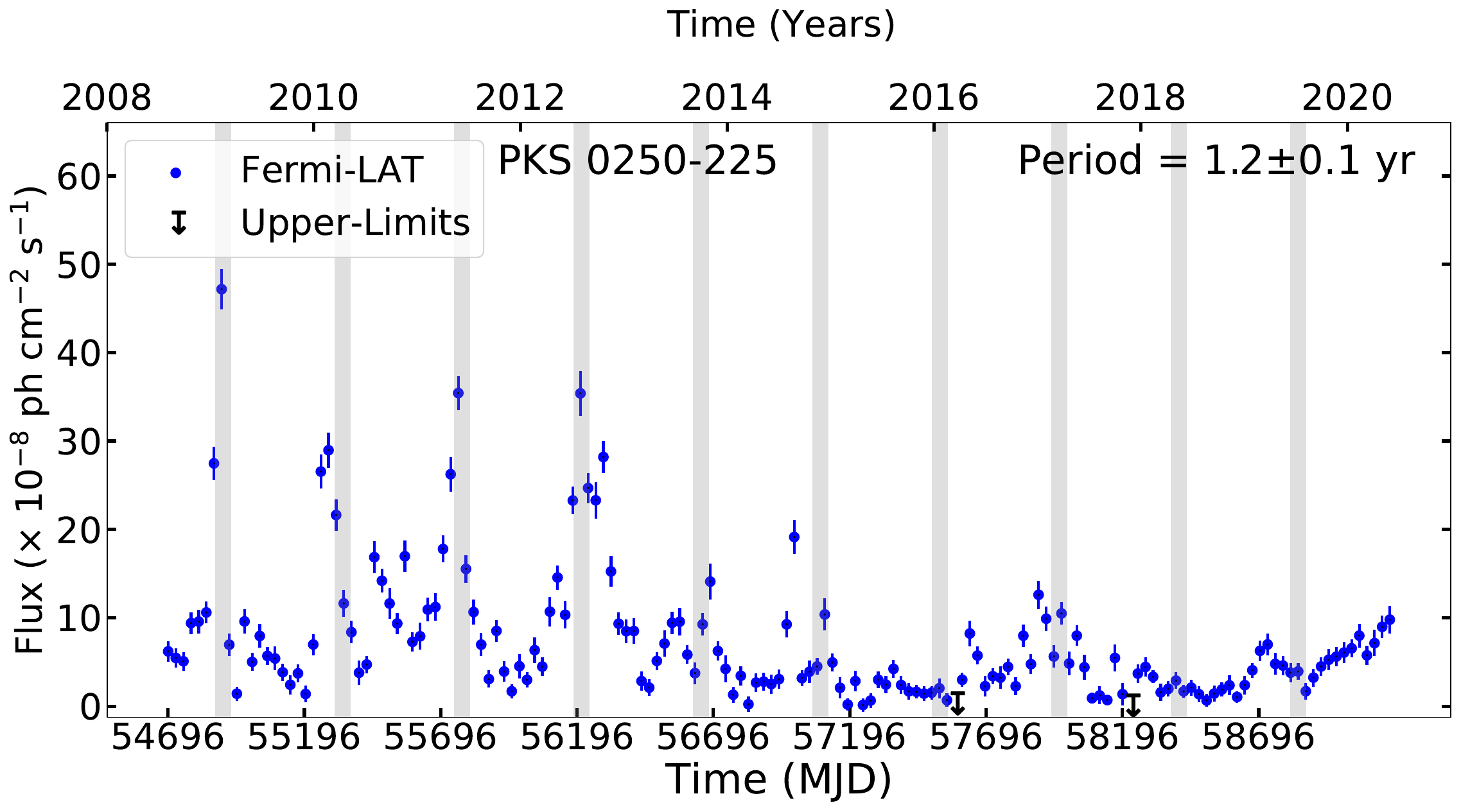}
        \includegraphics[scale=0.2195]{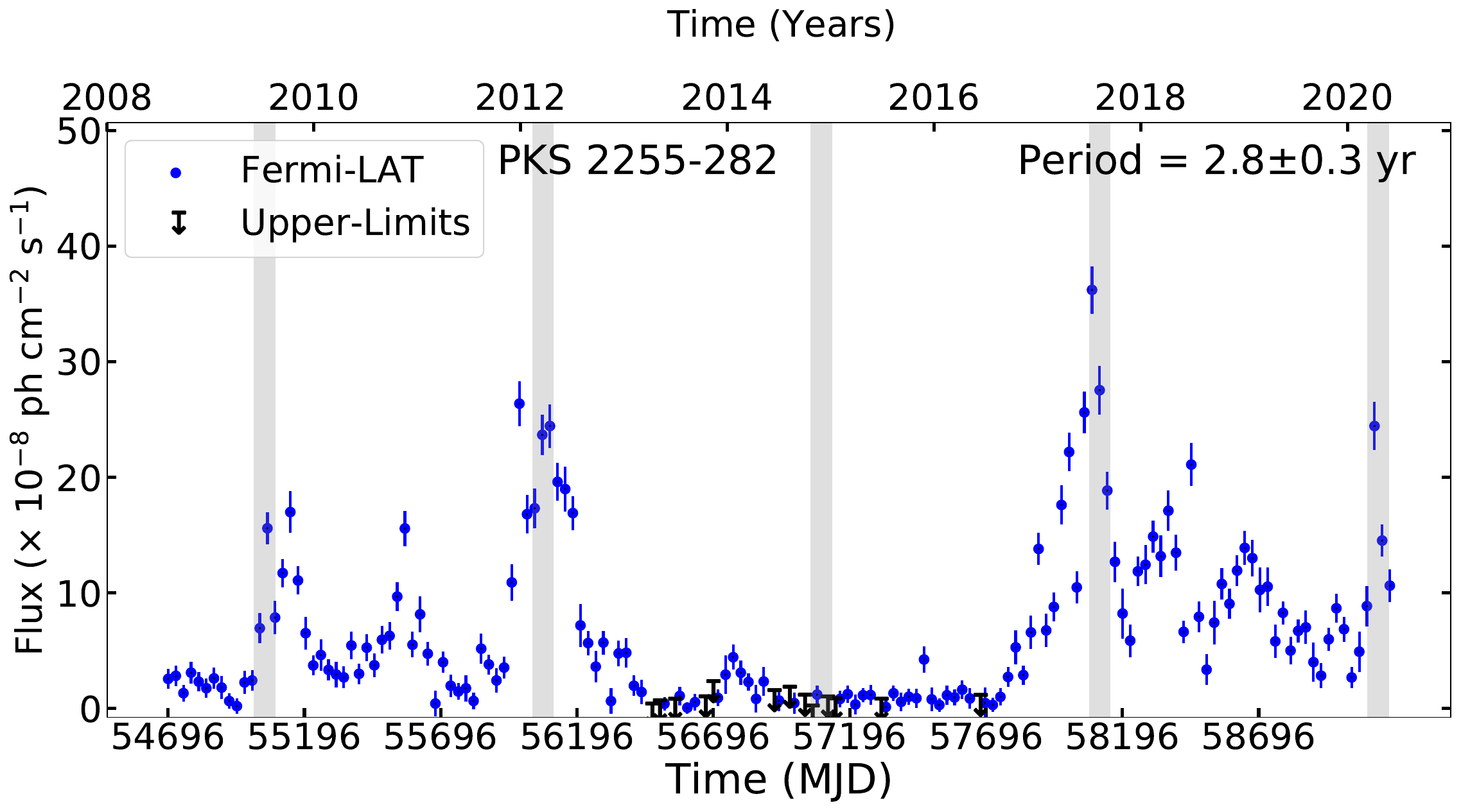}
        \includegraphics[scale=0.2195]{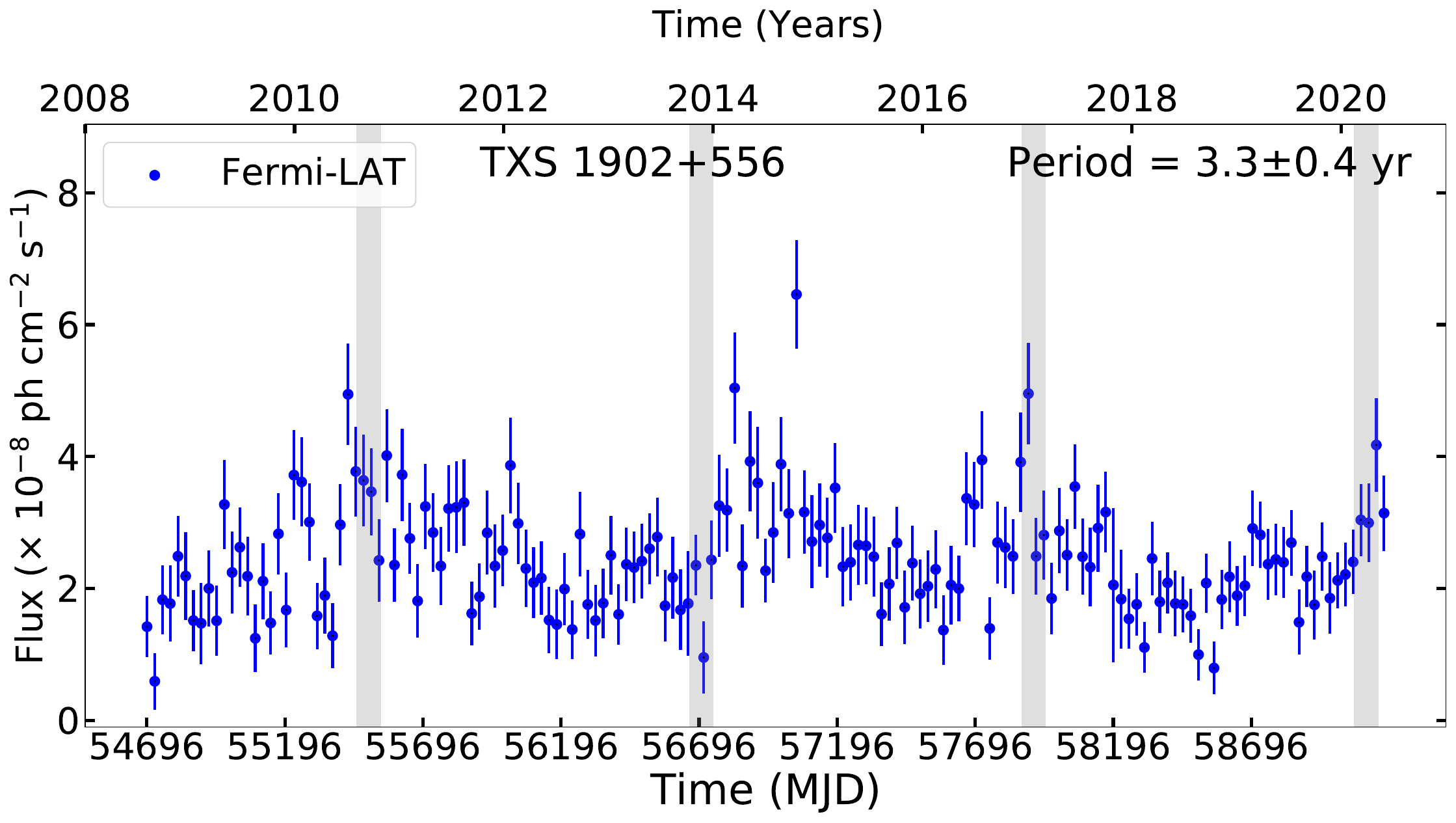}
  	\includegraphics[scale=0.2195]{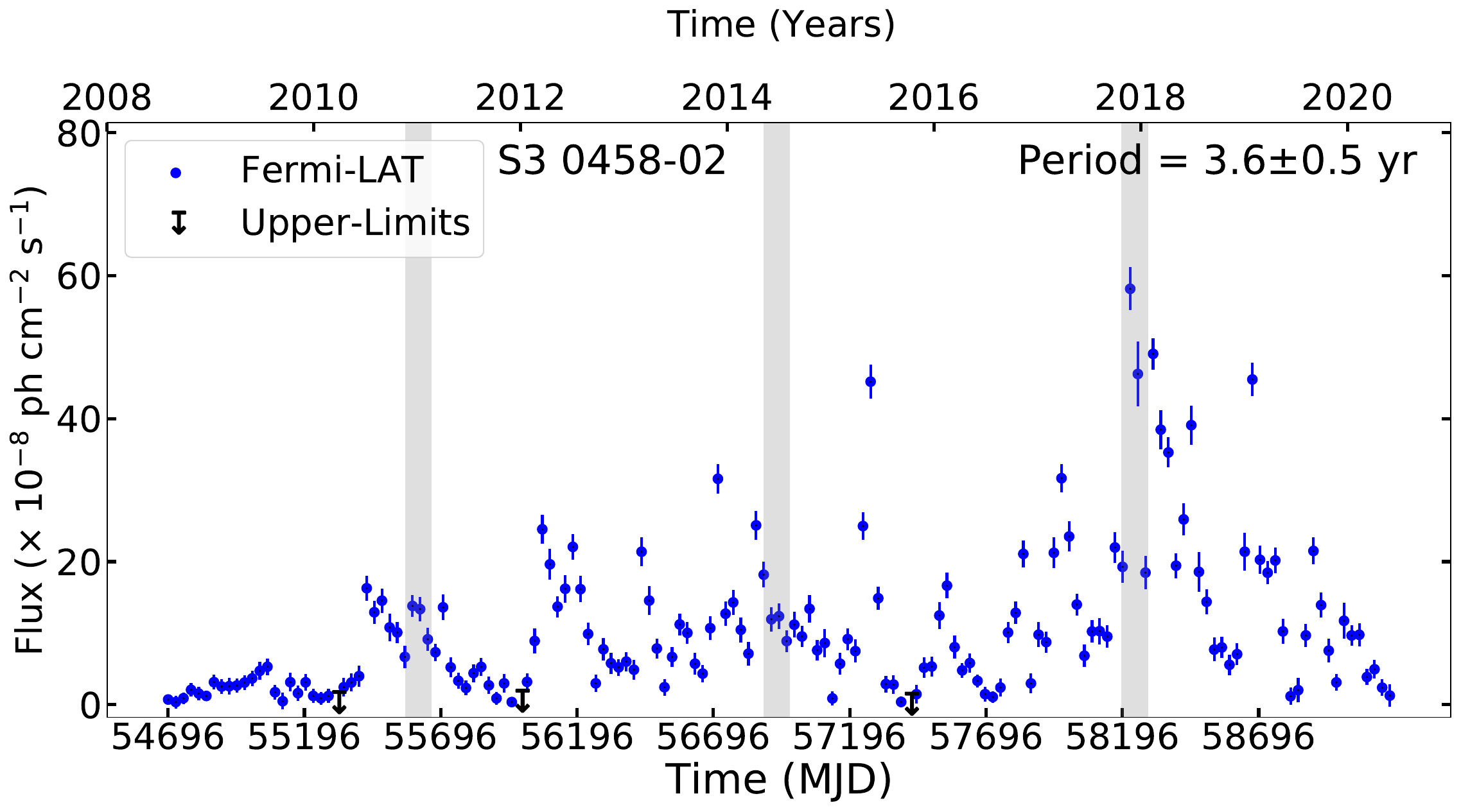}
 	\includegraphics[scale=0.2195]{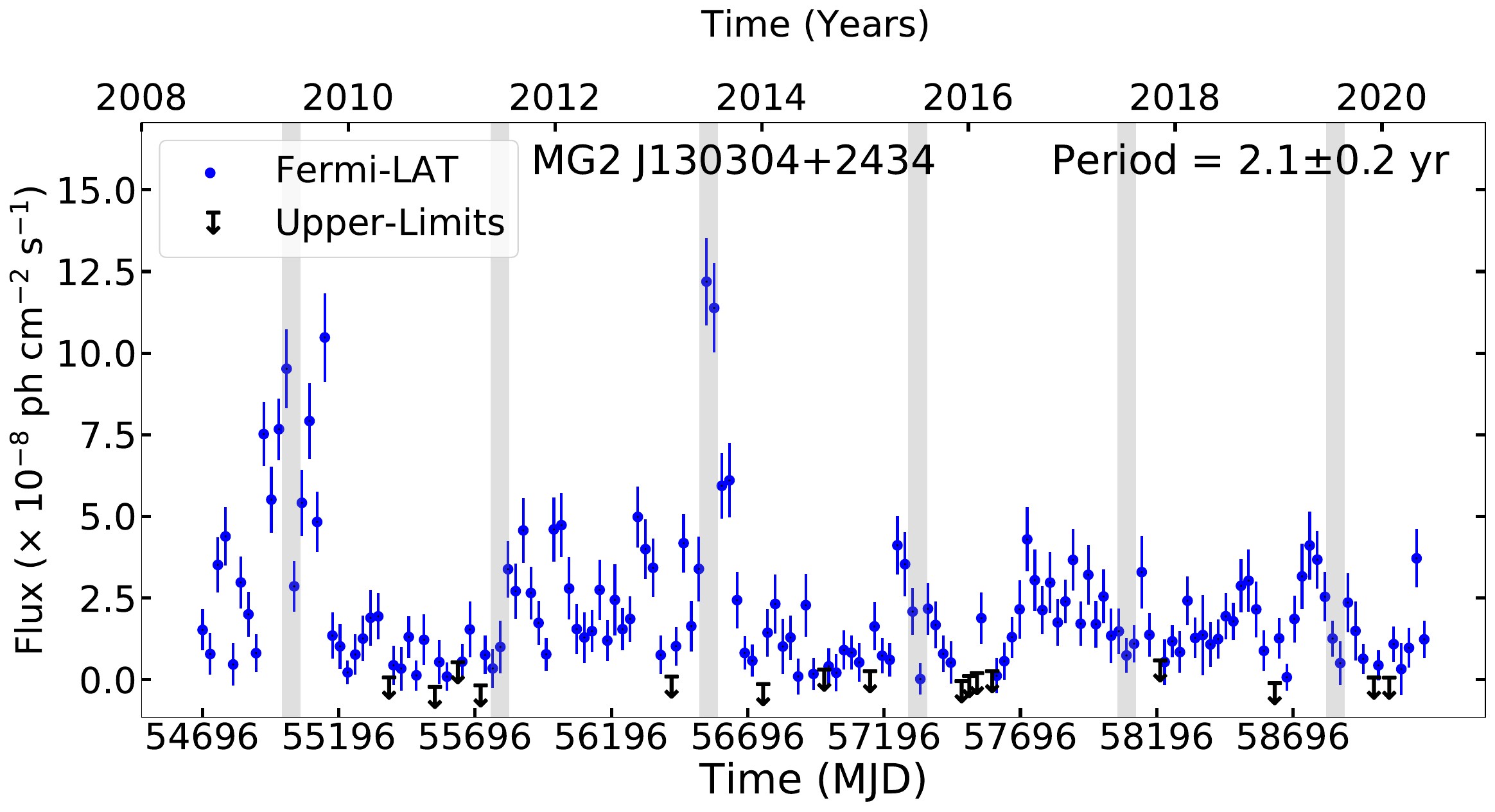}
	\caption{(Continued).}
\end{figure*}
\setcounter{figure}{1}
\begin{figure*}	
  	\includegraphics[scale=0.2165]{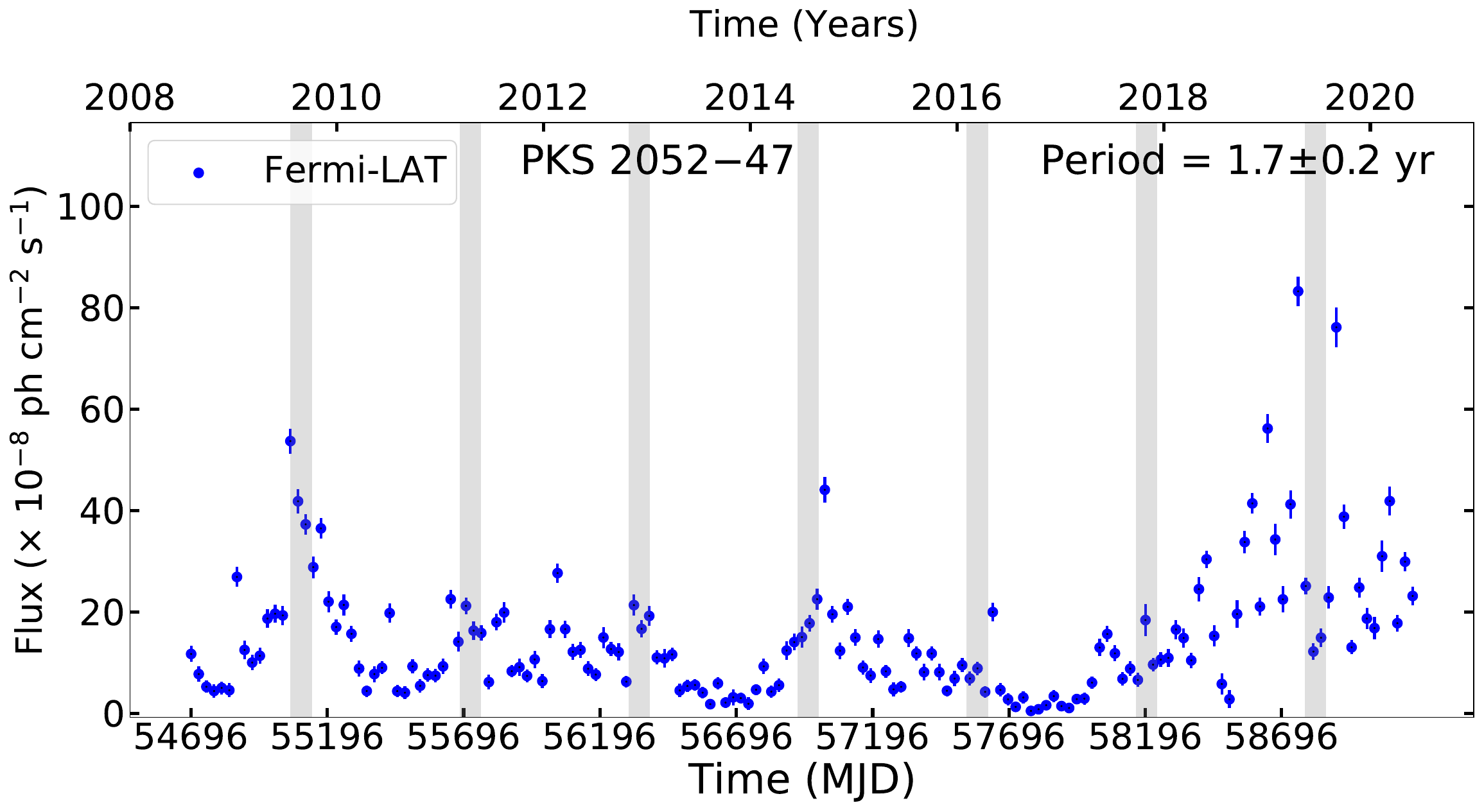}
	\includegraphics[scale=0.2195]{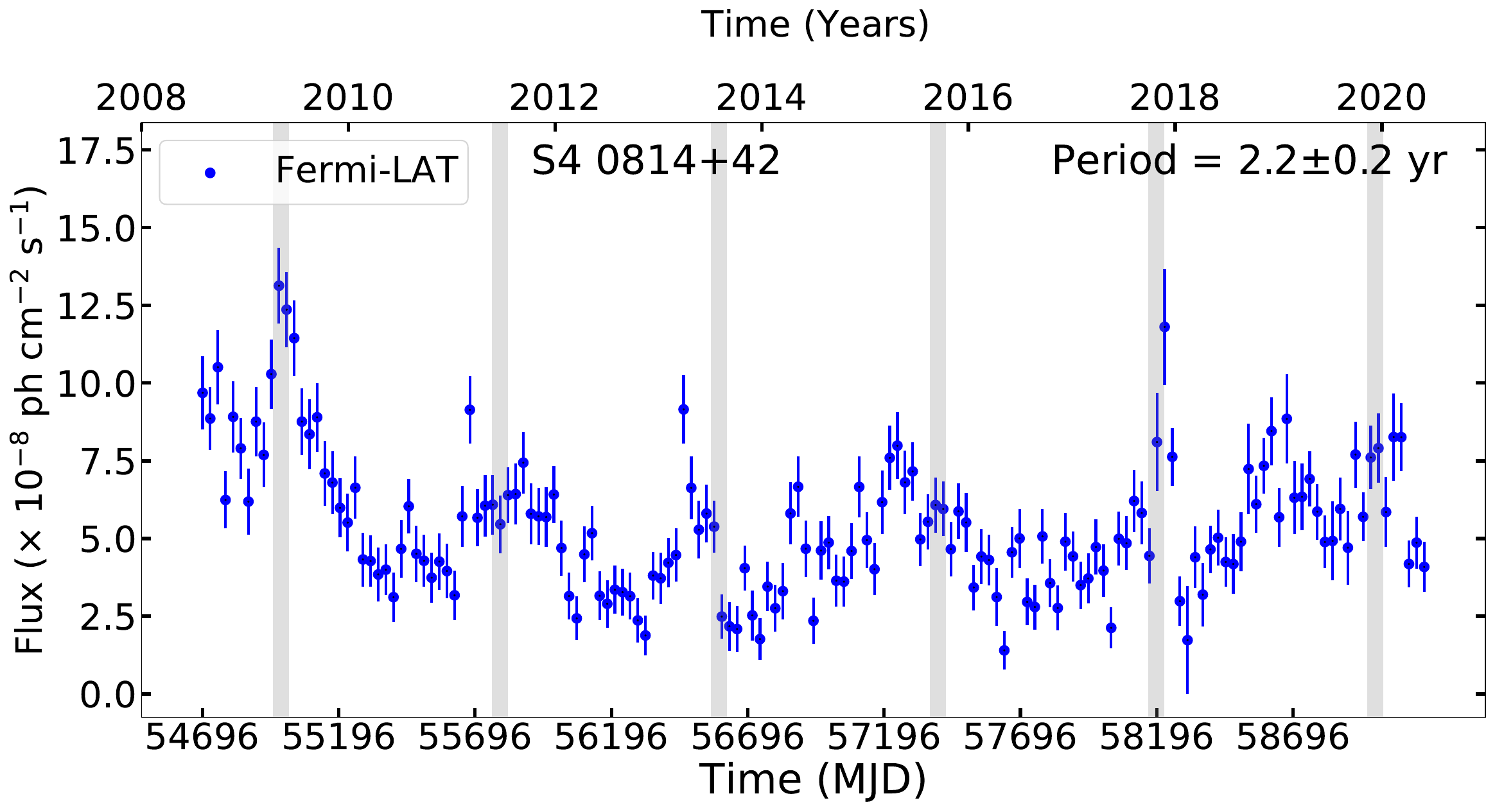}
 	\includegraphics[scale=0.2195]{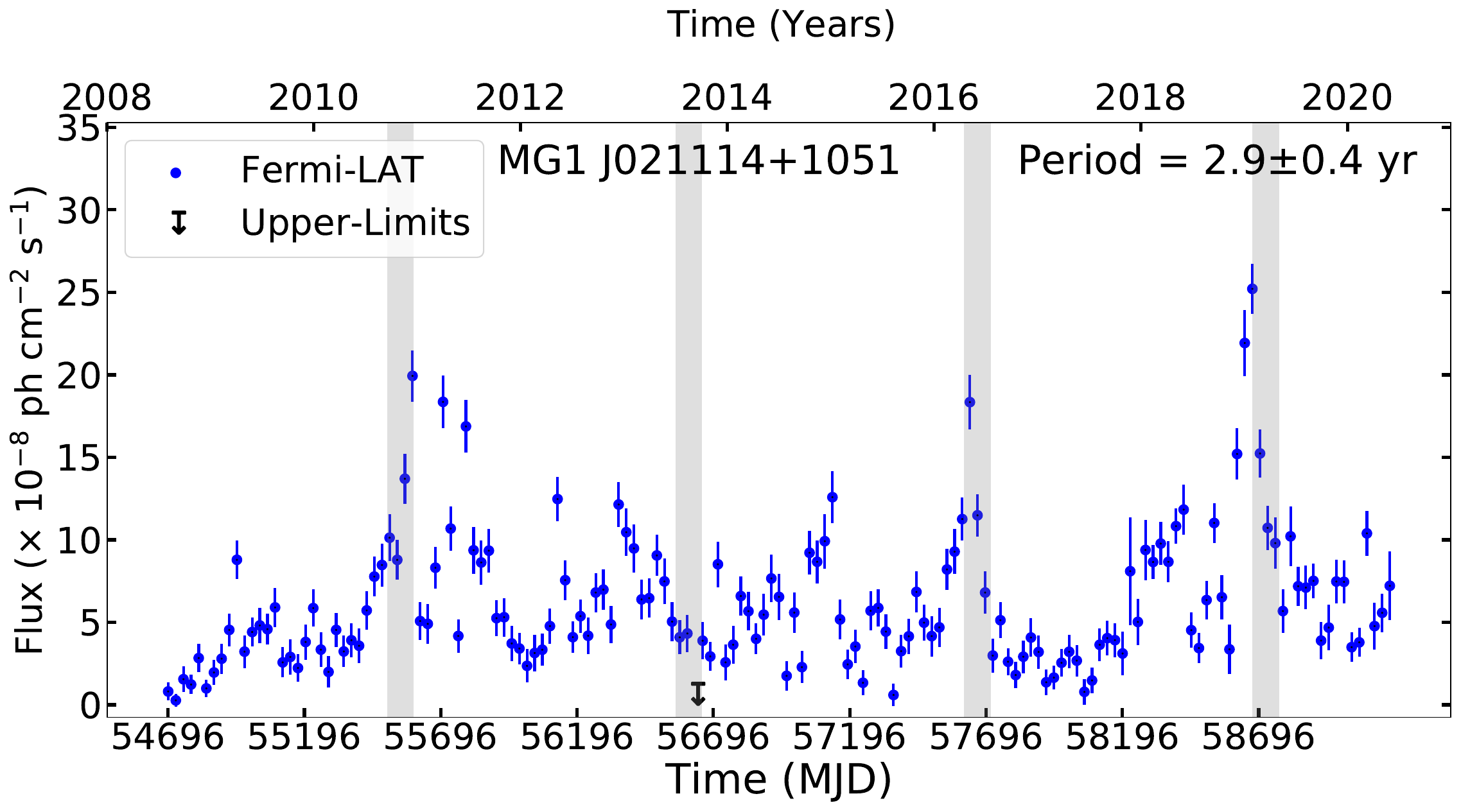}
	\includegraphics[scale=0.2195]{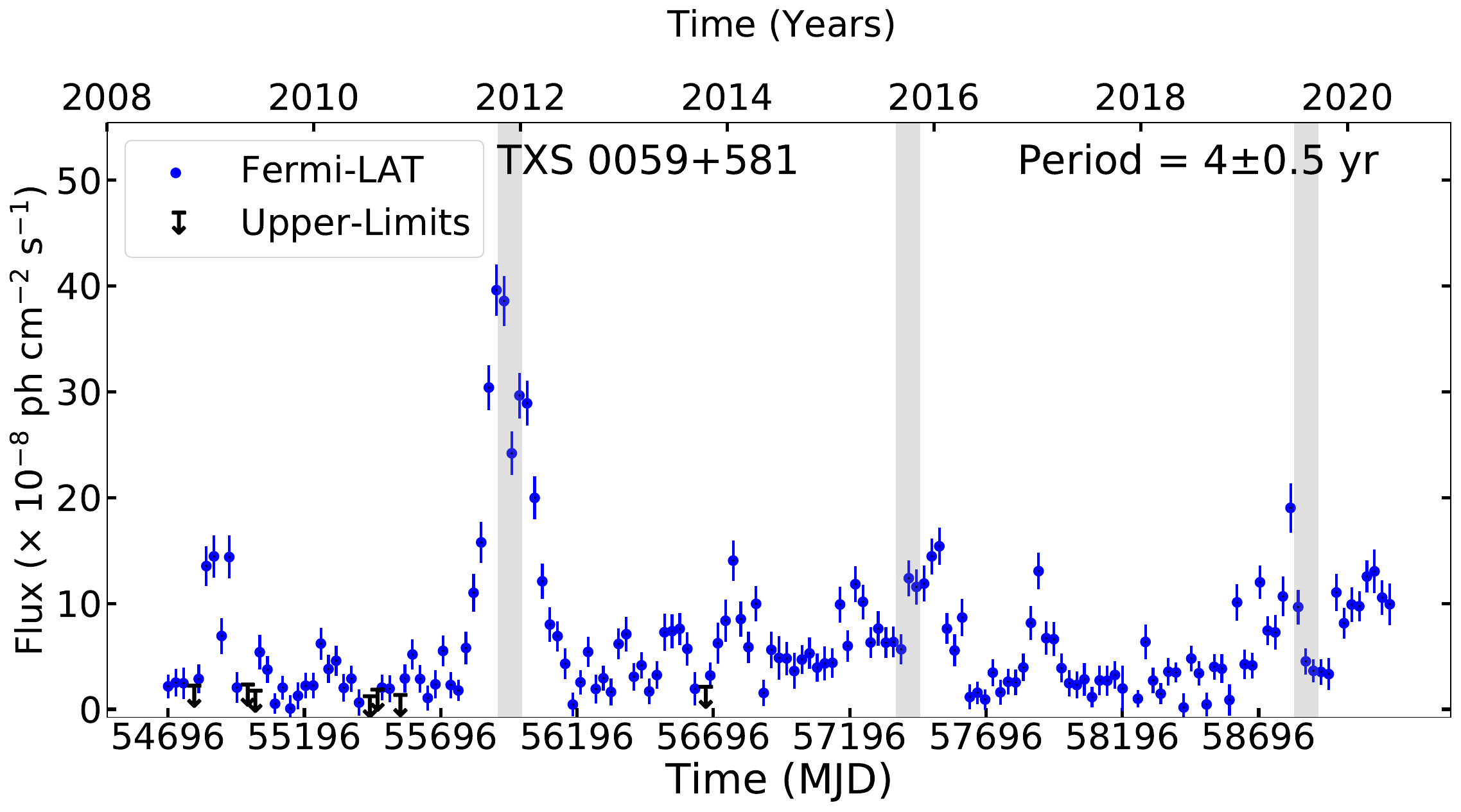}
 	\includegraphics[scale=0.2195]{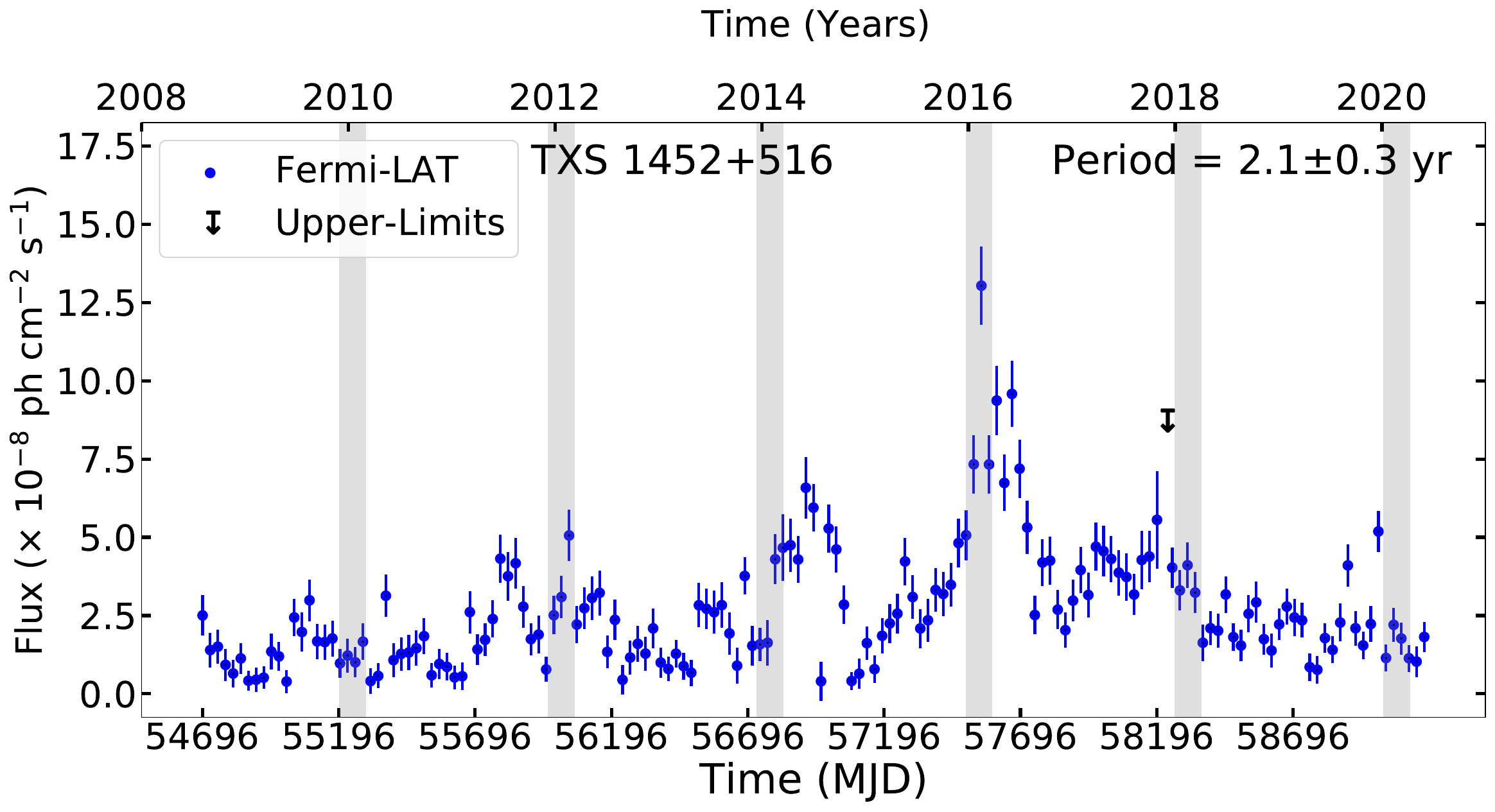}
  	\includegraphics[scale=0.2195]{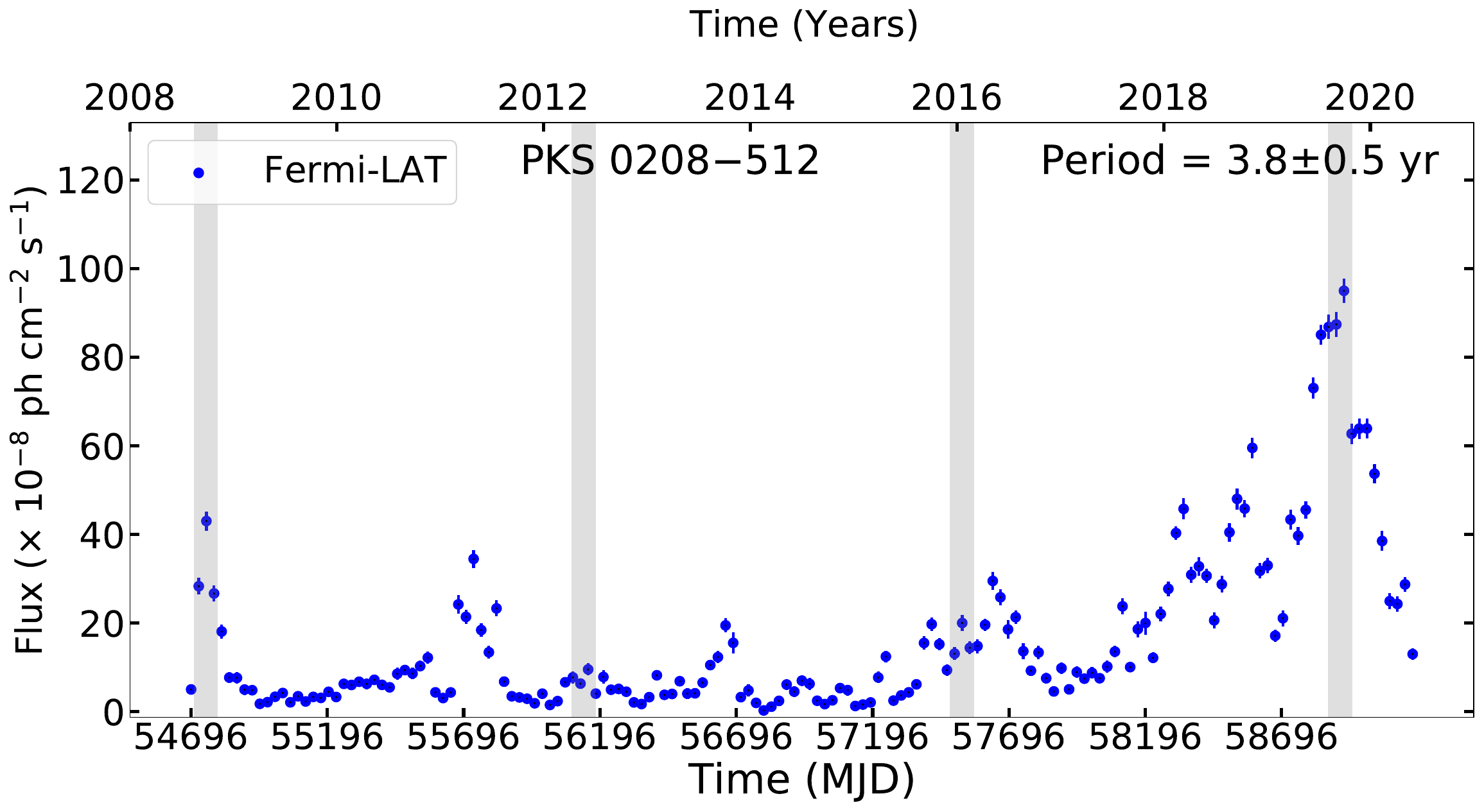}  	
	\caption{(Continued).}
\end{figure*}
\clearpage
\subsection{Power Spectral Densities}
Figure \ref{fig:psd_blazars} reports the PSD plots of the blazars in Table \ref{tab:slopes_poisson}. 
\setcounter{figure}{1}
\begin{figure*}
	\centering
        \includegraphics[scale=0.185]{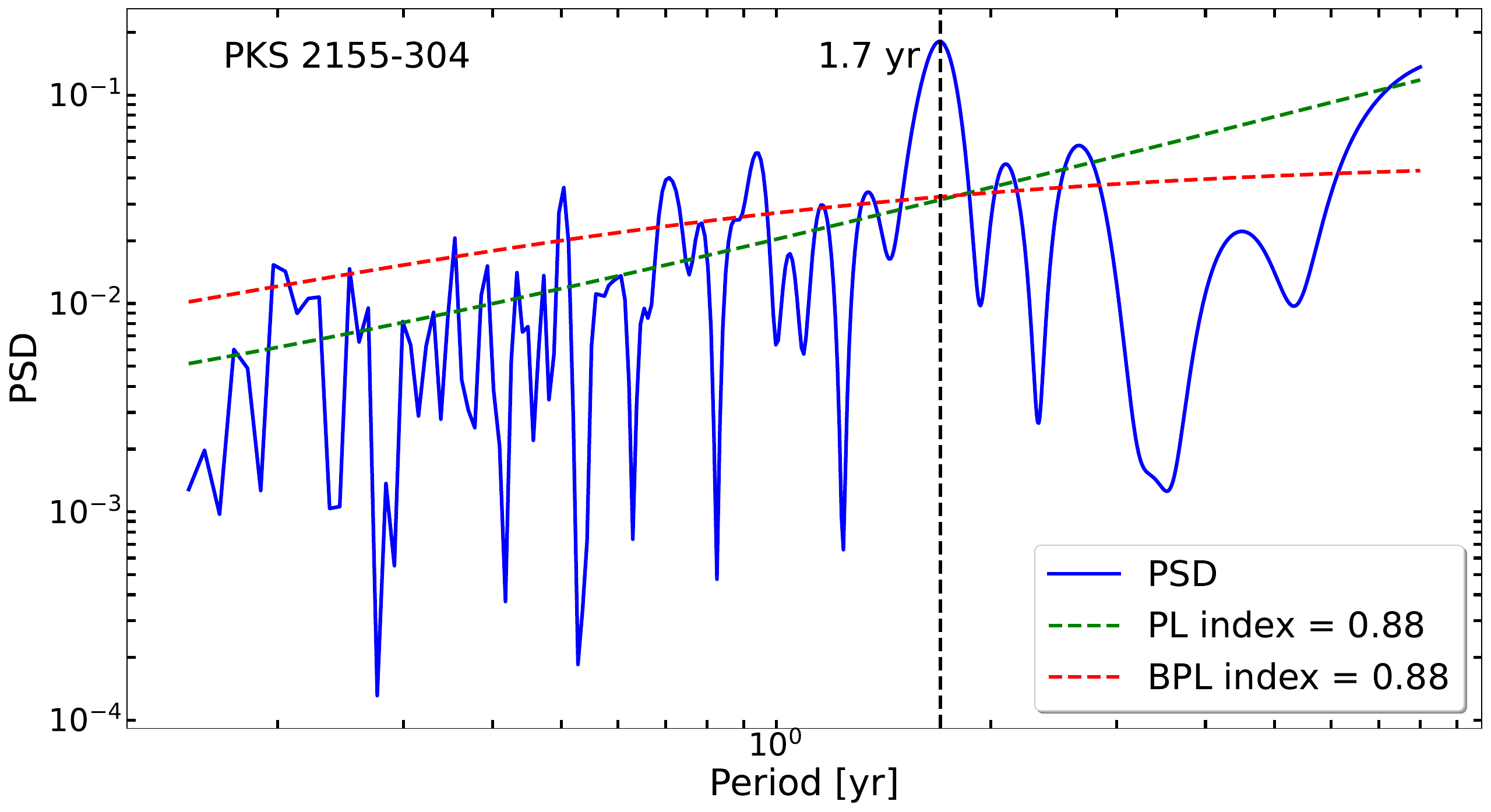}
        \includegraphics[scale=0.185]{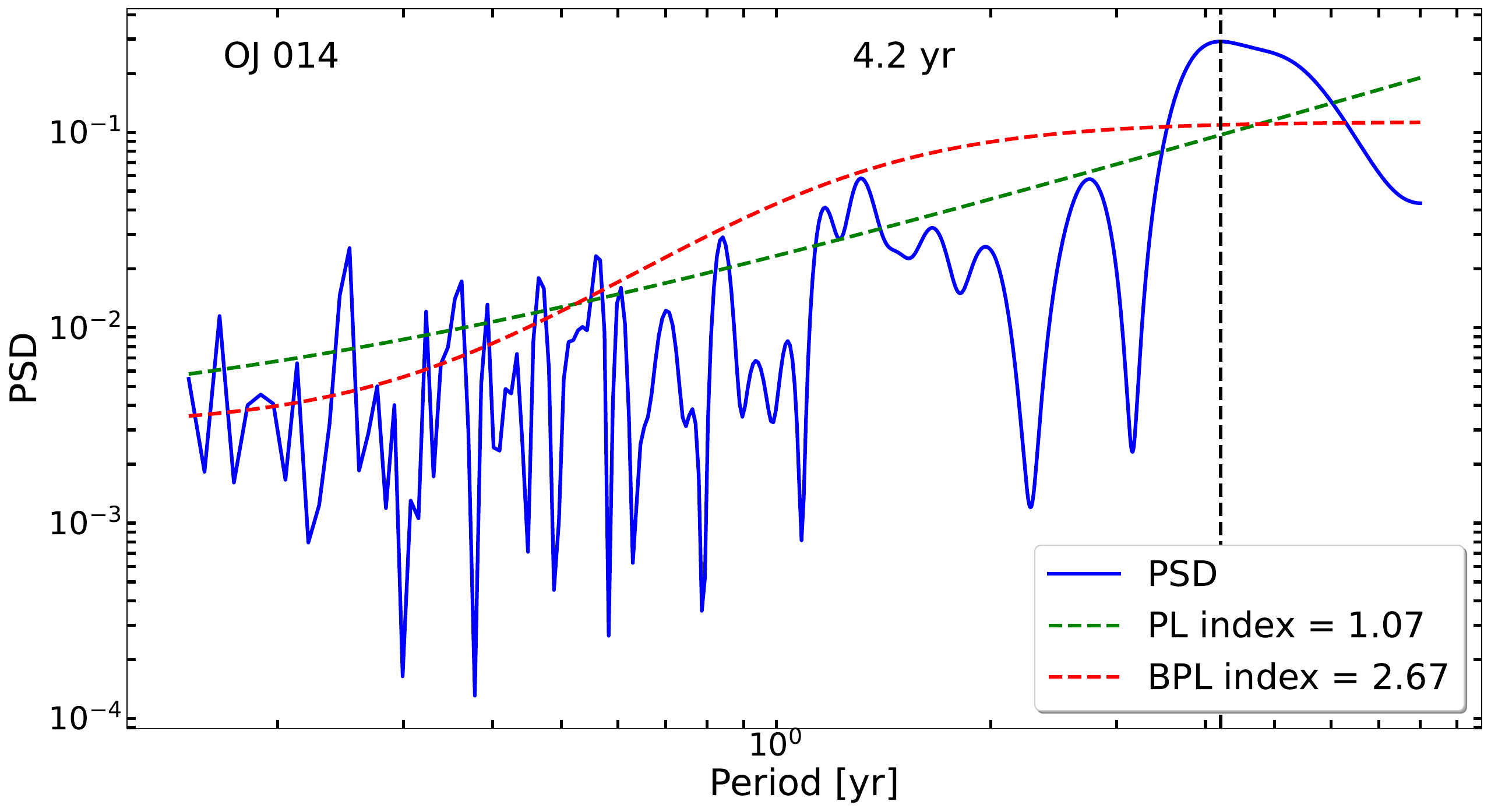} 
        \includegraphics[scale=0.185]{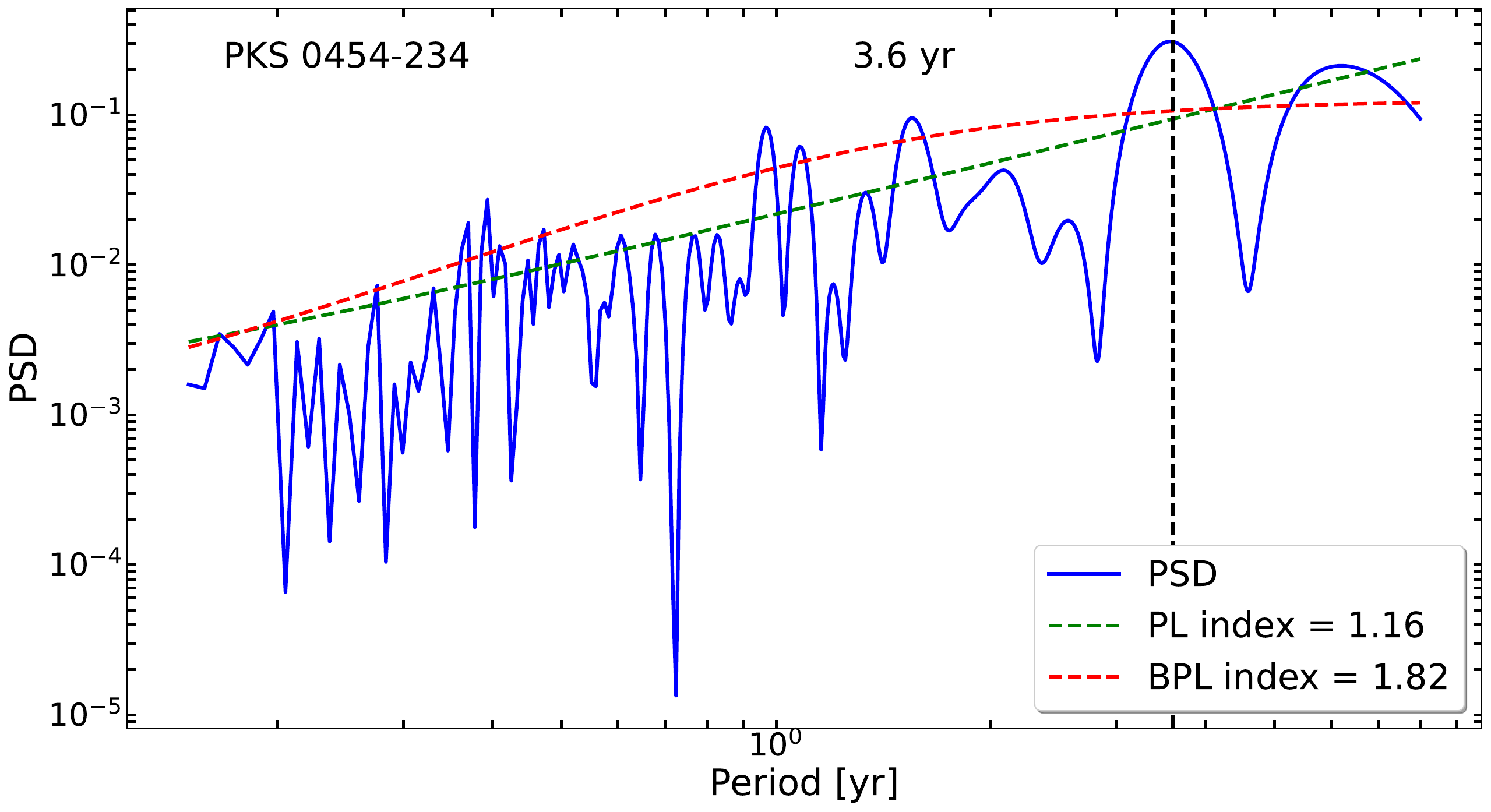}  
        \includegraphics[scale=0.185]{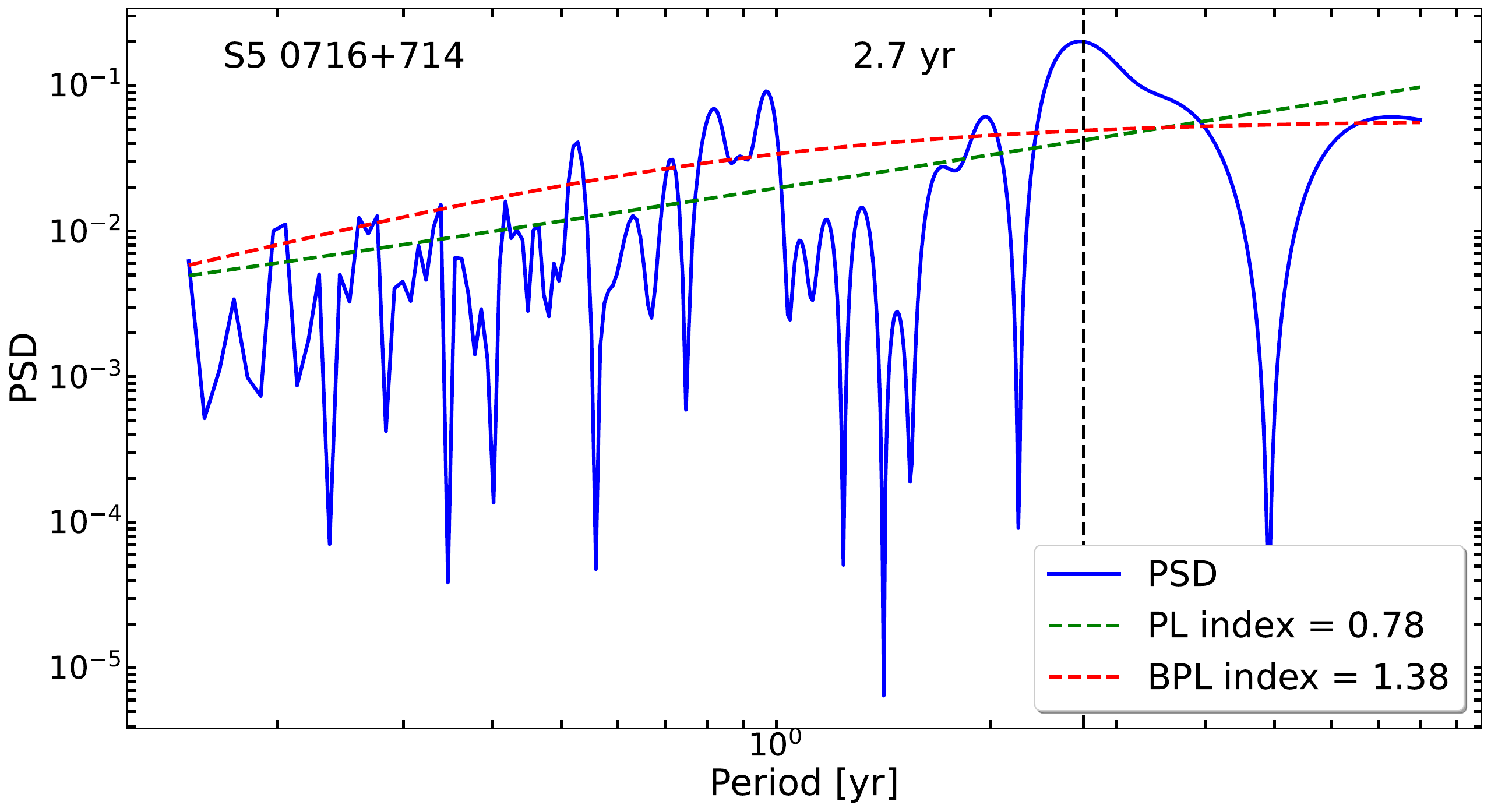}
        \includegraphics[scale=0.185]{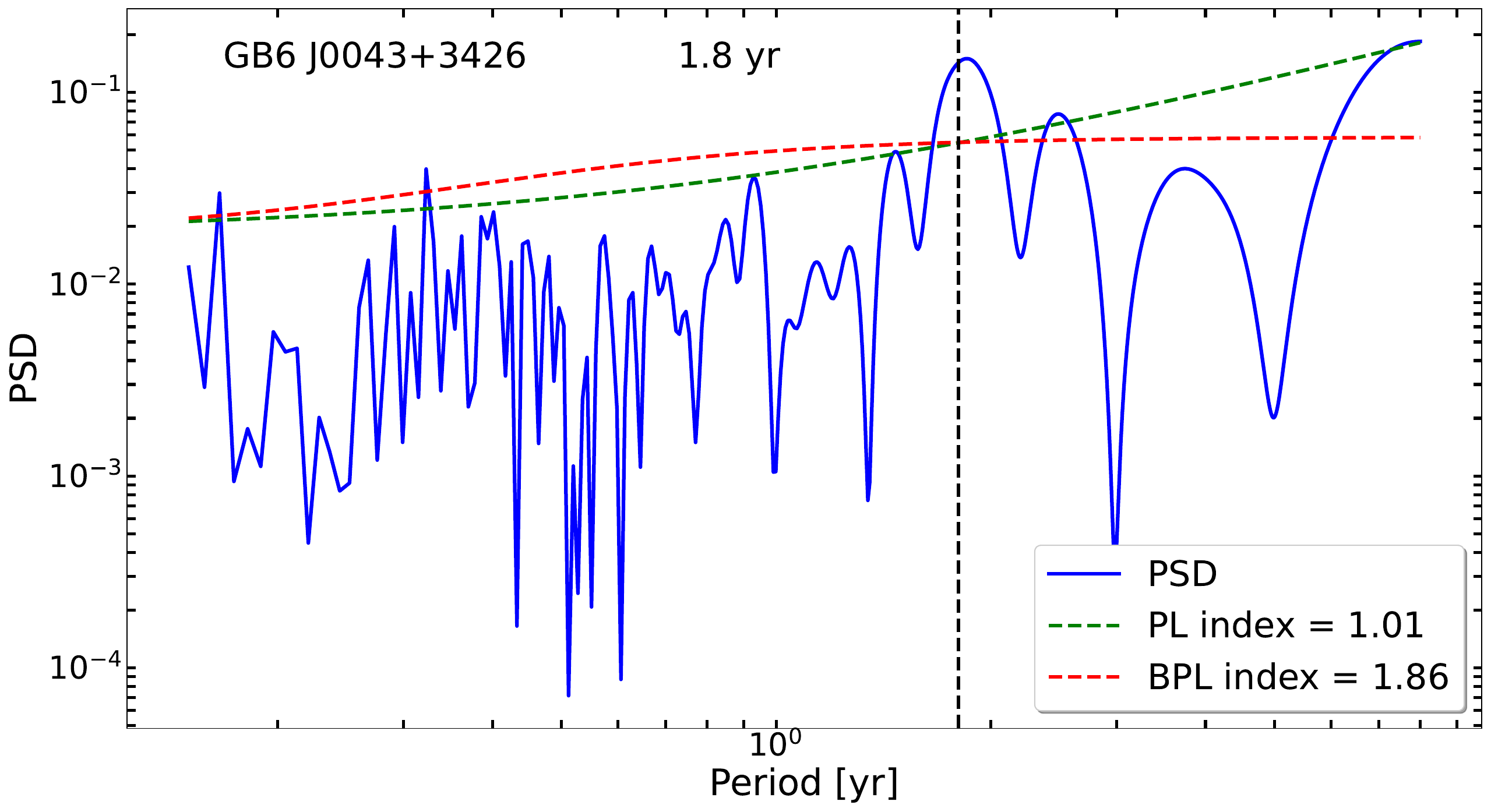}        
        \includegraphics[scale=0.185]{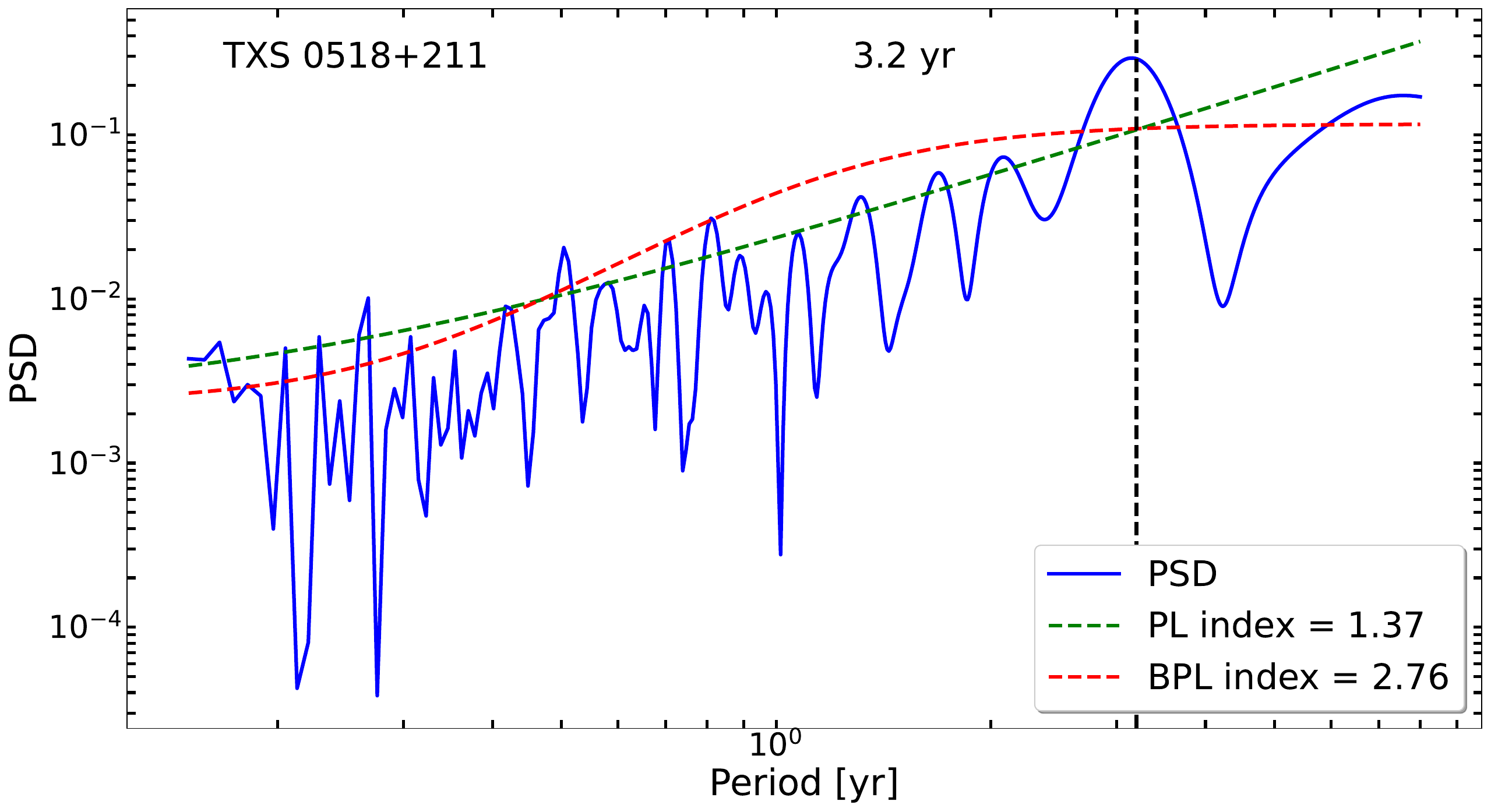}
        \includegraphics[scale=0.185]{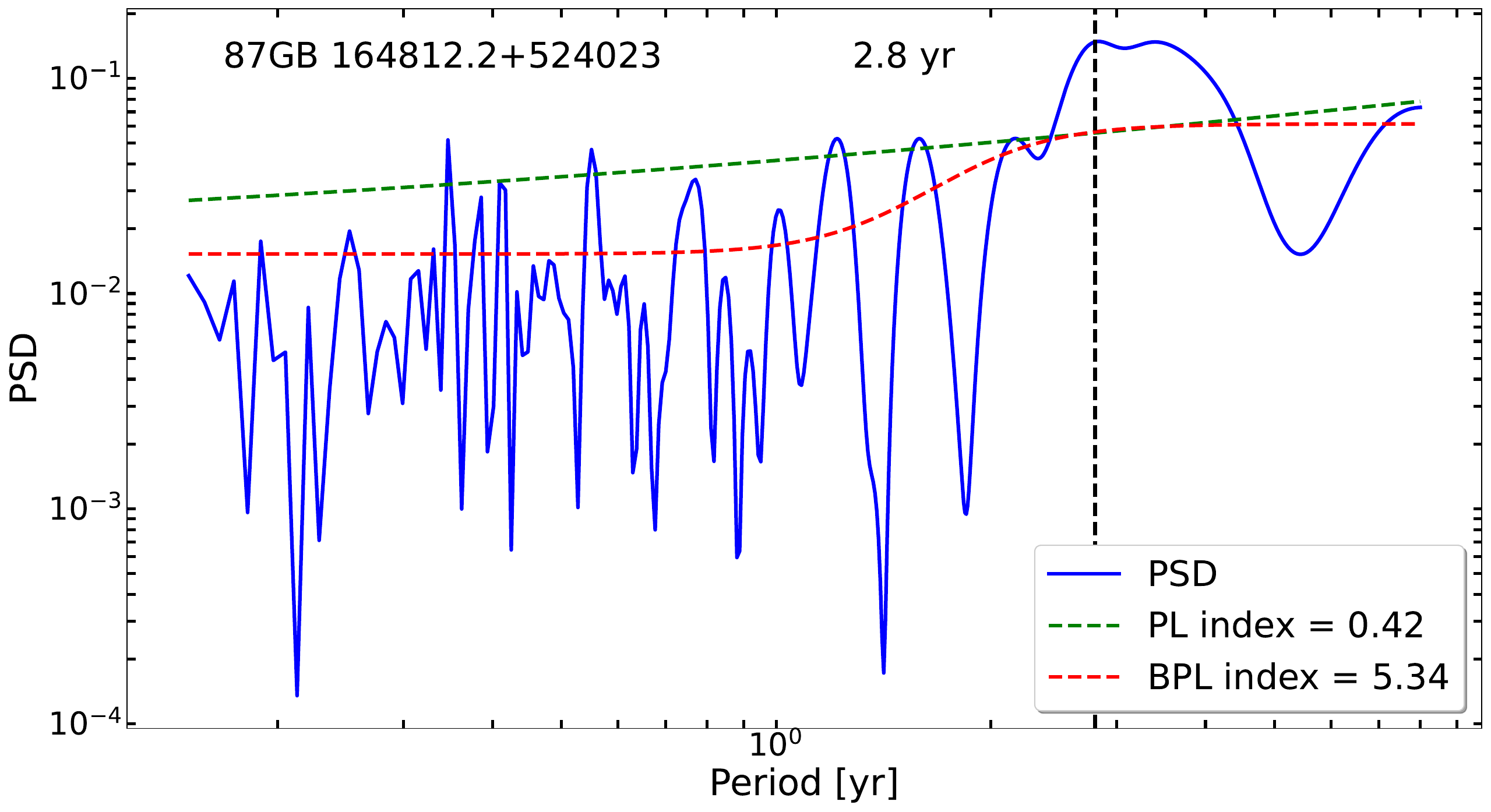}
        \includegraphics[scale=0.185]{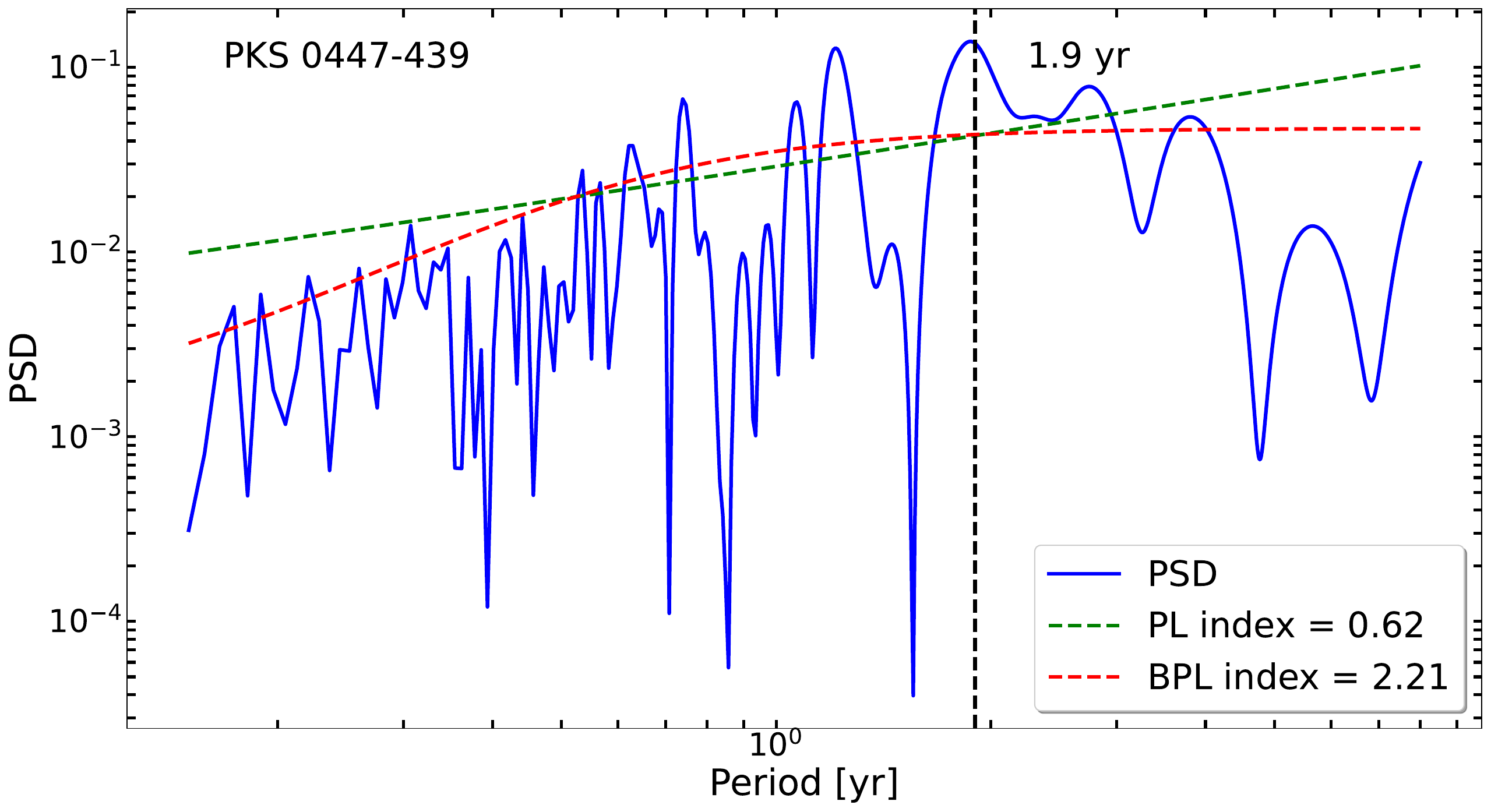}
	\caption{Power spectral densities according to the fit presented in Table \ref{tab:slopes_poisson}.}
	\label{fig:psd_blazars} 
\end{figure*}
\setcounter{figure}{1}
\begin{figure*}
	\centering
        \includegraphics[scale=0.185]{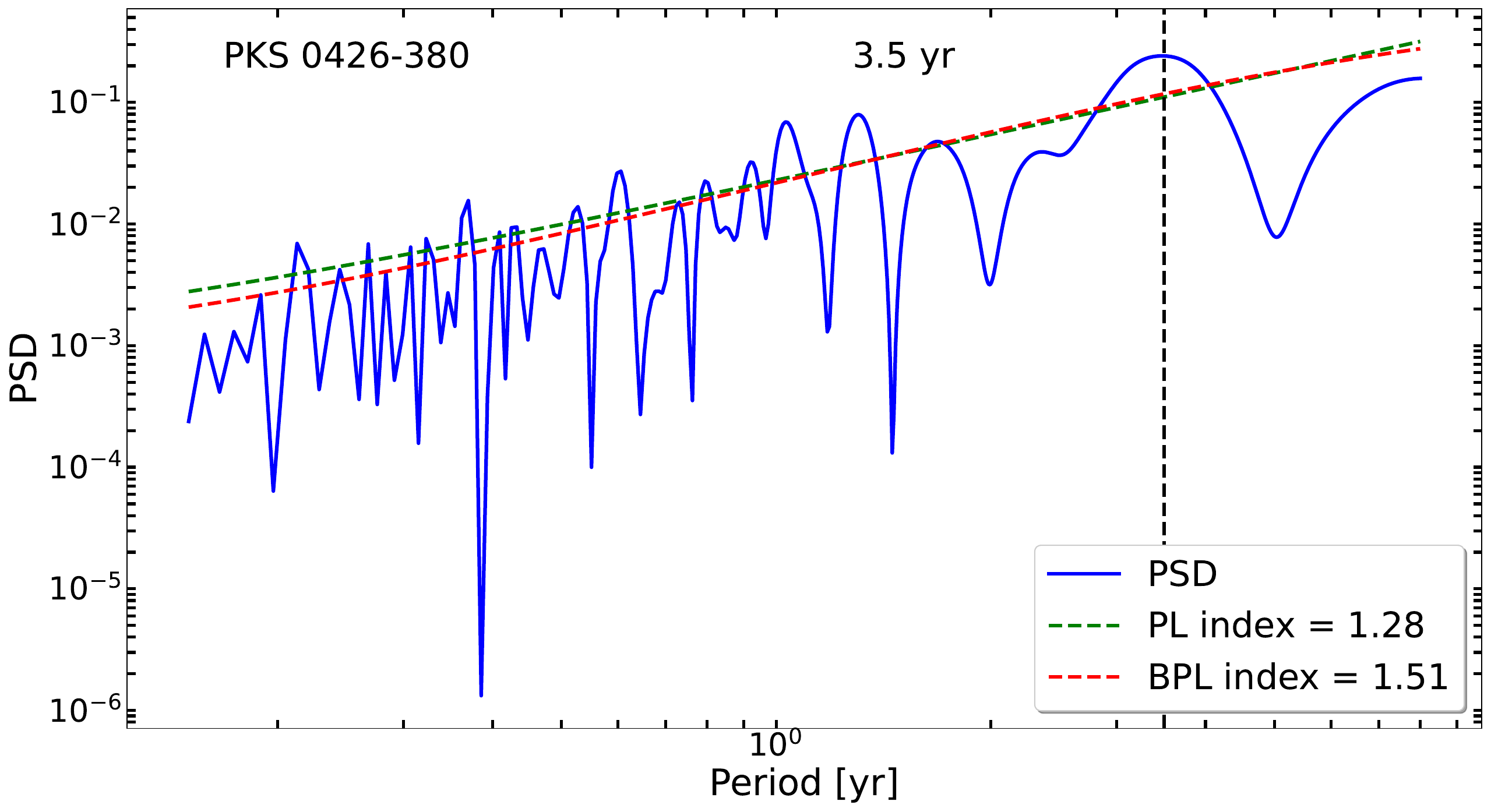}
        \includegraphics[scale=0.185]{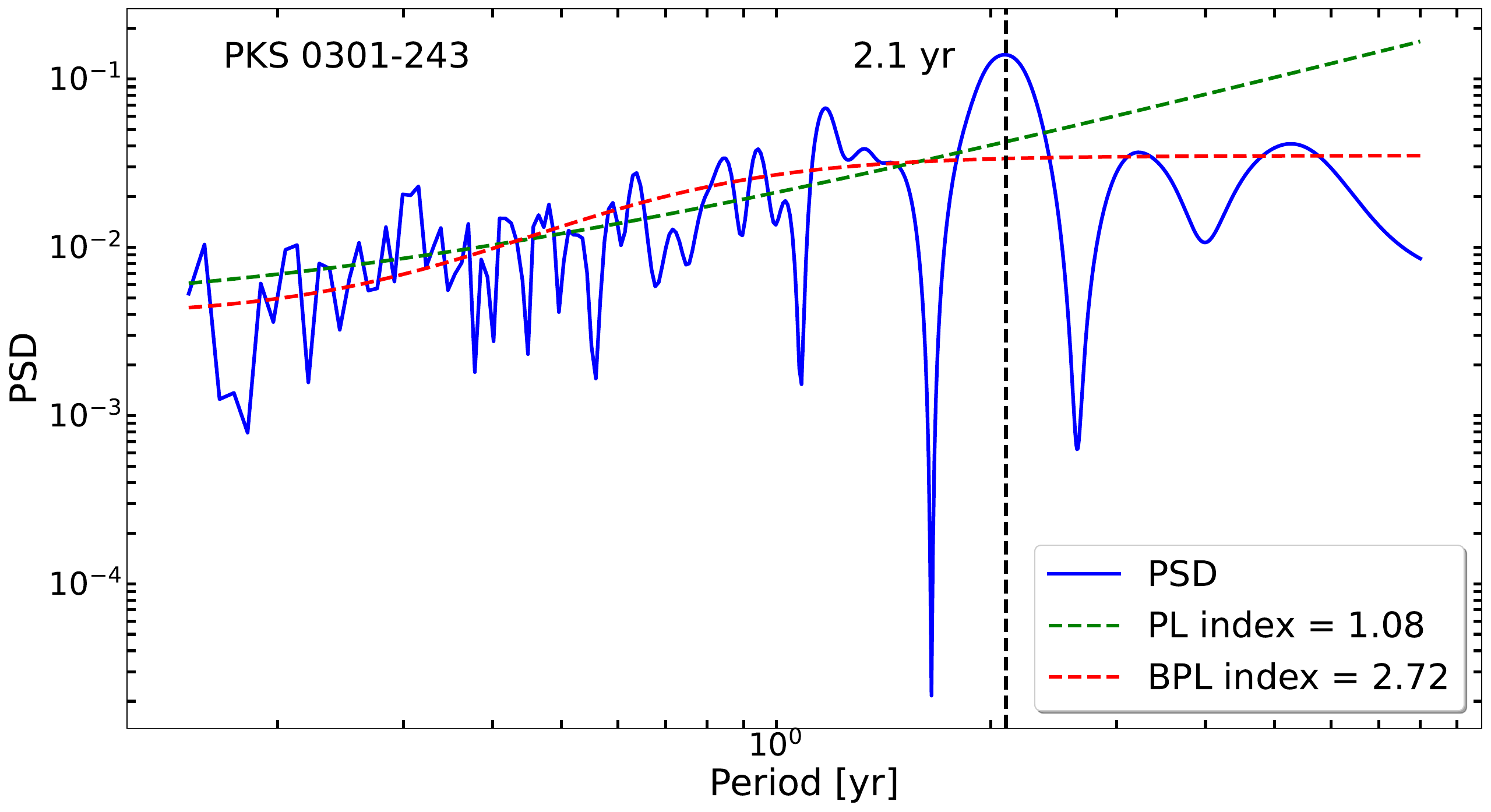}
        \includegraphics[scale=0.185]{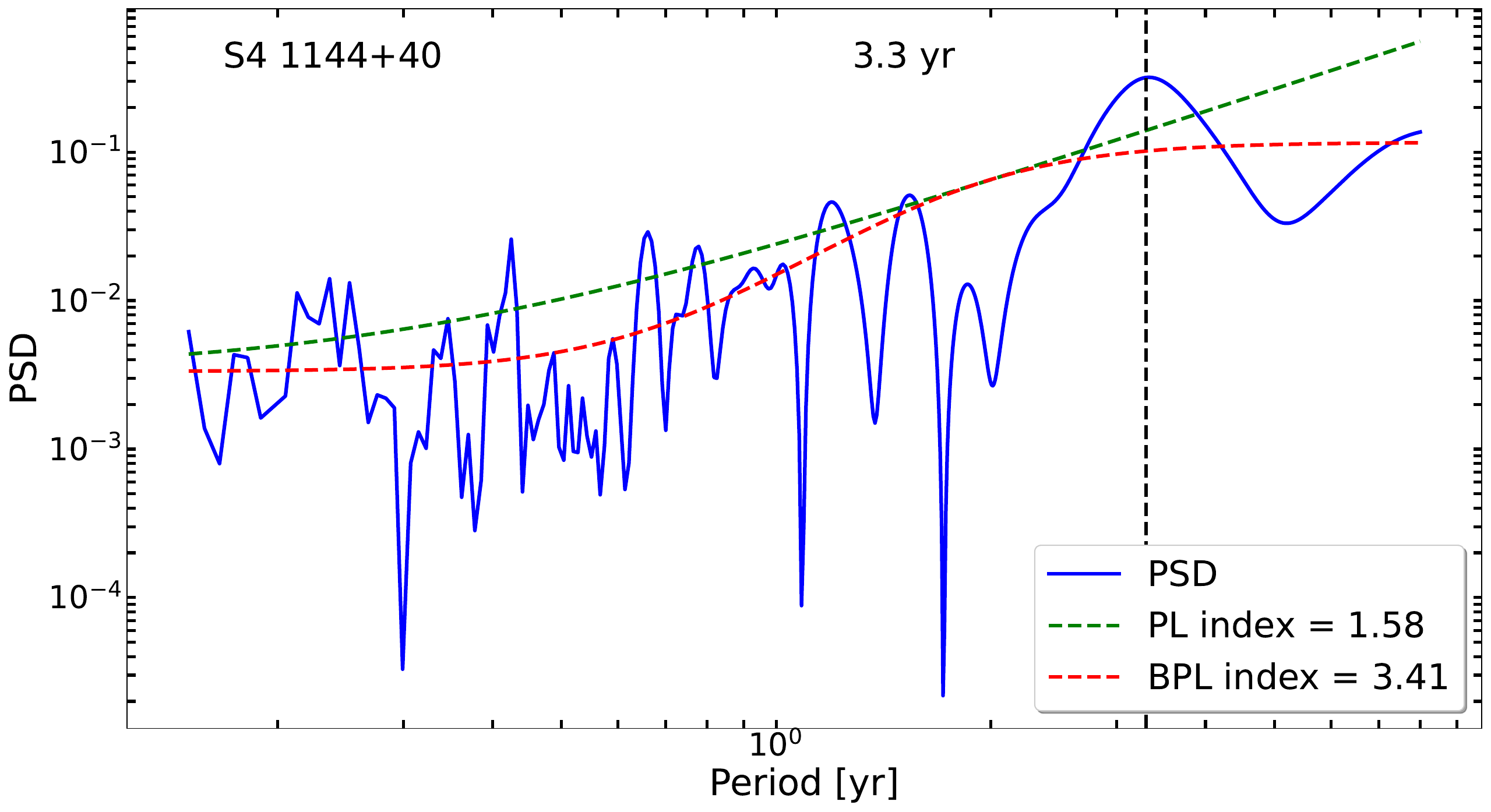}
        \includegraphics[scale=0.185]{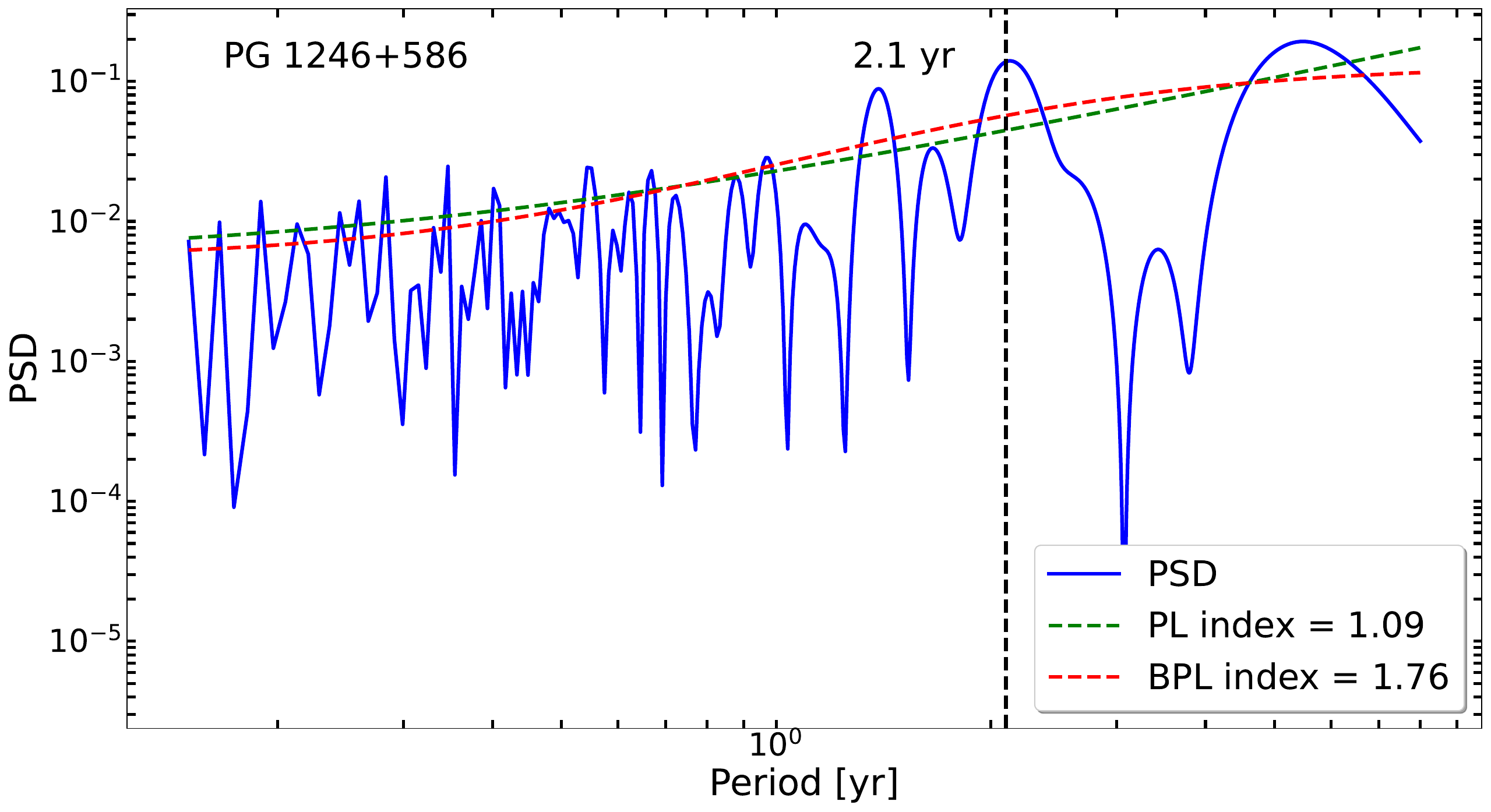}
        \includegraphics[scale=0.185]{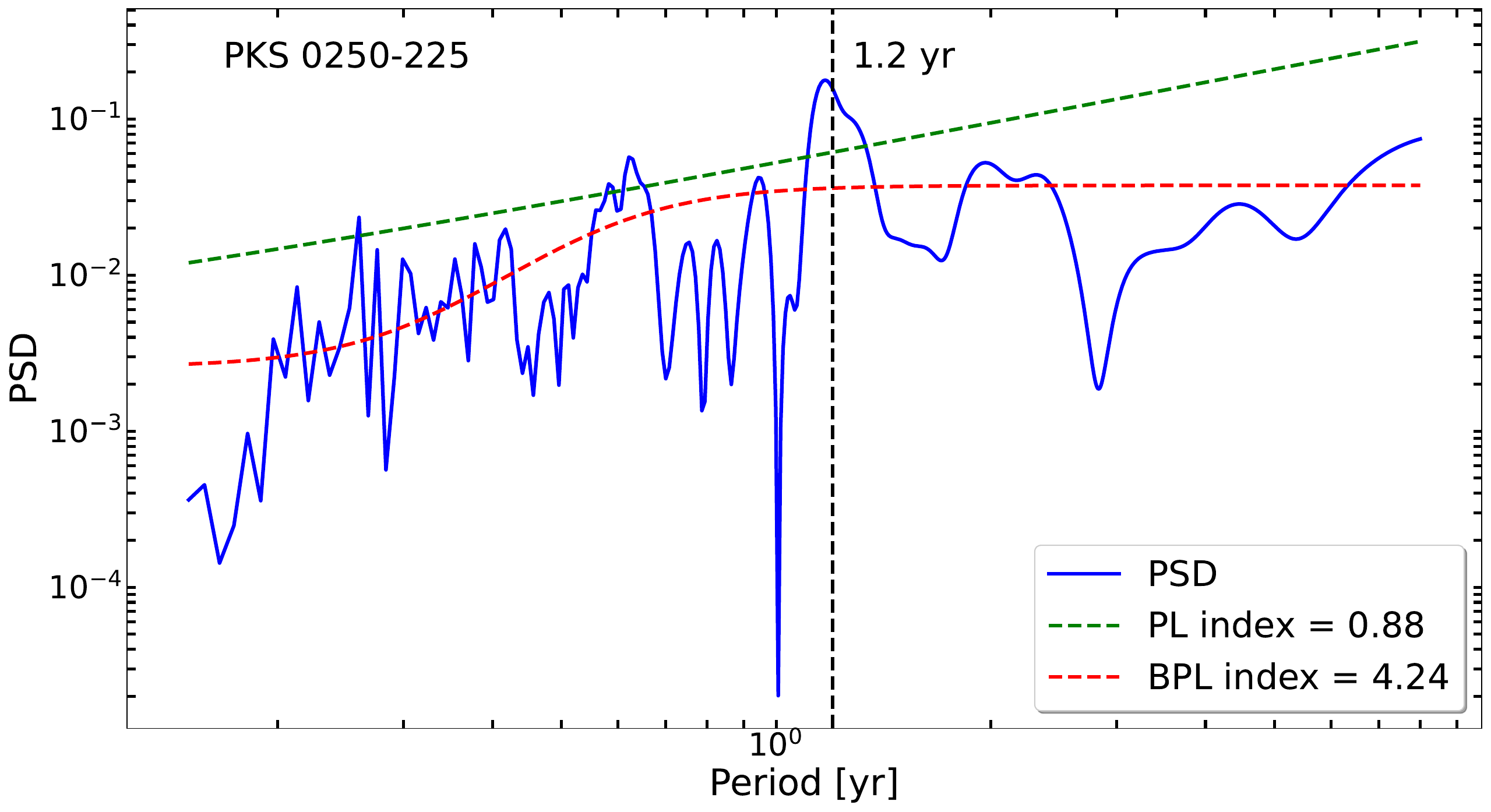}
        \includegraphics[scale=0.185]{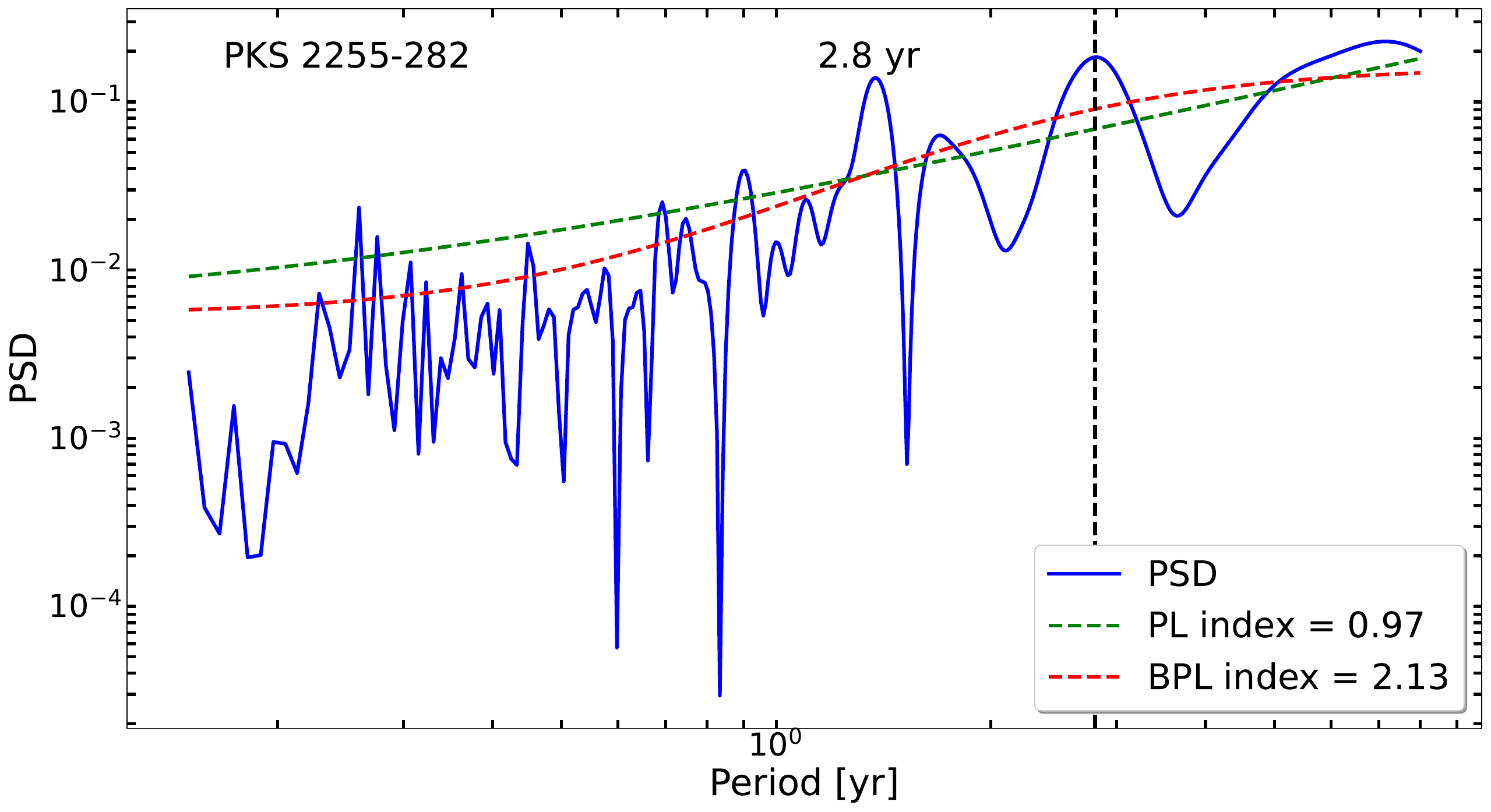}
        \includegraphics[scale=0.185]{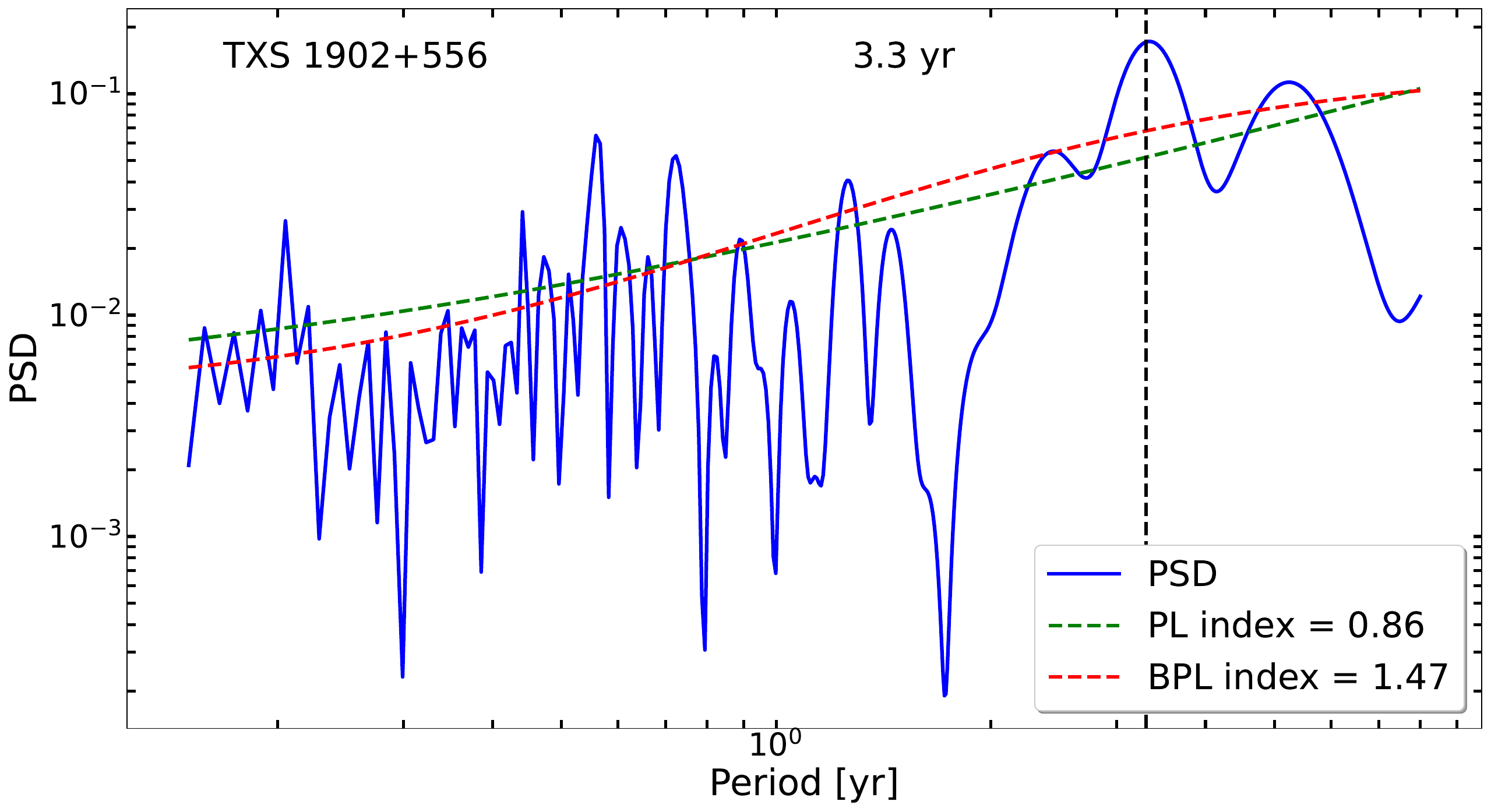}
        \includegraphics[scale=0.185]{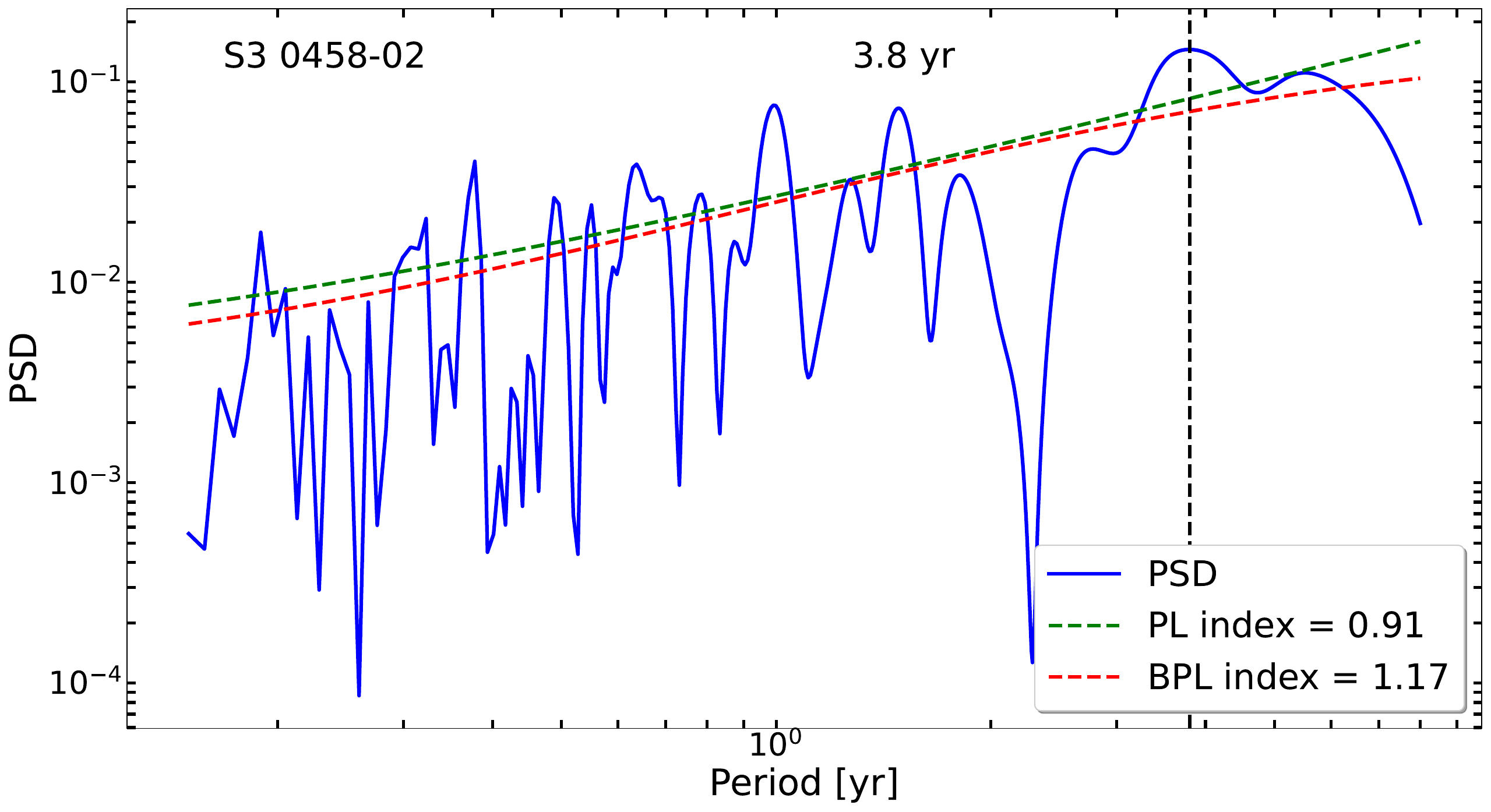}
	\caption{(Continued).}
\end{figure*}
\setcounter{figure}{1}
\begin{figure*}
	\centering
         \includegraphics[scale=0.185]{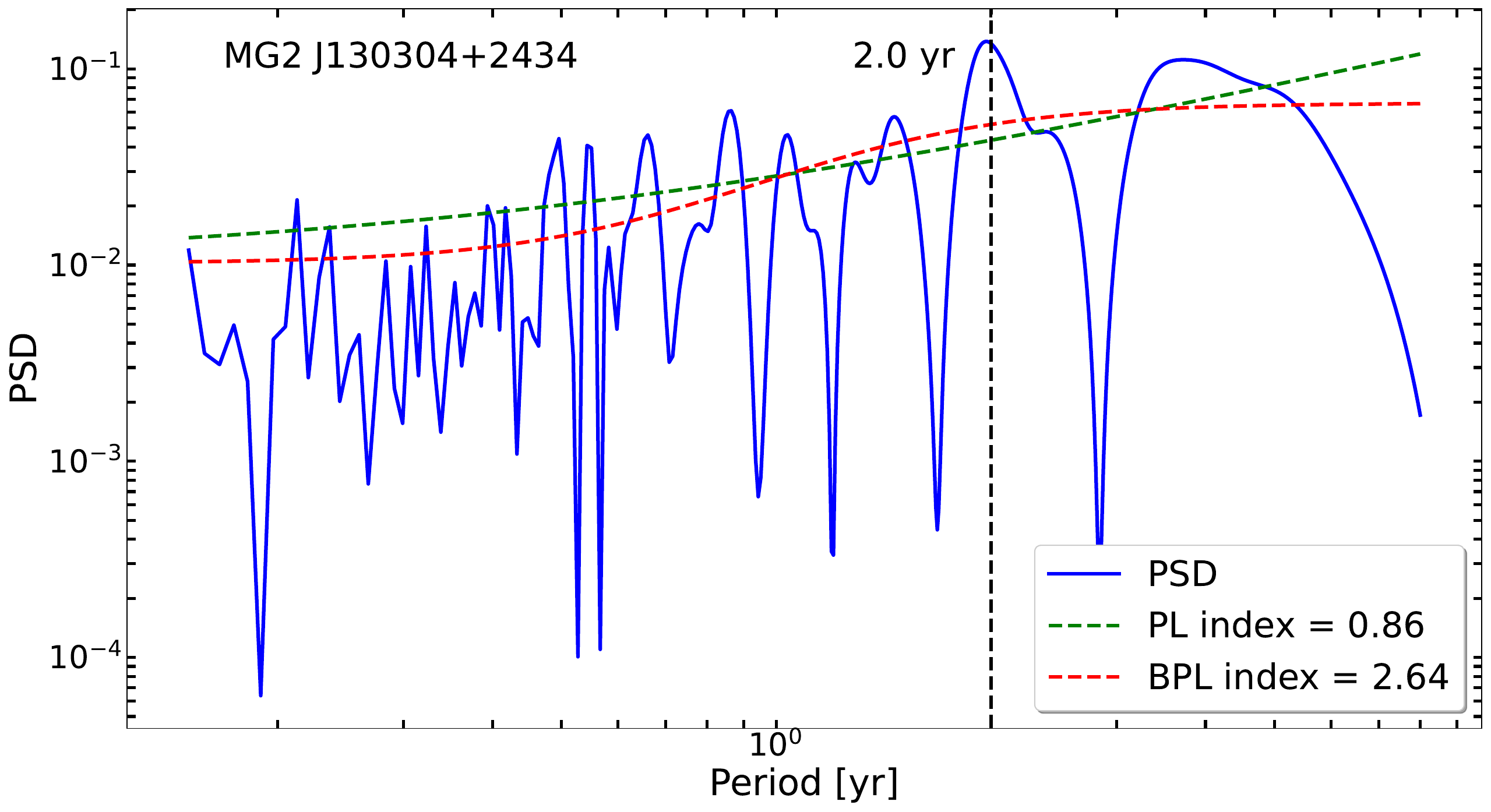}
         \includegraphics[scale=0.185]{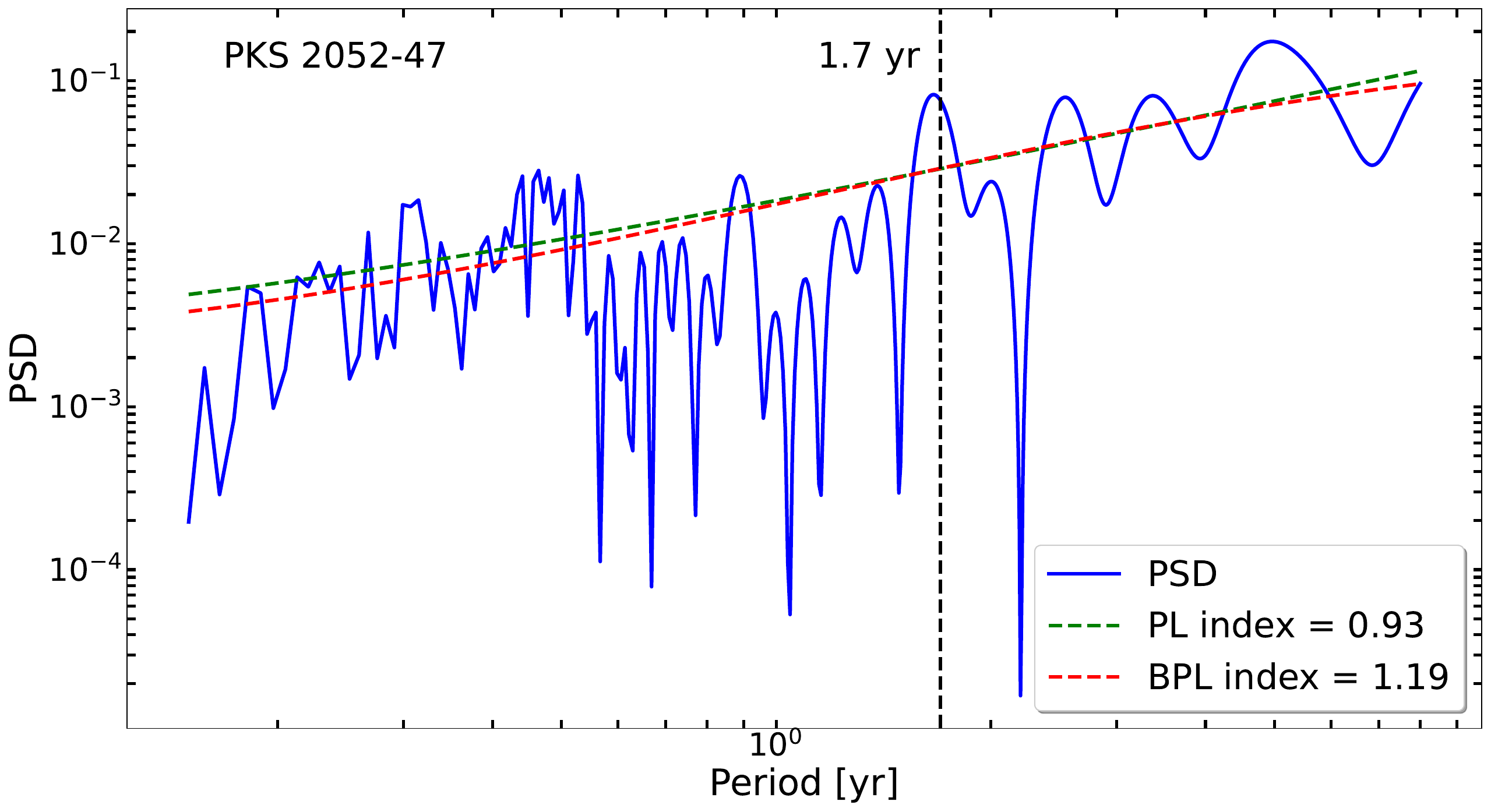}
        \includegraphics[scale=0.185]{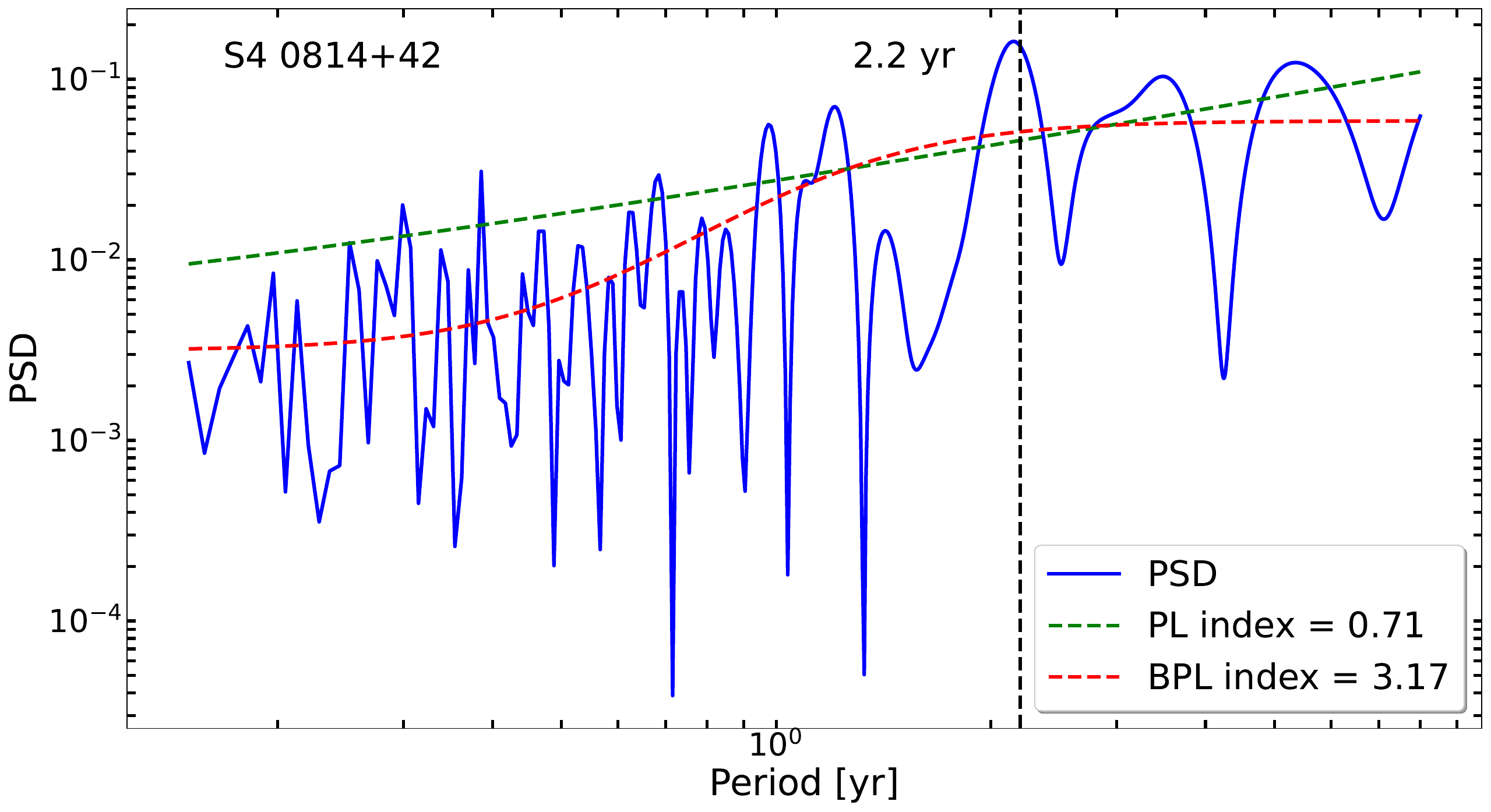}
        \includegraphics[scale=0.185]{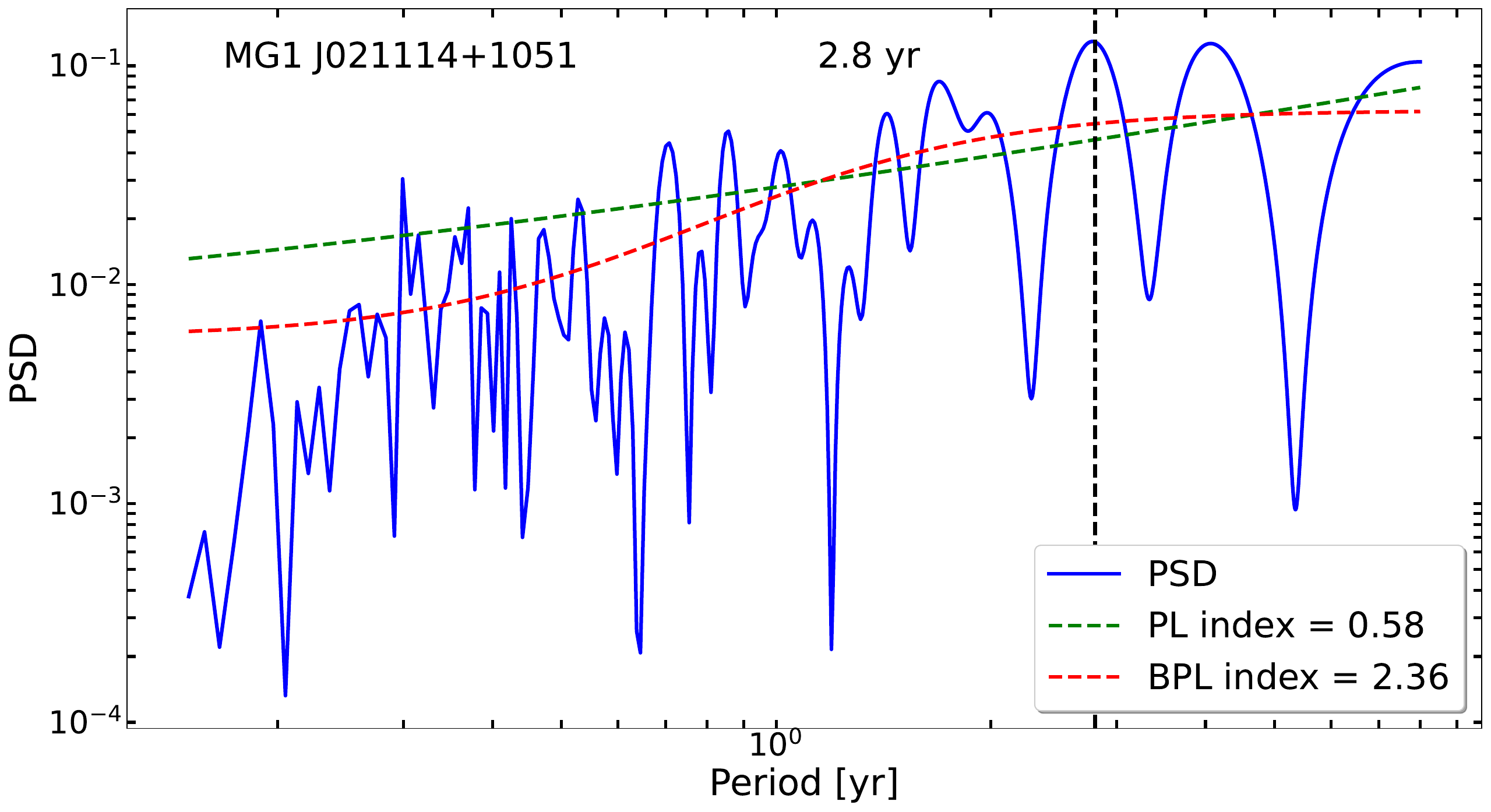}
        \includegraphics[scale=0.185]{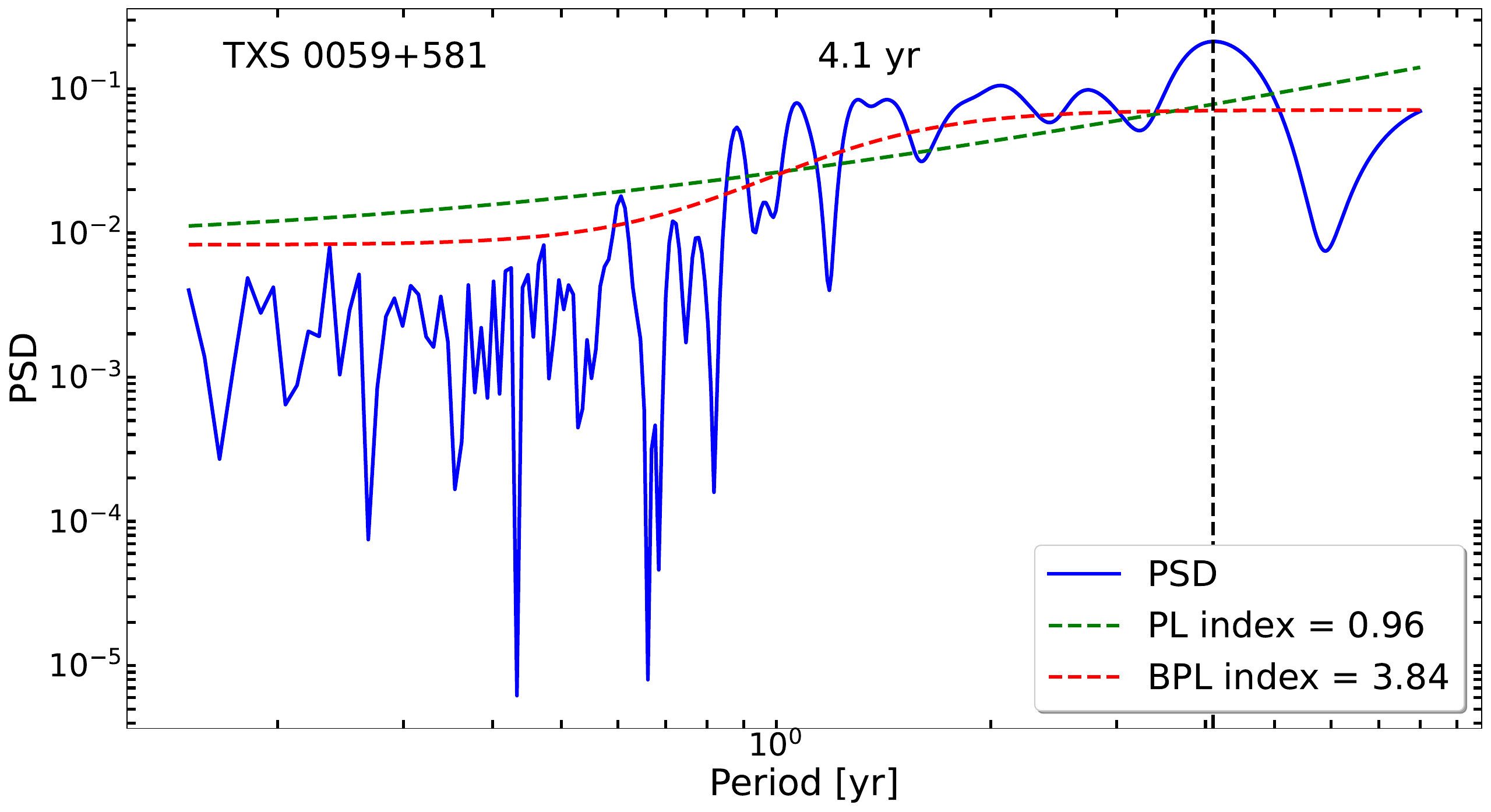}
        \includegraphics[scale=0.185]{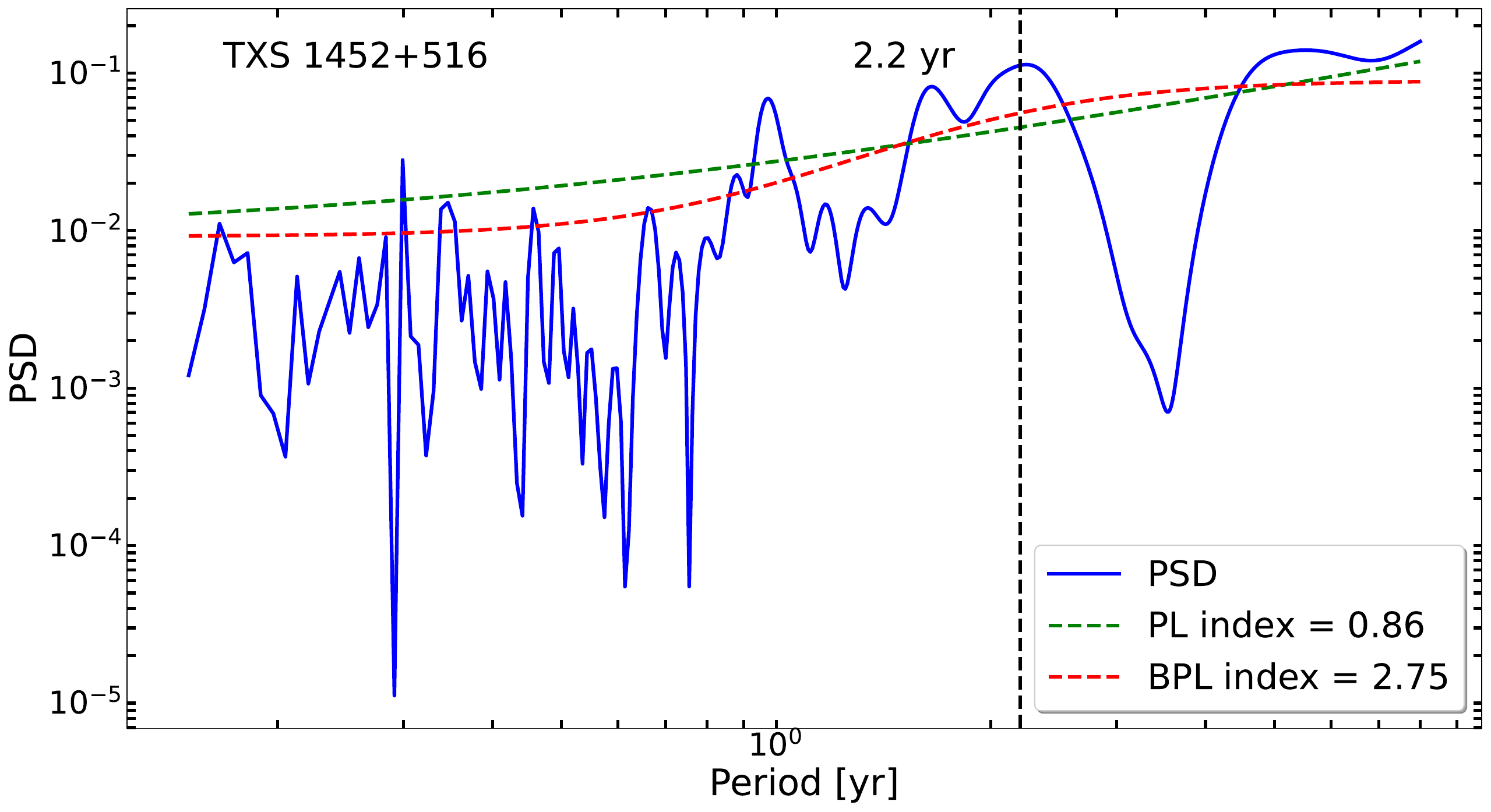}
        \includegraphics[scale=0.185]{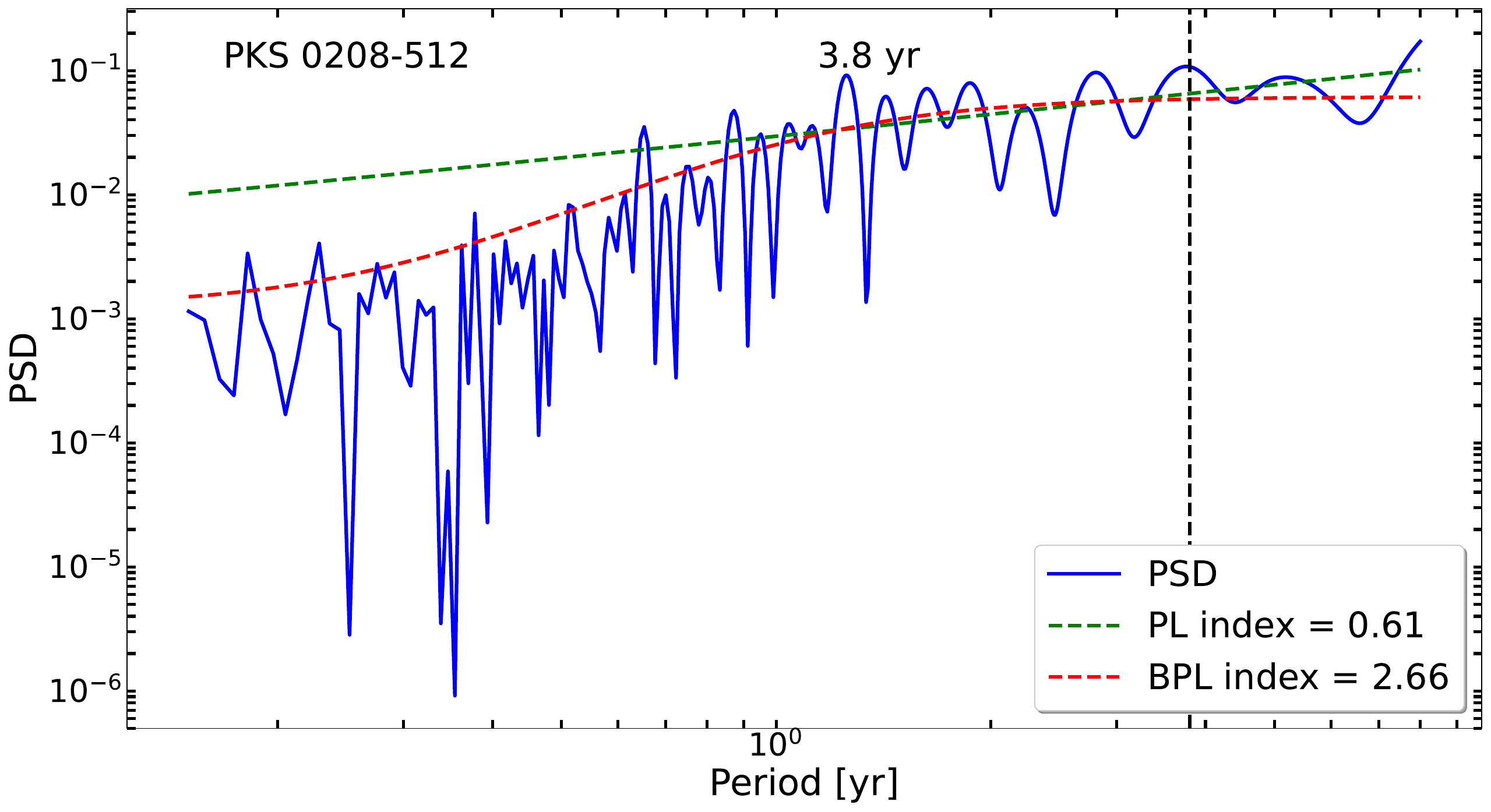}
	\caption{(Continued).}
\end{figure*}
\clearpage

\subsection{Tables}
This section presents Table \ref{tab:slopes_poisson}, Table \ref{tab:correction_power_law_new}, Table \ref{tab:methods_results}, and Table \ref{tab:mcmc_bayesian_arfima_results}, which summarize the various estimations used to determine the periods of the selected blazars.

\begin{table*}
\centering
\caption{
	List of periods and uncertainties (top) with their associated test statistics (bottom) for the periodic-emission candidates following a similar structure as Table~\ref{tab:candidates_list}. Note that there are some sources with two periods with high test statistics (organized by the amplitude of the peak), which are denoted by $\star$. The symbol $\dagger$ denotes the PDM results that present the harmonic effect described in \citet{penil_2020}. The symbol $\ddagger$ denotes periods that were presented in the CWT for 12 years of LC data. All periods are in years. \label{tab:methods_results}}
{%
\begin{tabular}{l|cccccccccc}
\hline
\hline
Association Name & LSP & GLSP & PDM & CWT \\	
\hline
PG 1553+113 & $2.1^{\pm0.2}_{5.1\sigma}$ & $2.1^{\pm0.2}_{4.6\sigma}$ & $2.2^{\pm0.1}_{4.2\sigma}$& $\ddagger2.2^{\pm0.4}_{4.4\sigma}$  \\
        PKS 2155$-$304 & $1.7^{\pm0.1}_{4.1\sigma}$ & $1.7^{\pm0.1}_{3.3\sigma}$ & $\dagger3.4^{\pm0.2}_{3.1\sigma}$ &$\ddagger1.7^{\pm0.3}_{3.3\sigma}$  \\ 
	\hline
	\hline	
        OJ 014 & $4.2^{\pm0.5}_{3.2\sigma}$ & $4.1^{\pm0.4}_{2.6\sigma}$ & $4.1^{\pm0.1}_{2.6\sigma}$ & $\ddagger4.4^{\pm0.9}_{3.6\sigma}$ \\ 	
	PKS 0454$-$234 & $3.6^{\pm0.2}_{3.2\sigma}$ & $3.5^{\pm0.4}_{2.8\sigma}$ & $3.5^{\pm0.4}_{2.4\sigma}$ & $\ddagger3.9^{\pm0.8}_{2.8\sigma}$ \\		
        S5 0716+714$\star$ & \makecell{$2.7^{\pm0.4}_{3.5\sigma}$ \\ $0.9^{\pm0.1}_{3.0\sigma}$} & $2.6^{\pm0.4}_{2.8\sigma}$ & $2.9^{\pm0.3}_{2.4\sigma}$ & \makecell{$\ddagger2.9^{\pm0.8}_{2.8\sigma}$ \\ $0.9^{\pm0.1}_{1.3\sigma}$} \\	
        GB6 J0043+3426 & $1.9^{\pm0.2}_{3.3\sigma}$ & $1.9^{\pm0.4}_{2.6\sigma}$ & $\dagger3.9^{\pm0.2}_{2.3\sigma}$ & $2.1^{\pm0.6}_{2.7\sigma}$ \\
        TXS 0518+211 & $3.0^{\pm0.4}_{3.0\sigma}$ & $3.1^{\pm0.4}_{2.4\sigma}$ & $3.1^{\pm0.2}_{2.4\sigma}$ & $\ddagger3.2^{\pm0.5}_{2.7\sigma}$ \\	
        87GB 164812.2+524 & $2.8^{\pm0.9}_{2.2\sigma}$ & $3.4^{\pm0.4}_{1.8\sigma}$ & $3.0^{\pm0.1}_{2.2\sigma}$ & $\ddagger3.8^{\pm0.8}_{2.6\sigma}$ \\
        PKS 0447$-$439 & $1.9^{\pm0.2}_{2.6\sigma}$ & $1.8^{\pm0.2}_{2.0\sigma}$ & $\dagger4.0^{\pm0.4}_{2.0\sigma}$ & $2.1^{\pm0.5}_{2.2\sigma}$ \\
        PKS 0426$-$380 & $3.5^{\pm0.6}_{2.6\sigma}$ & $3.6^{\pm0.5}_{2.0\sigma}$ & $3.4^{\pm0.5}_{1.7\sigma}$ & $\ddagger3.4^{\pm0.6}_{2.1\sigma}$ \\
        PKS 0301$-$243 & $2.1^{\pm0.2}_{2.7\sigma}$ & $2.1^{\pm0.2}_{1.9\sigma}$ & $\dagger4.0^{\pm0.1}_{1.3\sigma}$ & $2.2^{\pm0.2}_{2.1\sigma}$ \\
        S4 1144+40 & $3.3^{\pm0.2}_{2.7\sigma}$ & $3.3^{\pm0.5}_{1.5\sigma}$ & $3.2^{\pm0.5}_{2.0\sigma}$ & $\ddagger3.4^{\pm0.9}_{1.7\sigma}$ \\ 
        PG 1246+586 & $2.1^{\pm0.2}_{2.5\sigma}$  & $2.2^{\pm0.1}_{1.5\sigma}$ &  $\dagger4.2^{\pm0.1}_{1.1\sigma}$ & $\ddagger2.2^{\pm0.3}_{1.9\sigma}$ \\        
        PKS 0250$-$225  & $1.2^{\pm0.2}_{3.0\sigma}$ & $1.2^{\pm0.1}_{1.6\sigma}$ & $\dagger2.4^{\pm0.1}_{1.4\sigma}$ & $1.3^{\pm0.2}_{1.7\sigma}$ \\
        PKS 2255$-$282$\star$ & \makecell{$1.4^{\pm0.1}_{2.2\sigma}$ \\ $2.8^{\pm0.3}_{1.8\sigma}$} & \makecell {$2.8^{\pm0.4}_{1.8\sigma}$ \\ $1.4^{\pm0.1}_{1.0\sigma}$} & $2.7^{\pm0.2}_{1.7\sigma}$ & \makecell {$\ddagger3.7^{\pm0.8}_{1.7\sigma}$ \\ $1.4^{\pm0.3}_{1.1\sigma}$} \\
	TXS 1902+556 & $3.3^{\pm0.4}_{2.5\sigma}$ & $3.3^{\pm0.3}_{1.4\sigma}$ & $3.3^{\pm0.3}_{1.3\sigma}$ & $\ddagger3.4^{\pm0.7}_{1.6\sigma}$ \\	   
        S3 0458$-$02  & $3.7^{\pm0.9}_{1.8\sigma}$ & $4.1^{\pm0.6}_{1.2\sigma}$ & $3.8^{\pm0.3}_{1.3\sigma}$ & $4.4^{\pm0.7}_{1.5\sigma}$ \\
        MG2 J130304+2434 & $2.0^{\pm0.2}_{2.0\sigma}$ & $4.1^{\pm0.1}_{1.0\sigma}$ & $\dagger4.0^{\pm0.1}_{1.2\sigma}$ & $4.1^{\pm0.6}_{1.1\sigma}$ \\        
   	PKS 2052$-$47$\star$ & $1.6^{\pm0.2}_{2.5\sigma}$ & \makecell {$3.1^{\pm0.3}_{1.1\sigma}$ \\ $1.4^{\pm0.2}_{2.0\sigma}$} & $2.8^{\pm0.4}_{1.0\sigma}$ & $3.3^{\pm0.3}_{1.2\sigma}$ \\        	
        S4 0814+42 & $2.2^{\pm0.2}_{2.0\sigma}$ & $2.2^{\pm0.2}_{0.8\sigma}$ & $2.2^{\pm0.3}_{0.8\sigma}$ & $\ddagger2.3^{\pm0.3}_{0.8\sigma}$\\ 
	MG1 J021114+1051 & $2.8^{\pm0.3}_{2.0\sigma}$ & $1.7^{\pm0.1}_{0.8\sigma}$ & $2.8^{\pm0.2}_{0.8\sigma}$ & $\ddagger4.6^{\pm0.7}_{0.9\sigma}$ \\        	
	TXS 0059+581 & $4.0^{\pm0.6}_{1.7\sigma}$ & $4.1^{\pm0.6}_{0.7\sigma}$ & $4.2^{\pm0.1}_{0.8\sigma}$ & $\ddagger4.6^{\pm0.7}_{0.7\sigma}$ \\
        TXS 1452+516 & $2.0^{\pm0.4}_{1.4\sigma}$ & $2.0^{\pm0.2}_{0.7\sigma}$ & $\dagger4.3^{\pm0.4}_{0.5\sigma}$ & $2.1^{\pm0.6}_{0.7\sigma}$ \\ 
	PKS 0208$-$512 & $3.8^{\pm0.5}_{1.1\sigma}$ & $3.7^{\pm0.4}_{0.1\sigma}$ & $3.8^{\pm0.5}_{0.1\sigma}$ & $\ddagger4.6^{\pm0.9}_{0.2\sigma}$ \\
\hline
\hline
\end{tabular}%
}
\end{table*}
\begin{table*}
\centering
\caption{List of power-spectrum indices inferred by $ML-MCMC$ method for the PL and BPL of the PSD, considering the Poisson noise expressed by Eq. \ref{eq:p}.  The parameter ``A'' is the normalization ($rms^{2}/yr^{-1}$) and $\nu_{Bending}$ the bending frequency. The Poisson noise is represented by ``C'' ($rms^{2}/yr^{-1}$). The AIC values for both PSD models, as well as the RML for model comparison, are provided. \label{tab:slopes_poisson}}
{%
\begin{tabular}{l|cccccccccc}
\hline
\hline
Association Name & C & A / PL Index & AIC & A / BPL Index / $\nu_{Bending}$ & AIC & RML \\
\hline
PG 1553+113 & $2.97\mathrm{x}10^{-3}$ & \makecell{1.91$\mathrm{x}10^{-2}$$\pm$9.15$\mathrm{x}10^{-3}$\\1.25$\pm$0.13} & -1453.90 & \makecell{1.93$\mathrm{x}10^{-2}$$\pm$7.84$\mathrm{x}10^{-3}$ \\ 1.83$\pm$0.49 \\ 1.98$\pm$0.08} & -1519.19 & 6.62$\mathrm{x}10^{-15}$ \\
        PKS 2155$-$304 & $1.62\mathrm{x}10^{-3}$ & \makecell{1.87$\mathrm{x}10^{-2}$$\pm$8.59$\mathrm{x}10^{-3}$\\0.88$\pm$0.11} & -1328.91 & \makecell{4.75$\mathrm{x}10^{-2}$$\pm$1.03$\mathrm{x}10^{-2}$ \\ 0.88$\pm$0.09 \\ 1.19$\pm$0.22} & -1488.91 & 1.83$\mathrm{x}10^{-35}$ \\
        \hline
        \hline	
 OJ 014 & $3.42\mathrm{x}10^{-3}$ & \makecell{2.02$\mathrm{x}10^{-2}$$\pm$8.26$\mathrm{x}10^{-3}$\\1.07$\pm$0.09} & -1397.74  & \makecell{1.10$\mathrm{x}10^{-1}$$\pm$1.63$\mathrm{x}10^{-2}$ \\ 2.67$\pm$0.27 \\ 0.81$\pm$0.29} & -1690.12 & 3.23$\mathrm{x}10^{-64}$ \\	
	PKS 0454$-$234 & $7.37\mathrm{x}10^{-4}$ & \makecell{2.11$\mathrm{x}10^{-2}$$\pm$6.52$\mathrm{x}10^{-3}$\\1.16$\pm$0.07} & -1701.75 & \makecell{1.25$\mathrm{x}10^{-1}$$\pm$0.13$\mathrm{x}10^{-2}$ \\ 1.82$\pm$0.09 \\ 0.71$\pm$0.06} & -1790.97 & 4.22$\mathrm{x}10^{-20}$ \\	 
        S5 0716+714 & $5.97\mathrm{x}10^{-4}$ & \makecell{1.91$\mathrm{x}10^{-2}$$\pm$7.50$\mathrm{x}10^{-3}$\\0.78$\pm$0.08} & -1905.83 & \makecell{5.74$\mathrm{x}10^{-2}$$\pm$5.60$\mathrm{x}10^{-3}$ \\ 1.38$\pm$0.16 \\ 1.26$\pm$0.18} & -2009.53 & 3.02$\mathrm{x}10^{-23}$ \\ 
        GB6 J0043+3426 & $1.83\mathrm{x}10^{-2}$ & \makecell{$1.82\mathrm{x}10^{-2}$$\pm$6.61$\mathrm{x}10^{-3}$ \\1.01$\pm$0.11} & -2282.27 & \makecell{3.55$\mathrm{x}10^{-2}$$\pm$9.92$\mathrm{x}10^{-3}$ \\ 1.86$\pm$0.26 \\ 1.97$\pm$0.39} & -2384.97 & 5.01$\mathrm{x}10^{-23}$ \\ 
        TXS 0518+211 & $2.35\mathrm{x}10^{-3}$ & \makecell{2.13$\mathrm{x}10^{-2}$$\pm$1.25$\mathrm{x}10^{-3}$\\1.37$\pm$0.11} & -2221.78 & \makecell{1.14$\mathrm{x}10^{-1}$$\pm$2.93$\mathrm{x}10^{-2}$ \\ 2.76$\pm$0.18 \\ 0.82$\pm$0.09} & -2253.26 & 1.45$\mathrm{x}10^{-7}$ \\
        87GB 164812.2+524 & $1.53\mathrm{x}10^{-2}$ & \makecell{2.63$\mathrm{x}10^{-2}$$\pm$7.77$\mathrm{x}10^{-3}$\\0.42$\pm$0.07} & -1993.41 & \makecell{4.62$\mathrm{x}10^{-2}$$\pm$7.89$\mathrm{x}10^{-3}$ \\ 5.34$\pm$0.41 \\ 0.53$\pm$0.02} & -2042.38 & 2.32$\mathrm{x}10^{-11}$ \\
        PKS 0447$-$439 & $1.25\mathrm{x}10^{-3}$ & \makecell{2.78$\mathrm{x}10^{-2}$$\pm$8.15$\mathrm{x}10^{-3}$\\0.62$\pm$0.09} & -2349.50 & \makecell{4.55$\mathrm{x}10^{-2}$$\pm$7.14$\mathrm{x}10^{-3}$ \\ 2.21$\pm$0.21 \\ 1.62$\pm$0.18} & -2520.38 & 7.83$\mathrm{x}10^{-38}$ \\
        PKS 0426$-$380 & $8.33\mathrm{x}10^{-4}$ & \makecell{2.21$\mathrm{x}10^{-2}$$\pm$7.06$\mathrm{x}10^{-3}$\\1.28$\pm$0.12} & -1787.17 & \makecell{6.10$\mathrm{x}10^{-1}$$\pm$1.56$\mathrm{x}10^{-2}$ \\ 1.51$\pm$0.13 \\ 0.11$\pm$0.07} & -1921.43 & 7.03$\mathrm{x}10^{-30}$ \\
        PKS 0301$-$243 & $3.97\mathrm{x}10^{-3}$ & \makecell{1.73$\mathrm{x}10^{-2}$$\pm$6.78$\mathrm{x}10^{-3}$\\1.08$\pm$0.20} & -2411.60 & \makecell{3.12$\mathrm{x}10^{-2}$$\pm$8.13$\mathrm{x}10^{-3}$ \\ 2.72$\pm$0.34 \\ 1.47$\pm$0.15} & -2567.70 & 1.26$\mathrm{x}10^{-34}$ \\
        S4 1144+40 & $3.33\mathrm{x}10^{-3}$ & \makecell{2.07$\mathrm{x}10^{-2}$$\pm$6.48$\mathrm{x}10^{-3}$\\1.58$\pm$0.13} & -1516.63 & \makecell{1.13$\mathrm{x}10^{-1}$$\pm$2.38$\mathrm{x}10^{-2}$ \\ 3.41$\pm$0.28 \\ 0.53$\pm$0.05} & -1839.98 & 6.11$\mathrm{x}10^{-71}$ \\
        PG 1246+586 & $5.39\mathrm{x}10^{-3}$ & \makecell{1.75$\mathrm{x}10^{-2}$$\pm$6.13$\mathrm{x}10^{-3}$ \\ 1.09$\pm$0.05} & -1986.96 & \makecell{1.25$\mathrm{x}10^{-1}$$\pm$1.58$\mathrm{x}10^{-2}$ \\ 1.76$\pm$0.13 \\ 0.39$\pm$0.02} & -1998.73 & 2.78$\mathrm{x}10^{-3}$ \\
        PKS 2255$-$282 & $5.44\mathrm{x}10^{-3}$ & \makecell{2.34$\mathrm{x}10^{-2}$$\pm$7.14$\mathrm{x}10^{-3}$ \\0.97$\pm$0.08} & -1868.80 & \makecell{1.56$\mathrm{x}10^{-1}$$\pm$1.53$\mathrm{x}10^{-2}$ \\ 2.13$\pm$0.22 \\ 0.39$\pm$0.04} & -1895.91 & 1.29$\mathrm{x}10^{-6}$ \\
        PKS 0250$-$225 & $2.57\mathrm{x}10^{-3}$ & \makecell{2.63$\mathrm{x}10^{-2}$$\pm$6.13$\mathrm{x}10^{-3}$ \\ 0.88$\pm$0.087} & -2363.85 & \makecell{3.20$\mathrm{x}10^{-2}$$\pm$1.21$\mathrm{x}10^{-3}$ \\ 4.24$\pm$0.37 \\ 1.74$\pm$0.26} & -2601.55 & 2.43$\mathrm{x}10^{-52}$ \\
        TXS 1902+556 & $4.46\mathrm{x}10^{-3}$ & \makecell{1.69$\mathrm{x}10^{-2}$$\pm$7.18$\mathrm{x}10^{-3}$\\0.86$\pm$0.07} & -2130.05 & \makecell{1.25$\mathrm{x}10^{-1}$$\pm$2.73$\mathrm{x}10^{-2}$ \\ 1.47$\pm$0.08 \\ 0.31$\pm$0.01} & -2252.84 & 2.16$\mathrm{x}10^{-27}$ \\
\hline
\hline
\end{tabular}%
}
\end{table*}
\setcounter{table}{1}

\begin{table*}
\centering
\caption{(continued).}
{%
\begin{tabular}{l|ccccccccccc}
\hline
\hline
Association Name & C & A / PL Index & AIC & A / BPL Index / $\nu_{Bending}$ & AIC & RML \\
\hline
S3 0458$-$02 & $3.51\mathrm{x}10^{-3}$ & \makecell{2.35$\mathrm{x}10^{-2}$$\pm$1.28$\mathrm{x}10^{-3}$\\0.91$\pm$0.09} & -2221.87 & \makecell{1.56$\mathrm{x}10^{-1}$$\pm$5.74$\mathrm{x}10^{-2}$ \\ 1.17$\pm$0.10 \\ 0.21$\pm$0.09} & -2253.30 & 1.49$\mathrm{x}10^{-7}$ \\
        MG2 J130304+2434 & $1.10\mathrm{x}10^{-2}$ & \makecell{1.82$\mathrm{x}10^{-2}$$\pm$5.21$\mathrm{x}10^{-3}$\\0.86$\pm$0.13} & -2302.85 & \makecell{5.67$\mathrm{x}10^{-2}$$\pm$1.45$\mathrm{x}10^{-3}$ \\ 2.64$\pm$0.15 \\ 0.74$\pm$0.02} & -2388.63 & 2.36$\mathrm{x}10^{-19}$ \\
        PKS 2052$-$47  & $2.08\mathrm{x}10^{-3}$ & \makecell{1.63$\mathrm{x}10^{-2}$$\pm$6.56$\mathrm{x}10^{-3}$\\0.93$\pm$0.05} & -2282.80 & \makecell{1.75$\mathrm{x}10^{-1}$$\pm$9.33$\mathrm{x}10^{-2}$ \\ 1.19$\pm$0.12 \\ 0.14$\pm$0.07} & -2296.20 & 1.23$\mathrm{x}10^{-3}$ \\
        S4 0814+42 & $3.14\mathrm{x}10^{-3}$ & \makecell{2.44$\mathrm{x}10^{-2}$$\pm$2.11$\mathrm{x}10^{-3}$\\0.71$\pm$0.05} & -2157.11 & \makecell{5.58$\mathrm{x}10^{-2}$$\pm$2.11$\mathrm{x}10^{-3}$ \\ 3.17$\pm$0.37 \\ 0.81$\pm$0.05} & -2217.64 & 7.17$\mathrm{x}10^{-14}$ \\
        MG1 J021114+1051 & $5.78\mathrm{x}10^{-4}$ & \makecell{2.21$\mathrm{x}10^{-2}$$\pm$5.55$\mathrm{x}10^{-3}$\\0.58$\pm$0.06} & -2245.71 & \makecell{5.68$\mathrm{x}10^{-2}$$\pm$1.28$\mathrm{x}10^{-3}$ \\ 2.36$\pm$0.25 \\ 0.76$\pm$0.07} & -2336.78 & 1.67$\mathrm{x}10^{-20}$ \\      
        TXS 0059+581 & $8.48\mathrm{x}10^{-3}$ & \makecell{1.82$\mathrm{x}10^{-2}$$\pm$1.28$\mathrm{x}10^{-3}$\\0.96$\pm$0.07} & -2056.70 & \makecell{6.53$\mathrm{x}10^{-2}$$\pm$6.90$\mathrm{x}10^{-3}$ \\ 3.84$\pm$0.28 \\ 0.77$\pm$0.09} & -2110.71 & 1.87$\mathrm{x}10^{-12}$ \\
        TXS 1452+516 & $9.15\mathrm{x}10^{-3}$  & \makecell{1.83$\mathrm{x}10^{-2}$$\pm$1.23$\mathrm{x}10^{-3}$\\0.86$\pm$0.05} & -2102.64 & \makecell{8.06$\mathrm{x}10^{-2}$$\pm$5.38$\mathrm{x}10^{-3}$ \\ 2.75$\pm$0.21 \\ 0.51$\pm$0.02} & -2132.03 & 4.16$\mathrm{x}10^{-7}$ \\
        PKS 0208$-$512 & $6.14\mathrm{x}10^{-4}$ & \makecell{2.83$\mathrm{x}10^{-2}$$\pm$1.45$\mathrm{x}10^{-3}$\\0.61$\pm$0.06} & -2431.58 & \makecell{6.03$\mathrm{x}10^{-2}$$\pm$1.34$\mathrm{x}10^{-3}$ \\ 2.66$\pm$0.17 \\ 0.86$\pm$0.05} & -2617.92 & 3.44$\mathrm{x}10^{-41}$ \\ 
\hline
\hline
\end{tabular}%
}
\end{table*}
\setcounter{table}{2}

\begin{table*}
\centering
\caption{Results of the influence on the test statistics of the power-law approach. Three different models are studied for a power law, considering the slopes and uncertainties provided in Table \ref{tab:slopes_poisson}: [$\beta_{min}$, $A_{min}$], [$\beta$, $A$], [$\beta_{max}$, $A_{max}$]. \label{tab:correction_power_law_new}} 
{%
\begin{tabular}{l|cccccccccc}
\hline
\hline
Association Name & Model & LSP & GLSP & PDM & CWT \\
\hline
\multirow{2}{*}{PG 1553+113} & \makecell{\makecell{$\beta_{min}$=1.12 \\ $A_{min}$=$9.98\mathrm{x}10^{-3}$ \\ \\ } \\ \makecell{$\beta$=1.25 \\ $A$=$1.91\mathrm{x}10^{-2}$ \\ \\ } \\ \makecell{$\beta_{max}$=1.39 \\ $A_{max}$=$2.28\mathrm{x}10^{-2}$ \\ \\ } }
            &\makecell{4.4$\sigma$ \\ \\ \\ 3.7$\sigma$ \\ \\ \\ 3.1$\sigma$ \\ \\ \\ }
            &\makecell{4.0$\sigma$ \\ \\ \\ 3.5$\sigma$ \\ \\ \\ 3.0$\sigma$ \\ \\ \\ }
            &\makecell{3.5$\sigma$ \\ \\ \\ 3.2$\sigma$ \\ \\ \\ 2.8$\sigma$ \\ \\ \\ }
            &\makecell{3.2$\sigma$ \\ \\ \\ 2.7$\sigma$ \\ \\ \\ 2.5$\sigma$ \\ \\ \\ }
    \\
    \hline
    \multirow{2}{*}{PKS 2155$-$304} & \makecell{\makecell{$\beta_{min}$=0.76 \\ $A_{min}$=$1.01\mathrm{x}10^{-2}$ \\ \\ } \\ \makecell{$\beta$=0.88 \\ $A$=$1.87\mathrm{x}10^{-2}$ \\ \\ } \\ \makecell{$\beta_{max}$=0.99 \\ $A_{max}$=$2.73\mathrm{x}10^{-2}$ \\ \\ } \\}
            &\makecell{2.8$\sigma$ \\ \\ \\ 2.5$\sigma$ \\ \\ \\ 2.3$\sigma$ \\ \\ \\ }
            &\makecell{2.8$\sigma$ \\ \\ \\ 2.5$\sigma$ \\ \\ \\ 2.2$\sigma$ \\ \\ \\ }
            &\makecell{2.1$\sigma$ \\ \\ \\ 1.7$\sigma$ \\ \\ \\ 1.2$\sigma$ \\ \\ \\ }
            &\makecell{2.0$\sigma$ \\ \\ \\ 1.7$\sigma$ \\ \\ \\ 1.5$\sigma$ \\ \\ \\ } \\
\hline
\hline
\end{tabular}%
}
\end{table*}

\begin{table*}
\centering
\caption{Results of the influence on the test statistics of the bending power-law approach. Three different models are studied for the bending power law, according to the slope, the bending frequency, and the normalization, according to the associated uncertainties: [$\alpha_{min}$, $\nu_{Bending\_min}$, $A_{min}$], [$\alpha$, $\nu_{Bending}$, $A$], [$\alpha_{max}$, $\nu_{Bending\_max}$, $A_{max}$]. The symbol $\#$ indicates that the test statistic is constrained by the limited number of artificial LCs generated for the analysis.\label{tab:correction_bending_power_law_new}}
{%
\begin{tabular}{l|cccccccccc}
\hline
\hline
Association Name & Model & LSP & GLSP & PDM & CWT \\
\hline
\multirow{2}{*}{PG 1553+113} & \makecell{\makecell{$\alpha_{min}$=1.34 $\nu_{Bending\_min}$=1.89 \\ $A_{min}$=$1.15\mathrm{x}10^{-2}$ \\ \\ }\\ \makecell{$\alpha$=1.83 $\nu_{Bending}$=1.98 \\ $A$=$1.93\mathrm{x}10^{-2}$ \\ \\ } \\ \makecell{$\alpha_{max}$=2.32 $\nu_{Bending\_max}$=2.07 \\ $A_{max}$=$2.72\mathrm{x}10^{-2}$ \\ \\ } \\} &
        \makecell{$\#$4.9$\sigma$ \\ \\ \\ $\#$4.9$\sigma$ \\ \\ \\ 4.8$\sigma$ \\ \\ \\ } &
        \makecell{$\#$4.9$\sigma$ \\ \\ \\ 4.6$\sigma$ \\ \\ \\ 4.6$\sigma$ \\ \\ \\ } &
        \makecell{$\#$4.5$\sigma$ \\ \\ \\ 4.2$\sigma$ \\ \\ \\ 4.0$\sigma$ \\ \\ \\ } &
        \makecell{4.6$\sigma$ \\ \\ \\ 4.4$\sigma$ \\ \\ \\ 3.9$\sigma$ \\ \\ \\ } &
        \\
    \hline
    \multirow{2}{*}{PKS 2155$-$304} & \makecell{\makecell{$\alpha_{min}$=0.79 $\nu_{Bending\_min}$=0.96 \\ $A_{min}$=$4.65\mathrm{x}10^{-2}$ \\ \\ } \\ \makecell{$\alpha$=0.88 $\nu_{Bending}$=1.19 \\ $A$=$4.75\mathrm{x}10^{-2}$ \\ \\ } \\ \makecell{$\alpha_{max}$=0.98 $\nu_{Bending\_max}$=1.41 \\ $A_{max}$=$4.85\mathrm{x}10^{-2}$ \\ \\ } \\} &
        \makecell{2.9$\sigma$ \\ \\ \\ 2.9$\sigma$ \\ \\ \\ 2.8$\sigma$ \\ \\ \\ } &
        \makecell{3.4$\sigma$ \\ \\ \\ 3.3$\sigma$ \\ \\ \\ 3.1$\sigma$ \\ \\ \\ } &
        \makecell{2.9$\sigma$ \\ \\ \\ 3.1$\sigma$ \\ \\ \\ 3.0$\sigma$ \\ \\ \\ } &
        \makecell{3.5$\sigma$ \\ \\ \\ 3.3$\sigma$ \\ \\ \\ 3.3$\sigma$ \\ \\ \\ } &
        \\
\hline
\hline
\end{tabular}%
}
\end{table*}

\begin{table*}
\centering
\caption{List of periods provided by the Markov Chain Monte Carlo sine fits and the Bayesian quasi-periodic oscillation methods following a similar structure as Table~\ref{tab:candidates_list}. The \myhash~ in ARFIMA/ARIMA column indicates that the model used is ARIMA due to the LC being non-stationary (resulting from augmented Dickey-Fuller test, see $\S$\ref{sec:arima}). The period obtained with the ACF from the residuals generated from the original LC and the ARIMA/ARFIMA model is also listed. Note that there are some sources with two periods (organized by the test statistics), which are denoted by $\star$. Finally, Xs in the Dickey-Fuller and Box-Ljung columns indicate that the null hypothesis is rejected in such tests ($\S$\ref{sec:arima}). All periods are in years. \label{tab:mcmc_bayesian_arfima_results}} 
{%
\begin{tabular}{l|cccccccccc}
\hline
\hline
Association Name & MCMC & ARFIMA/ARIMA & ACF & Dickey-Fuller & Box-Ljung	\\
  & Sine Fitting & Residuals &  &  & \\
\hline
PG 1553+113$\star$ & 2.1 $\pm$ 0.1 & [17, 0.449, 18] &  \makecell{2.8 (2.0$\sigma$) \\ 2.0 (1.5$\sigma$)} & \redcheck & \redcheck \\
PKS 2155$-$304$\star$ & 1.7 $\pm$ 0.1 &  [10, 0.435, 10] & \makecell{1.6 (2.1$\sigma$) \\ 1.0 (1.6$\sigma$)} & \redcheck & \redcheck \\	
\hline
\hline
OJ 014 & 4.4 $\pm$ 0.1 & [19, 0.471, 15] & 3.8 (1.6$\sigma$) & \redcheck & \redcheck \\
PKS 0454$-$234$\star$ & 3.5 $\pm$ 0.1 & [20, 0.493, 20] & \makecell{2.0 (2.0$\sigma$) \\ 1.2 (1.6$\sigma$)} & \redcheck & \redcheck \\
S5 0716+714 & 2.7 $\pm$ 0.1 & [12, 0.447, 14] & 1.2 (1.6$\sigma$) & \redcheck & \redcheck \\
GB6 J0043+3426 & 1.8 $\pm$ 0.1 & [9, 0.472, 10]  & 2.0 (2.1$\sigma$) & \redcheck & \redcheck \\
TXS 0518+211 & $3.1^{+0.1}_{-0.8}$ & [15, 0.333, 13] & 2.9 (1.9$\sigma$) & \redcheck & \redcheck \\
87GB 164812.2+524023 & $2.9^{+0.8}_{-0.1}$ & [18, 0.445, 20] & 2.0 (1.2$\sigma$) & \redcheck & \redcheck \\
PKS 0447$-$439$\star$ & $1.8^{+0.1}_{-0.6}$ &  [15, 0.45, 14] & \makecell{1.5 (2.2$\sigma$) \\ 2.5 (1.5$\sigma$)}& \redcheck & \redcheck \\
PKS 0426$-$380$\star$ & 3.2 $\pm$ 0.1 & [14, 0.47, 14] &  \makecell{3.4 (2.3$\sigma$) \\ 2.2 (1.9$\sigma$)} & \redcheck & \redcheck \\
PKS 0301$-$243 & $1.4^{+0.8}_{-0.1}$ & [17, 0.451, 20] & 1 (1.7$\sigma$) & \redcheck & \redcheck \\
S4 1144+40 & 3.5 $\pm$ 0.1 & [18, 0.433, 19] & 2.5 (2.0$\sigma$) & \redcheck &  \redcheck \\
PG 1246+586 & 2.2 $\pm$ 0.1 & [16, 0.436, 20] & 3.5 (2.8$\sigma$) & \redcheck & \redcheck \\
PKS 0250$-$225 & 1.2 $\pm$ 0.1 &  \myhash[5, 1, 8] & 1.3 (1.5$\sigma$) & X & \redcheck \\
PKS 2255$-$282 & 3.1 $\pm$ 0.1 & [18, 0.441, 19] & 2.8 (1.6$\sigma$) & \redcheck & \redcheck \\     
TXS 1902+556 & 3.3 $\pm$ 0.1 & [20, 0.41, 15] & 3.4 (1.6$\sigma$) & \redcheck & \redcheck\\
S3 0458$-$02$\star$ & 1.5 $\pm$ 0.2 & [10, 0.453, 8] & \makecell{3.8 (2.2$\sigma$) \\ 2.1 (2.0$\sigma$)} & \redcheck & X \\
MG2 J130304+2434 & 1.2 $\pm$ 0.1 & [19, 0.471, 20] & 2.0 (2.1$\sigma$) & \redcheck & \redcheck \\ 
PKS 2052$-$47 & 2.6 $\pm$ 0.1 & [18, 0.441, 19] & 1.6 (1.9$\sigma$) & \redcheck & \redcheck \\                      
S4 0814+42 & 2.1 $\pm$ 0.1 & [14, 0.451, 14] & 3.1 (2.2$\sigma$) & \redcheck & \redcheck \\     
MG1 J021114+1051 & 1.7 $\pm$ 0.1 & [15, 0.414, 14] & 2.6 (1.95$\sigma$) & \redcheck & \redcheck \\        	
TXS 0059+581 & 4.1 $\pm$ 0.1 & [16, 0.391, 18] & 3.9 (1.4$\sigma$) & \redcheck & \redcheck \\		
TXS 1452+516 & 2.1 $\pm$ 0.1 &  [19, 0.448, 18] & 1.7 (1.9$\sigma$) & \redcheck & \redcheck \\   
PKS 0208$-$512 & $4.0^{+0.5}_{-1.0}$ & \myhash[2, 1, 8] & 1.2 (1.0$\sigma$) & X & X \\	
\hline
\hline
\end{tabular}%
}
\end{table*}

\bsp	
\label{lastpage}
\end{document}